\documentclass[
    reprint,
    preprintnumbers,
    superscriptaddress,
    nofootinbib,
     amsmath,amssymb,
     aps,
     prd,
    floatfix,
    ]{revtex4-2}
    
    \usepackage{graphicx}
    \usepackage[caption=false]{subfig}
    \usepackage[usenames,dvipsnames]{xcolor} 
    \usepackage{times, mathptmx} 
    \usepackage[utf8]{inputenc}
    \usepackage{color}
    \usepackage{hyperref}
    \usepackage[normalem]{ulem} 
    \usepackage{physics}
    \usepackage{comment}
    \usepackage{bm}

    \allowdisplaybreaks[1]
    \graphicspath{{figures/}}

\usepackage{float}

\newcommand{\eV}{\textup{eV}}
\newcommand{\pc}{\textup{pc}}

\newcommand{\mbh}{M_\mathrm{BH}}

\newcommand{\qmul}{Geometry, Analysis and Gravitation, School of Mathematical Sciences, Queen Mary University of London,
Mile End Road, London E1 4NS, United Kingdom}
\newcommand{\oxford}{Astrophysics, University of Oxford, Denys Wilkinson Building, Keble Road, Oxford OX1 3RH, United Kingdom}

\begin{document}

\title{Black hole merger simulations in wave dark matter environments}
\author{Jamie Bamber}
\email{james.bamber@physics.ox.ac.uk}
\affiliation{\oxford}
\author{Josu C. Aurrekoetxea}
\email{josu.aurrekoetxea@physics.ox.ac.uk}
\affiliation{\oxford}
\author{Katy Clough}
\email{k.clough@qmul.ac.uk}
\affiliation{\qmul}
\author{Pedro G. Ferreira}
\email{pedro.ferreira@physics.ox.ac.uk}
\affiliation{\oxford}

\begin{abstract}
The interaction of binary black hole mergers with their environments can be studied using numerical relativity simulations. These start only a short finite time before merger, at which point appropriate initial conditions must be imposed. A key task is therefore to identify the configuration that is appropriate for the binary and its environment at this stage of the evolution. In this work we study the behaviour of wave dark matter around equal mass black hole binaries, finding that there is a preferred, quasi-stationary profile that persists and grows over multiple orbits, in contrast to heavier mass dark matter where any overdensity tends to be dispersed by the binary motion. Whilst different initial configurations converge to the preferred quasi-stationary one after several orbits, unwanted transient oscillations are generated in the process, which may impact on the signal in short simulation runs. We also point out that naively superimposing the matter onto a circular binary results in artificially eccentric orbits due to the matter backreaction, which is an effect of the initial conditions and not a signature of dark matter. We discuss the further work required so that comparison of waveforms obtained with environments to vacuum cases can be done in a meaningful way.
\end{abstract}

\maketitle


\section{Introduction}

The detection of gravitational waves (GWs) from compact binary mergers \cite{LIGOScientific:2014pky,LIGOScientific:2016aoc,LIGOScientific:2017vwq,LIGOScientific:2021djp,VIRGO:2014yos,KAGRA:2020tym,Bailes:2021tot} allows us to constrain their astrophysical properties, which has important implications for populations studies and gives information about their formation and evolution \cite{Schmidt:2020ekt,Barack:2018yly,LIGOScientific:2018jsj,Nishizawa:2016jji,Barausse:2014oca,Kavanagh:2018ggo,Talbot:2017yur,Mapelli:2020vfa,Breivik:2016ddj,Fragione:2018vty,Cardoso:2019rvt}. In principle, GW observations can also provide a window on the environments of such binaries, since any non zero stress-energy tensor will modify the metric in their vicinity, resulting in changes to the character of the inspiral, merger and ringdown parts of the signal. These changes could capture the effects of standard baryonic matter like plasma-filled accretion disks, or dark matter overdensities, with the spatial distribution and physical nature of the matter giving rise to distinctive signatures at each stage \cite{Barausse:2014tra,Cardoso:2019rou,Yunes:2011ws,Yunes:2011ws,Kocsis:2011dr,Eda:2013gg,Macedo:2013qea,Coogan:2021uqv,Cardoso:2020lxx,Bertone:2018krk,Cole:2022ucw,Kavanagh:2020cfn,Hannuksela:2018izj,Baryakhtar:2022hbu,AlvesBatista:2021gzc,Zwick:2021dlg,Cardoso:2019upw}.

In practise, the energy densities required to give significant effects during the inspiral and ringdown parts of the signal are in most cases high relative to the expected astrophysical values \cite{Barausse:2014tra}. 
In the case of dark matter (DM), the average galactic densities as measured from observations of galactic rotation curves are at best of the order of $M_\odot/{\rm pc^3}$ or ${\rm GeV/cm}^3$, with the local density in the Solar neighbourhood of the order $\sim 0.01 M_\odot/{\rm pc^3}$ or $\sim 0.1 {\rm GeV/cm}^3$ \cite{Pato:2015dua,Nesti:2013uwa,Li:2020qva,DeMartino:2018zkx,Ablimit:2020gxw}. In the units of numerical relativity (NR) simulations, which are used to model the merger signals of compact objects, the effect of the density is measured relative to the Schwarzschild radius $R_s$ and scales with the compact object's mass. In such units average galactic DM densities are of the order
\begin{equation}
\rho_{DM} R_s^2 \sim 10^{-30} \left(\frac{\rho_{DM}}{M_{\odot}\pc^{-3}}\right) \left(\frac{\mbh}{10^6 M_{\odot}}\right)^2 ~. \label{eq-dmdensity}
\end{equation}
Clearly some enhancement in the density around a black hole (BH) or other compact object relative to this value is required in order for the effect to be above numerical error in a simulation (which is a minimum requirement for it to be modelled and measurable in observational data). 

Several mechanisms that create DM overdensities around isolated BHs do exist, with one well motivated one being the formation of dark matter spikes \cite{Gondolo:1999ef,Ferrer:2017xwm,Bertone:2007zz,Speeney:2022ryg,Lacroix:2018zmg,deJong:2021bbo}. These arise from the accretion and adiabatic redistribution of particle dark matter in the potential well around black holes as originally suggested by Gondolo and Silk \cite{Gondolo:1999ef}. For wave dark matter \cite{Urena-Lopez:2019kud,Hui:2016ltb,Hui:2021tkt,Ferreira:2020fam,Hu:2000ke,Niemeyer:2019aqm,Sikivie:2010bq,Hui:2020hbq,Arvanitaki:2009fg,Deng:2018jjz,Schive:2014dra,Grin:2019mub,Stott:2020gjj,El-Zant:2020god,Marsh:2015xka,Brito:2022lmd,Hui:2022sri,Traykova:2021dua,Brax:2019npi,Brito:2015yga,Marsh:2015wka,Hui:2016ltb,Khlopov:1985jw}, where light bosonic particles form a condensate with astrophysical scale de Broglie wavelengths, similar accretion effects occur for which the resulting profiles have been studied in \cite{Clough:2019jpm,Bamber:2020bpu,Hui:2019aqm,Vieira:2014waa,Santos:2020sut,Detweiler:1980uk,Baumann:2019eav,Sanchis-Gual:2020mzb,Clough:2018exo,Sanchis-Gual:2016jst}. The density of the cloud grown via such accretion depends strongly on the asymptotic dark matter environment, but for higher mass candidates gives a power-law enhancement close to the black hole that can be significant. At the other end of the scale, where the wavelength is significantly larger than the black hole, the gradient pressure of the field (sometimes also called the quantum pressure) resists any kind of overdensity and tends to smooth out the profile, suppressing any overdensity.

Another possible enhancement mechanism is the superradiant instability, in which a bosonic field can extract energy and angular momentum from a highly spinning black hole via repeated scattering in the ergoregion  (see \cite{Brito:2015oca} for a review).
Simulations with light massive vector fields suggest these superradiant clouds can grow to be up to $\sim10\%$ of the mass of the black hole \cite{East:2017ovw}, which takes the coefficient in equation \eqref{eq-dmdensity} to $\sim 10^{-5}$ in the best case, and a combination of both superradiance and accretion may lead to even higher densities \cite{Hui:2022sri}. The potential for such bound states to form around BH binaries has been studied in \cite{Ikeda:2020xvt}, as well as around neutron stars \cite{Day:2019bbh,Cardoso:2015zqa,Richartz:2013unq,Cardoso:2017kgn}.

A key question is whether overdensities that may form around isolated objects persist during a binary merger, which is what will be observed in GW data. For example, particle DM spikes have been shown with N-body simulations to disperse for equal mass mergers, meaning that objects close to merger or with a violent merger history are likely to have lost their DM environment \cite{Merritt:2002jz,Kavanagh:2018ggo,Bertone:2005hw}. For this reason, the key targets for detecting environments are extreme mass ratio inspirals (EMRIs) in LISA \cite{Barausse:2020rsu}, where the cloud may still be maintained during the inspiral. Another advantage of EMRIs is that there is a higher relative impact on the curvature for a given DM density for larger black hole masses, and the dephasing in the frequency of the signal during the inspiral accumulates over many orbits that can be potentially be observed in band or across ground based and space based detectors \cite{Maselli:2021men,Xie:2022uvp,Hannuksela:2019vip,Amaro-Seoane:2012lgq,Polcar:2022bwv,Amaro-Seoane:2007osp,Baumann:2022pkl,Cardoso:2022whc}. Studies of the impact of dark matter spikes \cite{Bertone:2018krk,Coogan:2021uqv,Cole:2022ucw,Kavanagh:2020cfn,Li:2021pxf,Yue:2019ozq,Yue:2017iwc} and superradiant clouds \cite{Hannuksela:2018izj,Baumann:2022pkl,Baumann:2021fkf,Baumann:2018vus,KumarPoddar:2019jxe,Zhang:2019eid,Barsanti:2022ana,Ferreira:2017pth} on EMRIs show that they are potentially detectable with LISA observations.

Beyond the inspiral regime of EMRIs, it is also interesting to consider whether the strong gravitational non-linearities present during a roughly equal mass merger may give rise to distinctive features in the gravitational wave signal or other electromagnetic emissions\footnote{For example, in cases where high densities are generated during the merger, certain dark matter models may have electromagnetic counterparts arising from self-annihilations, which would provide an alternative way of identifying particular candidates, such as the Peccei–Quinn QCD axion \cite{Edwards:2019tzf,Yuan:2020xui,Workman:2022ynf,Kouvaris:2022guf}.}. 
These signatures provide information in a different regime to the inspiral and thus their combination could confirm a detection or provide evidence for a particular candidate. 
Studies of similar mass binaries necessitate the use of NR simulations, in which the Einstein equations are solved numerically for the evolution of the binary and its environment, from some initial state prior to merger until after coalescence and ringdown. Due to the computational expense, such simulations can at best cover the last few (order 10) orbits before the merger.
Ultimately the goal is to generate waveform templates for binary mergers that include environmental effects, but a key question that ought to first be answered is whether, at such a late stage in the merger, such an environment will still be present at all, and if so, what spatial configuration it will have. In other words, {\it what is the correct initial data for the matter environment?}

In this paper we study this question for the case of wave dark matter accreting onto an equal mass, non spinning BH binary, focussing on the regime where the wavelength of the scalar is of the same order as the Schwarzschild radii of the individual black holes (which is also similar to their separation at the start of an NR simulation). We begin in Sec. \ref{sec-fixed_orbits} by using toy simulations of fixed BH orbits to show that in this case the accretion of dark matter onto the binary is not disrupted by the orbital motion as it is for higher mass particle candidates. There is instead a quasi-stationary profile that builds up over time, providing a well-motivated initial configuration for such matter in NR simulations.

Having identified a well-motivated profile, we then study the impact of using different initial profiles on the DM evolution in Sec. \ref{sec-matter_evolution}. We study the profile of the matter on the background of a binary merger simulated in full general relativity, initially neglecting the backreaction of the matter onto the binary motion. We compare cases that start with our quasi-stationary profile to more arbitrary configurations such as gaussians, and study the effect of cutting off the accretion at some finite radius.

Finally, in Sec. \ref{sec-backreaction} we use recently developed techniques \cite{Aurrekoetxea:2022mpw} to construct constraint satisfying initial data, and turn on backreaction to study the effect of superimposing the different matter profiles on a circular vacuum inspiral.
Some background information on the set up is given in Sec. \ref{sec-background}, and throughout this work we use geometric units where $G = c = 1$.

Our work builds on a number of earlier related investigations into DM environments of compact object binaries.
The interaction of a black hole binary and a scalar field environment in the early-inspiral regime, where the separation is large and almost constant with time, has been explored via effective field theory \cite{Wong:2019kru,Wong:2020qom}, weak field approximations \cite{Khmelnitsky:2013lxt,Blas:2016ddr,Blas:2019hxz,Ribeiro:2022ohq,Annulli:2020lyc,Liu:2021llm,Kavic:2019cgk,Boskovic:2018rub,Dave:2021lcv}; perturbative schemes \cite{Berti:2019wnn,Baumann:2018vus,Horbatsch:2011ye,Brito:2017zvb,Zhang:2019eid,Takahashi:2021eso,Takahashi:2021yhy,Su:2021dwz}, and N-body and mesh numerical simulations \cite{Kavanagh:2020cfn,Ficarra:2021qeh,Ikeda:2020xvt,Rozner:2019gba,Cardoso:2020hca}. The effect of a non-vacuum environment on the post-merger ``ringdown" regime, particularly the effect on the frequencies of the characteristic quasi-normal modes has also been explored by a number of authors \cite{Leung:1997was,Medved:2003pr,Bamber:2021knr,Zhang:2022roh,Zhang:2021bdr}.
To model the highly relativistic and dynamical merger one must use full numerical relativity.
An axion-like scalar field environment was considered by Yang et al. \cite{Yang:2017lpm}, simulating the effect of a thin shell of matter on the binary merger. Choudhary et al. \cite{Choudhary:2020pxy} go further, starting from the last orbit before merger and using a Gaussian as an initial profile for the scalar field. They find a change in the post-merger ringdown signal caused by the increase in effective mass of the final black hole. Ikeda et. al. \cite{Ikeda:2020xvt} studied the bound states that may form around binaries via superradiance, and find they can be well described with a perturbative ``gravitational molecule" description. Most recently Zhang et al. \cite{Zhang:2022rex} simulated binary mergers with a spherical scalar field shell, and examined the effect on the gravitational recoil of the binary and the scalar and gravitational radiation. They found that the scalar cloud accelerated the merger, and increased the recoil kick. Related work has also been done in modified gravity in the context of scalar-tensor theories
\cite{Healy:2011ef,Berti:2013gfa,Cao:2013osa,Witek:2018dmd,Okounkova:2020rqw,Okounkova:2022grv}. We note that in works where the scalar field grows due to superradiance or a non-minimal coupling to gravity the state of an isolated BH often depends only on the properties of the BH, and not the surrounding DM environment. However, we emphasise that similar issues to those identified in this work regarding the ambiguity of the initial state could still arise if the individual scalar clouds interact and form a common cloud prior to the point at which the simulation is started -- in most (if not all) cases, no analytic form for a common binary cloud is known. 

\section{Modelling wave dark matter around black holes}
\label{sec-background}

For sub eV dark matter the occupation number of the particles in each state is high, with the de Broglie wavelength much larger than the particle separation \cite{Hui:2021tkt}. We can then treat it as a classical field, in particular, a scalar field for spin-0 bosonic dark matter, which results in wave-like behaviour on astrophysical scales \cite{Schive:2014dra} (see \cite{Hui:2021tkt,Urena-Lopez:2019kud,Niemeyer:2019aqm} for reviews).

Specifically, the system we consider is Einstein gravity with a minimally coupled massive complex scalar field $\varphi$, described by the action
\begin{equation}
    S = \int \dd^4 x \sqrt{-g}\left(\frac{R}{16\pi G} - \frac{1}{2}\left(\nabla_{\mu}\varphi\right)^* \left(\nabla^{\mu}\varphi\right) - V(\varphi,\varphi^*)\right),
\end{equation}
with a simple quadratic potential
\begin{equation}
    V(\varphi,\varphi^*) = \frac{1}{2}\mu^2 \varphi^* \varphi.
\end{equation}
The dynamics of the scalar field is thus governed by the Klein-Gordon equation on a curved background
\begin{equation}
\left[\nabla^{\alpha}\nabla_{\alpha} - \mu^2\right]\varphi = 0,
\end{equation}
where $\mu$ is a parameter related to the scalar field mass\footnote{$\mu$ is the inverse length scale $\mu = 2\pi/\lambda_c = m_s c/\hbar$ associated with the scalar field mass $m_s$. In Planck units $\mu = m_s$, so it is common to refer to $\mu$ simply as ``the scalar mass", but in geometric units $\hbar \neq 1$ and so they differ.} and the field has an associated Compton wavelength $\lambda_c = 2\pi/\mu$. In this work we consider a regime where the scalar field wavelength is comparable in size to the black hole radius, corresponding to masses of $\sim 10^{-9}-10^{-17}\eV$ for BH masses between $10-10^9 M_{\odot}$.

The interaction of such massive scalar fields with isolated black holes has been extensively studied, showing that long-lived scalar clouds can grow around black holes either from simple gravitational accretion from the environment \cite{Hui:2016ltb,Clough:2019jpm,Bamber:2020bpu,Sanchis-Gual:2016jst}, or via the mechanism of superradiance for spinning black holes \cite{Brito:2015oca}.  
In the former case, the solution in the asymptotically flat region far from the black holes is a spatially homogeneous oscillatory solution of the form $\varphi = \varphi_0 e^{-i\mu t}$, which describes a fluid of roughly constant density and zero pressure on average - i.e. dark matter. The solution closer to the black hole is described by the Heun functions \cite{Hui:2019aqm,Santos:2020sut,Vieira:2014waa,Hortacsu:2011rr}, with characteristic oscillations in the spatial profile on length scales set by the scalar wavelength.

In this paper we go beyond the single black hole spacetime and consider the simplest possible black hole binary with total ADM mass $M$: two equal mass ($\mbh\approx 0.5M$) non-spinning black holes with an initial separation of $d\approx 12 M$ on roughly circular orbits. The exact parameters are given in Table \ref{table:BH_params}, which result in an inspiral of about $10$ orbits before merger with an initial orbital period $T \sim 270M$.

We study the interaction between the scalar field and the binary black hole solving the Klein-Gordon equation with different levels of approximation in the background metric:
\begin{itemize}
    \item[1.-] Fixed orbit simulations - the metric background is the superposition of two isotropic BH solutions, moving on circular orbits. We evolve the scalar field on this background to test whether a dark matter profile accumulates or disperses over time. See Sec. \ref{sec-fixed_orbits}.
    \item[2.-] Fully general relativistic evolution with $G=0$ - the background is now evolved in full general relativity, but we neglect the backreaction of the matter onto the metric to focus on the impact of the binary on the matter evolution. See Sec. \ref{sec-matter_evolution}. 
    \item[3.-] Fully general relativistic evolution with $G=1$ - finally we turn on backreaction to study the impact of the DM environment on the binary motion. See Sec. \ref{sec-backreaction}.
\end{itemize}

In each case we take the scalar mass $\mu= 0.34 M^{-1}$, corresponding to a scalar wavelength of around $\lambda_c \sim 18 M$, slightly larger than the black hole separation but much smaller than the binary period. For black holes in the LVK frequency band, this corresponds to a mass of $\sim 10^{-11}\eV$, while for supermassive binaries detectable with LISA it can go down to $\sim 10^{-19}\eV$ for a binary with ADM mass $10^9 M_{\odot}$. This choice was motivated by a brief study of the angular momentum flux as described in \cite{Clough:2021qlv,Croft:2022gks}, where it was identified as the value that gave the largest exchange of angular momentum with the binary\footnote{A more in depth study of the accretion behaviour for different masses is given in \cite{Ficarra:2021qeh}.}. 
Our simulations therefore represent a ``best case'' for the impact of the dark matter on the binary.

Our results are presented in the following sections, with details of the code set up and validation contained in the Appendix.

\section{Fixed orbit simulations: finding the quasi-stationary profile}
\label{sec-fixed_orbits}

\begin{figure*}[t]
    \centering
    \includegraphics[width=\linewidth]{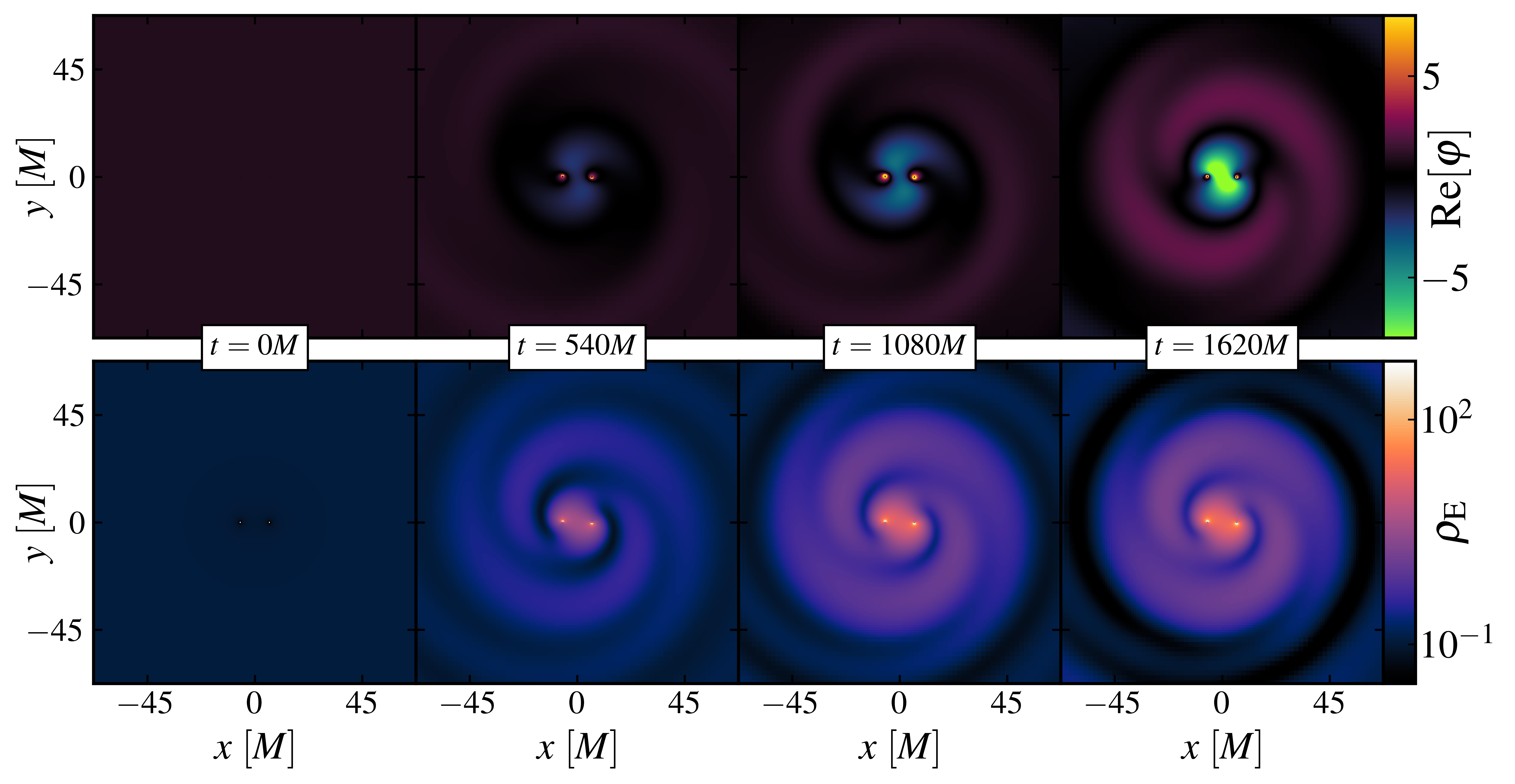}
\caption{Evolution of the real part of the scalar field (top) and energy density (bottom) on the fixed orbit binary background after $0,2,4$ and $6$ orbits respectively. The binary generates a scalar cloud that quickly settles into a quasi-stationary spiralling profile that grows in amplitude over time. In the centre around each black hole spikes form in the scalar field amplitude, resulting in an enhanced energy density, with an additional accumulation of matter in the potential well in the middle of the binary. Further out we see regions of both higher and lower density forming, with the patterns on a length scale related to the binary separation and scalar wavelength. Movie can be found in \href{https://youtu.be/XevfJKLO9ec}{https://youtu.be/XevfJKLO9ec}.}
\label{fig:2D_rhophi_panel}
\end{figure*}

During the early stages of a binary merger, the black holes are widely separated and follow approximately Keplerian orbits, with emission of gravitational waves tending to circularise the orbits over time \cite{Peters:1964zz,Krolak:1987ofj,Cardoso:2020iji,Will:2019lfe}. During this phase, the dark matter profiles of the two BHs will evolve largely independently, generating nontrivial density profiles via accretion. A common DM cloud will also tend to circularise the orbits via the effect of dynamical friction \cite{Becker:2021ivq}.
As the binary separation decreases, at some point the DM clouds will merge and interact
\footnote{See \cite{Ikeda:2020xvt}\cite{Baumann:2022pkl}\cite{Baumann:2021fkf}\cite{Baumann:2018vus} for studies in the context of superradiance.}, eventually forming a common cloud which is ``stirred up'' by the binary. It will not, therefore, have a smooth gaussian profile and will carry some angular momentum due to its interaction with the spacetime curvature.
To investigate the resulting configuration, we construct an approximate toy model for the late inspiral where we model the gravitational field as a superposition of two isotropic black hole metrics
\begin{equation}
    \dd s^2 = -\left(\frac{1 + \Phi/2}{1 - \Phi/2}\right)^2 \dd t^2 + \left(1 - \Phi/2\right)^4(\dd r^2 + r^2 \dd \Omega^2),
\end{equation}
where $\dd \Omega^2 = \dd \theta^2 + \sin^2\theta \dd \phi^2$ and 
\begin{equation}
    \Phi(t,\boldsymbol{r}) = -\frac{G m_1}{\vert \boldsymbol{r}-\boldsymbol{r}_1(t) \vert} - \frac{G m_2}{\vert \boldsymbol{r}-\boldsymbol{r}_2(t)\vert}
\end{equation}
is an effective gravitational potential. We impose that the black holes ($m_1=m_2=\mbh$) follow circular Keplerian orbits of radius $d/2$ and frequency 
\begin{equation}
    \omega_\mathrm{BBH} = \sqrt{\frac{2 G \mbh}{d^3}},
\end{equation}
with their centres located at
\begin{align}
    \boldsymbol{r}_{1}(t) = \left( + \frac{d}{2}\cos(\omega_\mathrm{BBH}~t),~ + \frac{d}{2}\sin(\omega_\mathrm{BBH}~t),~ 0 \right),  \\
    \boldsymbol{r}_{2}(t) = \left( - \frac{d}{2}\cos(\omega_\mathrm{BBH}~t),~ -\frac{d}{2}\sin(\omega_\mathrm{BBH}~t),~ 0 \right).
\end{align}
Close to each of the black holes this metric tends to a Schwarzschild metric in isotropic coordinates, while far from the black holes it tends towards a weak field limit. This metric is not a solution of the Einstein equations, but is merely designed to study the way in which a common quasi-stationary profile can form in a period where the orbits are not yet rapidly decaying as at merger.

We start the simulations with the binary immersed in a homogeneous scalar field, choosing $\varphi(t=0)=\varphi_0$ and $\Pi(t=0)=-i\mu \varphi_0$, and study the evolution of the scalar field over several orbits. The (real) parameter $\varphi_0$ controls the asymptotic density of the infinite reservoir, but since we neglect backreaction for this simulation, we can rescale the results to any physical asymptotic density we choose.

\begin{figure}[t]
    \centering
    \includegraphics[width=\linewidth]{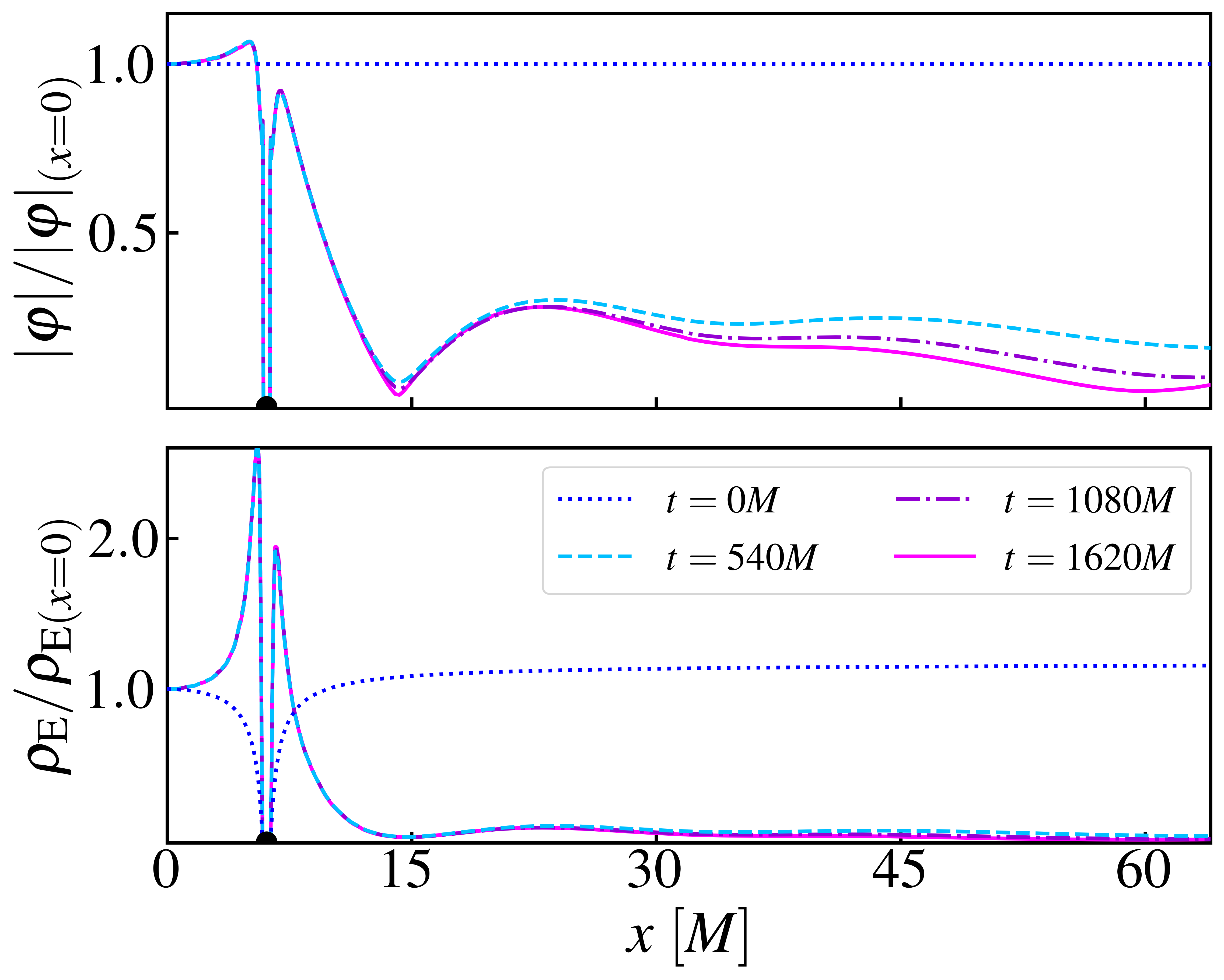}
    \caption{Normalised density and scalar field profiles along half of the fixed orbit binary axis. The black dot illustrates the location of one of the black holes. We see that after a relatively short amount of time (approximately two orbits) the field has gone from a homogeneous profile to a quasi stationary profile peaked around and between the BHs (the zero values within the horizon are due to the choice of gauge and excision conditions for the evolution). This results in a region of enhanced density in the potential well between the binary BHs.}
    \label{fig:Newt_rhophi_prof}
\end{figure}

We find that the scalar field rapidly accretes from its asymptotic value into a cloud around the two black holes and forms a persistent spiral profile within a few orbits that co-rotates with the binary, see Fig. \ref{fig:2D_rhophi_panel}. The scalar field (top panel) is pushed towards large amplitudes, such that the energy density (bottom panel) around and between the black holes increases by several orders of magnitude. Once the profile forms it grows over time homogeneously, fed by the asymptotic reservoir of dark matter imposed at the boundaries.

This quasi-stationary profile can be studied in more detail in Fig. \ref{fig:Newt_rhophi_prof}, where we plot the evolution of the density and scalar field profiles along the axis of the binary, normalised relative to their central values. Even within the first two orbits, both quantities have already settled into the persistent profile that grows steadily in amplitude over time. We observe density spikes around each of the black holes, an accumulation of scalar matter in the potential well between the black holes, and smaller amplitude density peaks further away on length-scales that depend on the binary separation and scalar wavelength. The scalar cloud is significantly enhanced compared to a superposition of two of the scalar field profiles found around isolated black holes (like those studied in \cite{Bamber:2020bpu,Clough:2019jpm,Hui:2019aqm,Vieira:2014waa,Santos:2020sut,Brax:2019npi}), due to the non-linear effect of the combined gravitational potential.

We note that this persistent profile contrasts with higher mass particle-like dark matter, where dark matter density spikes have been shown to disperse under the influence of the binary motion \cite{Merritt:2002jz,Kavanagh:2018ggo,Bertone:2005hw}. 
The enhanced density is important both for its effect on the gravitational wave signal and also for potential direct detection for models with standard model couplings \cite{Yuan:2020xui}.

Even with this simple fixed orbit model, the finite size of the numerical domain prevents us from evolving the system for many orbits. However, the persistent and consistently growing scalar field profile shown here allows us to predict the qualitative behaviour of the cloud at late times -- it seems that a common spiral shaped cloud would continue to grow until the dark matter reservoir is eventually exhausted, at which point the binary would be left with an isolated spiral shaped cloud that would gradually decay away. Assuming that the reservoir is not exhausted before merger, the cloud should continue to grow and not disperse.\footnote{Analytical and numerical studies have suggested that, for the typical scalar field masses used here and reasonable assumptions, clouds can survive for cosmological times \cite{Cardoso:2022nzc,Hui:2022sri}.}

\section{GR evolution with $\mathbf{G=0}$: the impact of the binary on the matter evolution}
\label{sec-matter_evolution}

\begin{figure*}[t!]
    \centering
    \includegraphics[width=\linewidth]{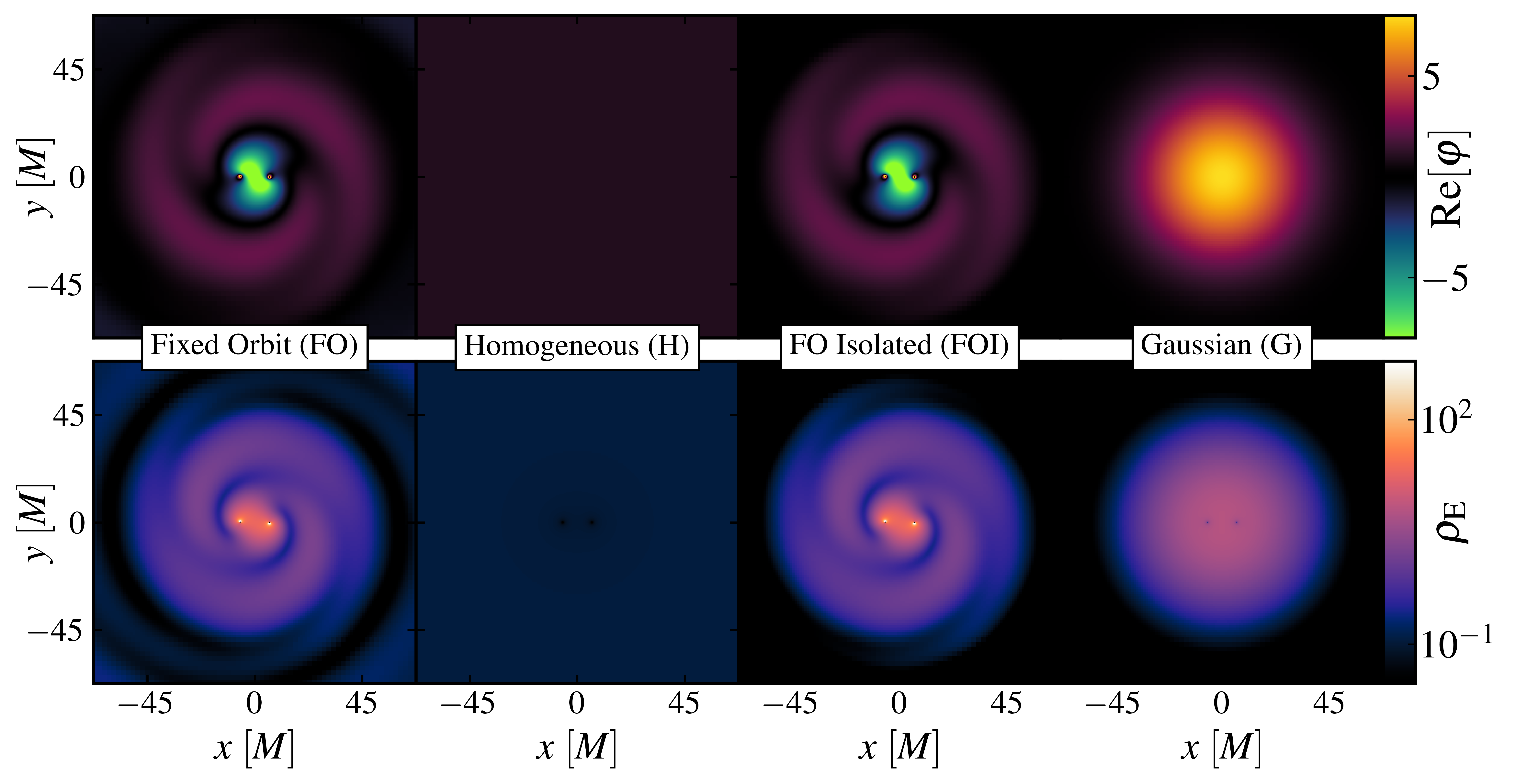}
\caption{Here we show a comparison of the initial data profiles. From left: FO, H, FOI and G. The top row shows the real part of the scalar field, while the bottom shows the energy density on a log scale. Details on the scaling of these profiles is given in the main text and in Fig. \ref{fig:SF_profile_panel}.}
\label{fig:2D_initial_profiles}
\end{figure*}

We now study the evolution of the wave dark matter on a fully general relativistic binary merger, with the Klein-Gordon equation evolved on a dynamical spacetime described by the Einstein's field equations
\begin{equation}\label{eq:EFE}
    R_{\mu\nu} - \frac{1}{2}g_{\mu\nu}R = 8\pi G T_{\mu\nu},
\end{equation}
where $R_{\mu\nu}$ is the Ricci tensor, and $T_{\mu\nu}$ is the energy momentum tensor of the scalar field.
To first isolate the impact of the binary on the matter evolution, we turn off the backreaction by setting Newton's constant $G=0$ in Eqn. \eqref{eq:EFE}. This ensures that different cloud configurations evolve in the same BBH background, described by the black holes' vacuum trajectories.

The BH initial conditions are Bowen-York data \cite{AlcubierreMiguel2008It3N,Brandt:1994ee,Brandt:1996si,Okawa:2014nda,Dennison:2006nq,Cook:2000vr,Bowen:1979,Bowen:1980yu,York:1978gql} for the parameters in table \ref{table:BH_params} and we solve the Hamiltonian constraint for the correction to the conformal factor arising from the non zero boosts.

The main goal is to compare the evolution of several initial scalar field configurations. First, we observe that the profile found in our toy model with fixed orbits is close to the stationary profile for the fully GR binary solution, as we see very little transient evolution when we impose it on the binary and begin the evolution. This can be contrasted with other possible choices for the initial scalar field profile, such as a Gaussian, where we find significant initial transients before the preferred profile is reached.

The scalar profiles we examine can be classified into two types: extended and isolated clouds. In the former class, the energy density of the cloud reaches the boundary of our simulated domain, allowing for continued accretion from spatial infinity. (This is imposed using extrapolating boundary conditions as described in \cite{Radia:2021smk}.) Within this class we study:
\begin{itemize}
    \item \textbf{Fixed Orbit (FO):}\\
    We take as initial data the resulting quasi-stationary scalar field profile after evolving the binary in the homogeneous dark matter halo for $6$ orbits $\varphi_\mathrm{FO}$, see Fig. \ref{fig:2D_rhophi_panel}. This can be considered the ``most correct'' initial condition for a scalar field that is still accreting up to the merger.\\
    \item \textbf{Homogeneous (H):}\\
    We take the same homogeneous initial profile described in the previous section by setting $\varphi(t=0) = \varphi_0$ and $\Pi(t=0) = -i\mu \varphi_0$. In this profile we still need to grow the quasi-stationary profile, so expect some transient evolution.
\end{itemize}

In the second class of initial conditions, the scalar cloud has a sufficiently large radius to cover the BBH, but the energy density goes to zero at the boundaries, reproducing an isolated cloud which has exhausted its dark matter reservoir. (For these cases we use Sommerfeld radiative boundary conditions as described in \cite{Radia:2021smk}.) Here we study two cases:
\begin{itemize}
    \item \textbf{Fixed Orbit Isolated (FOI):}\\
    We apply a $f(r)\sim\tanh[r-60M]$ envelope to the FO scalar field profile, so that $\varphi_\mathrm{FOI}(t=0) =f(r)\varphi_\mathrm{FO}(t=0)$ and the configuration reproduces an isolated cloud which has exhausted its dark matter reservoir. Again this is a ``correct'' profile around the black holes, but the cut off we introduce is rather arbitrary, and therefore introduces some transient evolution in the overall shape of the cloud as it settles into a quasi-stationary, isolated profile.
    \item \textbf{Gaussian (G):}\\
    We choose a Gaussian profile for $\varphi(t=0) = \varphi_0 \exp[- r^2/\sigma^2]$ and $\Pi(t=0) = -i\mu \varphi(t=0)$, where $r$ is the distance to the centre of the binary. We use $\sigma = 1/0.03 M$ and choose $\varphi_0$ so that the total mass of the initial scalar cloud is the same as the isolated fixed orbit cloud described above. We expect transient evolution in this case before we arrive at the quasi-stationary profile.
\end{itemize}
We plot 2D slices of the initial density configurations perpendicular to the binary orbital axis in Fig. \ref{fig:2D_initial_profiles}.\\

\begin{figure*}[t]
    \centering
    \includegraphics[width=\linewidth]{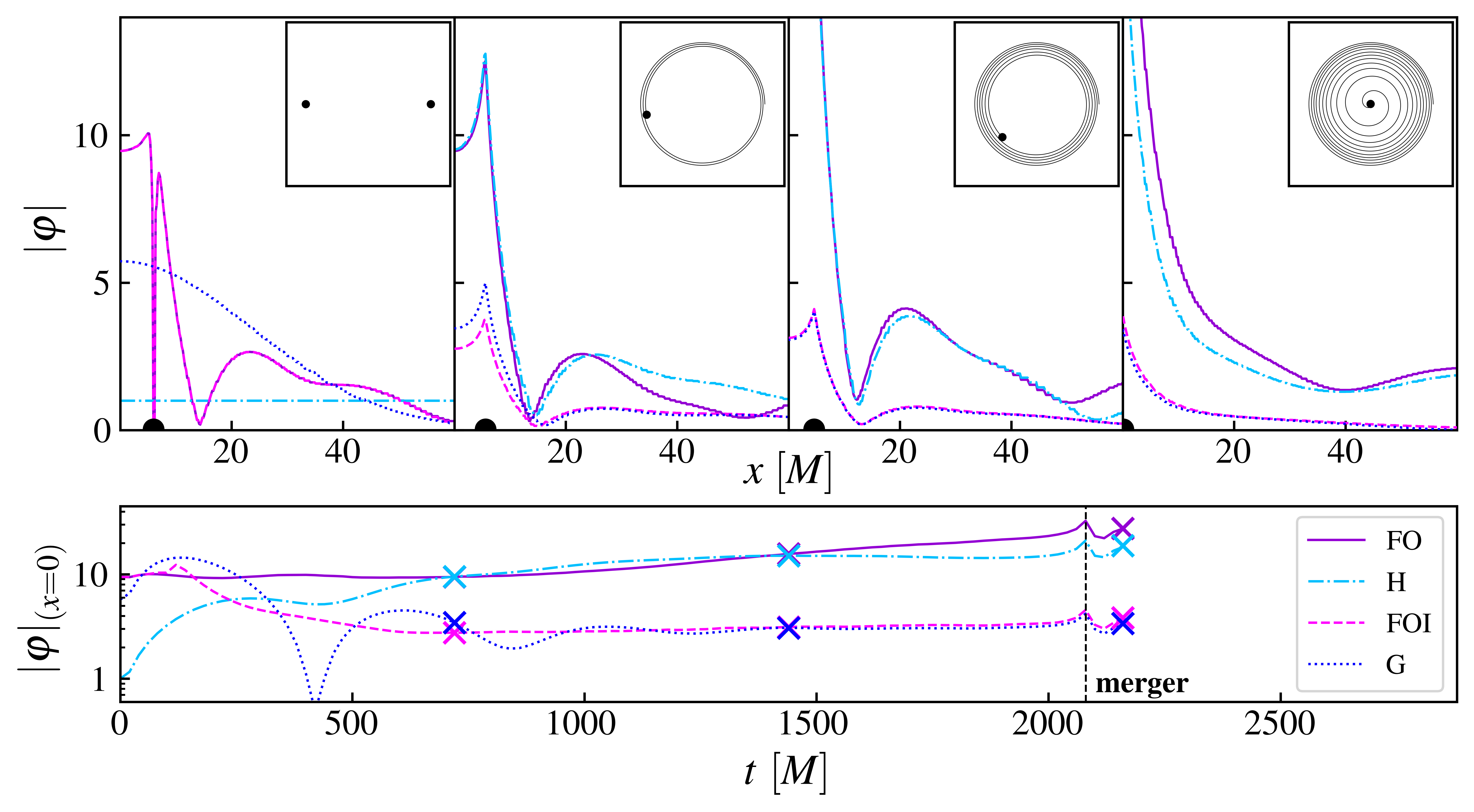}
    \caption{We plot the $|\varphi|$ profile along the axis of the binary for different initial data profiles. The amplitudes of the fixed orbit isolated (FOI) and Gaussian profiles are normalised so that the integrated mass of the whole cloud is the same. The amplitude of the homogeneous profile is chosen to match the initial asymptotic value of the fixed orbit profile and is set to unity in these plots. The black dot illustrates the location of one of the black holes. The bottom panel plots the evolution of the scalar field amplitude at the centre of the binary with time. Markers illustrate the time of the upper panels. A movie showing the evolution of these profiles can be found in \href{https://youtu.be/wGUUUutuiyU}{https://youtu.be/wGUUUutuiyU}.}
    \label{fig:SF_profile_panel}
\end{figure*}

To quantify the differences we extract the value of the scalar field profile along the line joining the BHs, see the top panel of Fig. \ref{fig:SF_profile_panel}. In the bottom panel of Fig. \ref{fig:SF_profile_panel} we track the value of the scalar field at the centre of the binary, which provides a reasonable indication of the amount of transient evolution. We see as expected that the scalar field needs some time to settle down to the quasi-stationary configuration, with the most transient evolution in the case G and the least in FO.
\footnote{Note that the scalar field in the FO and FOI initial data goes to zero inside the horizon. However when we evolve it in the full GR evolution in the moving puncture gauge \cite{Liu:2009al,Hannam:2006xw} it quickly relaxes to give a continuous non-zero density spike. This difference is mainly due to the different choice of lapse between the fixed orbit metric and the moving punctures gauge used for the full relativistic evolution. One should thus consider the change in the sub-horizon initial scalar field profile to be an artefact of the change in gauge choice.}
However, even when starting from very different profiles the scalar field evolves into the same kind of spiral shaped, quasi-stationary, configuration as described in section \ref{sec-fixed_orbits} within a few orbits. The final state post merger is a single density spike with power law tails, of the kind studied in \cite{Clough:2019jpm,Bamber:2020bpu,Hui:2019aqm,Brax:2019npi,Vieira:2014waa,Santos:2020sut,Sanchis-Gual:2016jst}, with solutions that can be approximated by confluent Heun functions \cite{Hortacsu:2011rr}. 

The main difference between the extended (FO, H) and isolated (FOI, G) clouds is that the former can continue to accrete from infinity, so we see the central amplitude continuing to grow throughout the merger, increasing the density and size of the final scalar cloud.
The homogeneous data accretes rapidly at the centre, and in only a few orbits reaches the fixed orbit quasi-stationary configuration. The difference in amplitude between the final cloud from the homogeneous data and FO data is expected given that the fixed orbit data has effectively been accreting for an extra six orbital periods during our fixed-orbit simulation.

For the isolated configurations (FOI and G), the cloud does not continue to grow as there is no reservoir of DM at the boundary. In the case of FOI, because there is no asymptotic density to confine it, the cloud settles into a more distributed (but qualitatively similar) shape - this is a consequence of imposing an arbitrary cut off in the density at finite radius. The Gaussian case, on the other hand, shows strong transient behaviour with several large oscillations around the quasi-stationary profile, on a timescale roughly corresponding to the period of the binary. After $\sim 2$-$3$ orbits the same configuration is reached - the fact that the final shapes and amplitudes match is a result of starting with the same cloud mass, and implies that the two cases have radiated equal amounts, and been accreted equally by the BHs, during their transient evolution.

These results demonstrate that within only a few orbits the transients in the cloud will die away and a quasi-stationary configuration will be reached. However, the initial profile can make a significant difference to the transient evolution of the scalar cloud during this time, and therefore may have an impact for shorter simulations. We have not yet considered the backreaction of matter onto the metric, but where this is included it is possible that these transient effects could impact on the black hole trajectories, as well as creating additional radiation of gravitational waves. We will discuss this further in the following section. 

\section{GR evolution with $\mathbf{G=1}$: the impact of the matter on the binary evolution}
\label{sec-backreaction}

The ultimate goal of NR simulations with environments is to quantify the impact of the matter on the binary evolution, and resulting gravitational wave emission, so as to learn about the properties of the DM. To do so we must include the backreaction of the evolving matter on the metric background.
We therefore restore $G=1$ and solve the Hamiltonian and momentum constraints using the novel CTTK method \cite{Aurrekoetxea:2022mpw}. In particular, we use the \textit{hybrid} CTTK approach, where we choose a spatially varying mean curvature that depends on the energy density distribution $K^2=24\pi G\rho$, where $\rho$ is the ADM energy density, and solve the constraints obtaining corrections for both the conformal factor and traceless conformal extrinsic curvature, on top of an initial Bowen-York solution for a boosted black hole binary in vacuum. This means the matter environment introduces corrections to the effective initial momenta and bare masses of the black holes.

Now that the amplitude of our field has a physical meaning (it is related to the matter density roughly as $\rho \sim \varphi_0^2$), we quantify this by specifying the relative cloud mass to (vacuum ADM) binary mass ratio.

\begin{figure}[t]
    \centering
    \includegraphics[width=\linewidth]{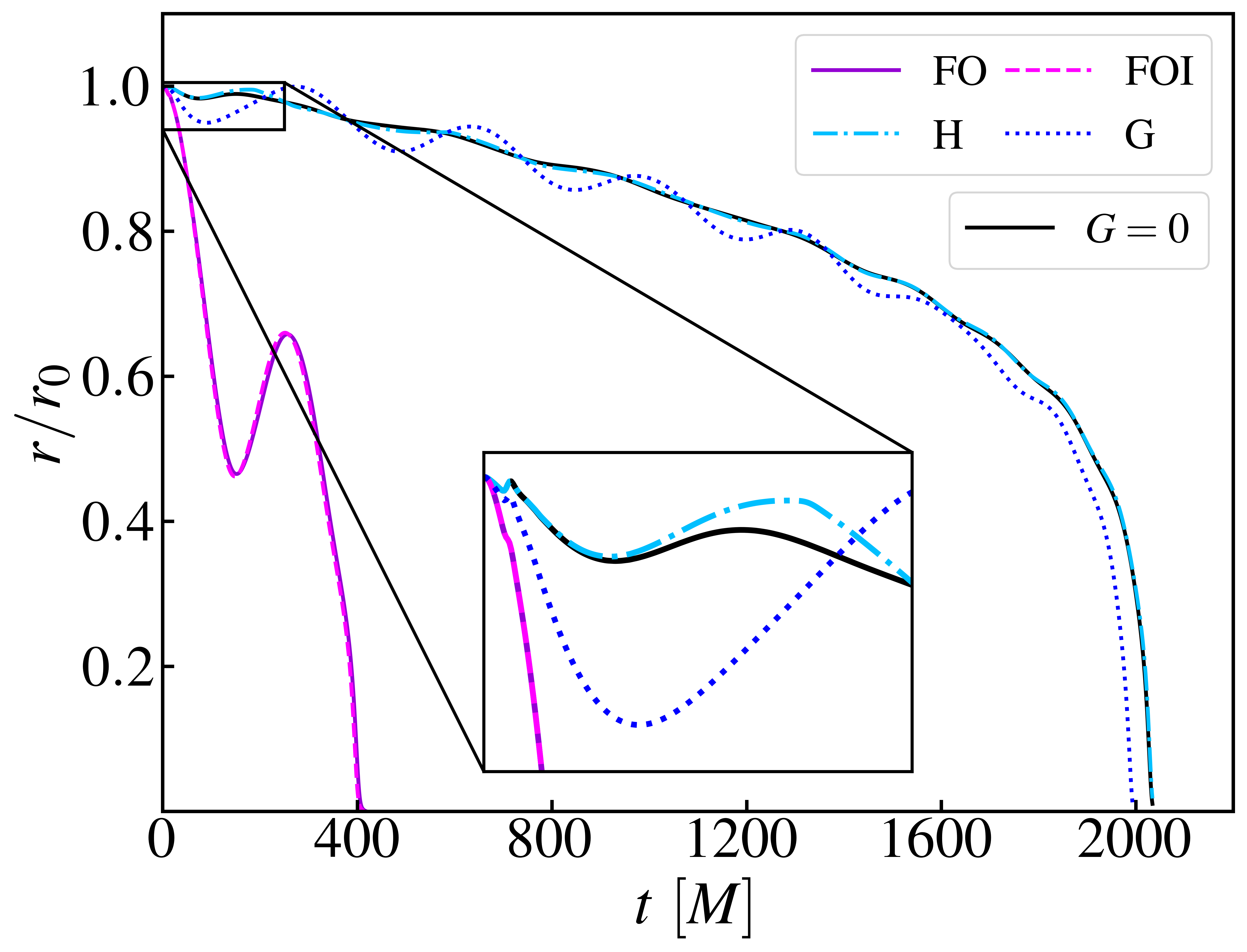}
    \caption{Orbital trajectory of one of the black holes in simulations of a vacuum binary (black line), and in binaries with total cloud mass of $M_\mathrm{cloud}/M = 10^{-3}$. Different colored lines depict the effect of the different scalar field profiles with the same black hole parameters.}
    \label{fig:orbit}
\end{figure}

We superpose our profiles onto the vacuum BH parameters for the masses and momenta, and evolve the metric plus matter to merger, which we find happens faster in the presence of the DM clouds. Some dephasing of the signal compared to the vacuum evolution is expected due to the effects of dynamical friction, radiation of the scalar cloud, and backreaction onto the metric, but the impact we see is unexpectedly large. Examining the black hole trajectories in Fig. \ref{fig:orbit} reveals that, even for light clouds where $M_{cloud}/M\approx 0.01\%$, the black holes are pushed into eccentric orbits, giving rise to large deviations from the circular trajectories that we would expect at this late stage of the inspiral and a prompt merger. We find that this effect is proportional to the energy density near the black holes, being the largest for both the extended and isolated fixed orbit profiles (FO and FOI) due to the presence of density spikes near the horizons. For the more artificial initial conditions H and G where the initial energy density near the black holes is smaller, the effect is smaller but still non-negligible. The use of the CTTK method means that in general we obtain larger corrections to the effective initial black hole momenta, and smaller corrections to the effective bare masses, compared to the CTT method. However the unwanted eccentricity is an unavoidable consequence of the non-trivial matter environment, and is particularly significant where there are high energy and momentum flux densities near the black hole horizons, as in our preferred quasi-stationary solution.  

Because the deviations are significant, we cannot simply extract waveforms and compare them to the vacuum case to determine how the presence of a scalar cloud impacts the gravitational wave signal, as we would not be able to conclude that such signatures are caused by the presence of the scalar field rather than by the induced (physical but unwanted) eccentricity. Even if one tunes the initial momenta and masses of the black hole punctures so that their initial coordinate velocities and accelerations match the vacuum case, this is not sufficient to remove the effect.
Further work is required to remove this effect. One should follow a similar method to those used to obtain low-eccentricity initial data for binary black holes in vacuum spacetimes \cite{Pfeiffer:2007yz,Habib:2020dba,Buonanno:2010yk,Ramos-Buades:2018azo}. In these methods, one measures the trajectory and eccentricity over $2$-$3$ full orbits for an initial choice of input parameters, then uses a high order post-Newtonian \cite{Blanchet:2006zz} approximation or gradient descent to estimate the correction to these input parameters (typically the initial momenta and masses of the black hole punctures) needed to reduce the eccentricity. One then iterates this procedure until the desired eccentricity is achieved. Our case is more complicated because the black holes are perturbed by the backreaction from the matter distribution, which we solve for numerically. However, one could in principle obtain a similar scheme by incorporating the effect of a matter cloud into a new post-Newtonian binary model, something which is beyond the scope of this paper\footnote{Post-Newtonian dynamics have been explored for binaries in scalar-tensor theories \cite{Bernard:2018hta,Julie:2019sab,Yagi:2011xp,Shiralilou:2020gah,Shiralilou:2021mfl,Brax:2021qqo} but not, as far as we are aware, for minimally coupled massive scalar field environments.}. 

\section{Discussion}

One of the key challenges in constructing reliable gravitational wave templates for BH binary mergers with environments is establishing the correct initial data for numerical relativity simulations, which begin only a relatively short time before the merger. 

Here we conducted numerical simulations of the accretion of wave dark matter around binary black holes in fixed orbits, choosing a scalar mass that gives a large interaction with the BH binary (i.e., the regime where the wavelength of the field is similar to the Schwarzschild radii of the BHs). We found that the scalar field quickly converges to a persistent non-trivial profile, peaked around each black hole, which grows in amplitude as the accretion continues. 

We then explored how different choices of initial data affected the evolution of the DM cloud during the binary merger. Our results suggest that the profile found in the fixed orbit simulations is an attractor solution, as other choices of initial data converged to the same distribution over the course of several orbits. This means that for long numerical simulations the precise choice of initial scalar profile may be unimportant, as the field will quickly converge to the quasi-stationary distribution. However for numerical simulations which only consist of a small number of orbits, the transients and the resulting loss of control over the initial data from using a non stationary profile may affect the final results. 
We also saw that allowing continued accretion from a scalar dark matter reservoir over the course of the merger results in a larger scalar cloud around the final black hole, confirming the importance of the local DM environment for the system's evolution. 

Lastly, we obtained constraint satisfying initial data for the scalar profiles obtained from our fixed orbit simulations. To the best of our knowledge, this is the first time that environments with non zero angular momentum have been studied in the initial conditions for DM environments around binaries. We found that naively using the parameters for a quasi-circular binary merger in vacuum will not usually produce low eccentricity orbits once matter is added, and because of this one cannot simply compare the resulting waveforms to establish the signatures of dark matter environments. For example, one may find that the merger happens sooner merely because of the increased eccentricity, and not due to effects like dynamical friction and gravitational radiation.

We conclude that a more comprehensive approach is needed to find initial data that both satisfies the Einstein constraints and produces realistic low-eccentricity inspirals, similar to that developed for vacuum binaries \cite{Pfeiffer:2007yz,Habib:2020dba,Buonanno:2010yk,Cook:2000vr,Dennison:2006nq}. This should ideally incorporate the quasi-stationary profile as part of the process of solving for the initial data, as is done in neutron star simulations \cite{Tichy:2012rp,Kyutoku:2014yba,Tsokaros:2015fea}, or match to a post-Newtonian model \cite{Blanchet:2006zz,Bernard:2018hta,Julie:2019sab,Yagi:2011xp,Shiralilou:2020gah,Shiralilou:2021mfl} for the matter and black holes in the late inspiral regime. 

Only once the problem of initial conditions is under control can the degeneracies between the effects of a non-trivial dark matter environment and changes in other parameters of the binary be quantified, so that we can examine how to disentangle the two. We may then also extend the parameter space to include unequal mass and spinning black hole binaries, where kicks may occur \cite{Cardoso:2020lxx}, as well as exploring a wider range of scalar field masses, and the impact of self-interactions. 

\section*{Acknowledgements}

\noindent We would like to thank Lam Hui and Macarena Lagos for their input and discussions in the early stages of this project. We also thank Giuseppe Ficarra, Eugene Lim and Helvi Witek for helpful conversations. We thank the GRChombo collaboration (\href{www.grchombo.org}{www.grchombo.org}) for their support and code development work. JB acknowledges funding from a UK Science and Technology Facilities Council (STFC) studentship. JCA acknowledges funding from the Beecroft Trust and The Queen’s
College via an extraordinary Junior Research Fellowship (eJRF). This project has received funding from the European Research Council (ERC) under the
European Union’s Horizon 2020 research and innovation programme (Grant Agreement No 693024). KC acknowledges funding from the UKRI Ernest Rutherford Fellowship (grant number ST/V003240/1). 
For the purpose of Open Access, the author has applied a CC BY public copyright licence to any Author Accepted Manuscript version arising from this submission.

Part of this work was performed using the DiRAC Data Intensive service at Leicester, operated by the University of Leicester IT Services, which forms part of the STFC DiRAC HPC Facility (www.dirac.ac.uk). The equipment was funded by BEIS capital funding via STFC capital grants ST/K000373/1 and ST/R002363/1 and STFC DiRAC Operations Grant ST/R001014/1. This work also used the DiRAC@Durham facility managed by the Institute for Computational Cosmology on behalf of the STFC DiRAC HPC Facility (www.dirac.ac.uk). The equipment was funded by BEIS capital funding via STFC Capital Grants ST/P002293/1, ST/R002371/1 and ST/S002502/1, Durham University and STFC Operations Grant ST/R000832/1. DiRAC is part of the National e-Infrastructure. The authors also thank Thomas Helfer, and acknowledge the Texas Advanced Computing Center (TACC) at The University of Texas at Austin for providing HPC resources \cite{10.1145/3311790.3396656} that have contributed to the research results reported within this paper. URL: \href{http://www.tacc.utexas.edu}{http://www.tacc.utexas.edu}

\bibliography{biblio}

\begin{thebibliography}{189}%
\makeatletter
\providecommand \@ifxundefined [1]{%
 \@ifx{#1\undefined}
}%
\providecommand \@ifnum [1]{%
 \ifnum #1\expandafter \@firstoftwo
 \else \expandafter \@secondoftwo
 \fi
}%
\providecommand \@ifx [1]{%
 \ifx #1\expandafter \@firstoftwo
 \else \expandafter \@secondoftwo
 \fi
}%
\providecommand \natexlab [1]{#1}%
\providecommand \enquote  [1]{``#1''}%
\providecommand \bibnamefont  [1]{#1}%
\providecommand \bibfnamefont [1]{#1}%
\providecommand \citenamefont [1]{#1}%
\providecommand \href@noop [0]{\@secondoftwo}%
\providecommand \href [0]{\begingroup \@sanitize@url \@href}%
\providecommand \@href[1]{\@@startlink{#1}\@@href}%
\providecommand \@@href[1]{\endgroup#1\@@endlink}%
\providecommand \@sanitize@url [0]{\catcode `\\12\catcode `\$12\catcode
  `\&12\catcode `\#12\catcode `\^12\catcode `\_12\catcode `\%12\relax}%
\providecommand \@@startlink[1]{}%
\providecommand \@@endlink[0]{}%
\providecommand \url  [0]{\begingroup\@sanitize@url \@url }%
\providecommand \@url [1]{\endgroup\@href {#1}{\urlprefix }}%
\providecommand \urlprefix  [0]{URL }%
\providecommand \Eprint [0]{\href }%
\providecommand \doibase [0]{https://doi.org/}%
\providecommand \selectlanguage [0]{\@gobble}%
\providecommand \bibinfo  [0]{\@secondoftwo}%
\providecommand \bibfield  [0]{\@secondoftwo}%
\providecommand \translation [1]{[#1]}%
\providecommand \BibitemOpen [0]{}%
\providecommand \bibitemStop [0]{}%
\providecommand \bibitemNoStop [0]{.\EOS\space}%
\providecommand \EOS [0]{\spacefactor3000\relax}%
\providecommand \BibitemShut  [1]{\csname bibitem#1\endcsname}%
\let\auto@bib@innerbib\@empty
\bibitem [{\citenamefont {Aasi}\ \emph {et~al.}(2015)\citenamefont {Aasi} \emph
  {et~al.}}]{LIGOScientific:2014pky}%
  \BibitemOpen
  \bibfield  {author} {\bibinfo {author} {\bibfnamefont {J.}~\bibnamefont
  {Aasi}} \emph {et~al.} (\bibinfo {collaboration} {LIGO Scientific}),\ }\href
  {https://doi.org/10.1088/0264-9381/32/7/074001} {\bibfield  {journal}
  {\bibinfo  {journal} {Class. Quant. Grav.}\ }\textbf {\bibinfo {volume}
  {32}},\ \bibinfo {pages} {074001} (\bibinfo {year} {2015})},\ \Eprint
  {https://arxiv.org/abs/1411.4547} {arXiv:1411.4547 [gr-qc]} \BibitemShut
  {NoStop}%
\bibitem [{\citenamefont {Abbott}\ \emph {et~al.}(2016)\citenamefont {Abbott}
  \emph {et~al.}}]{LIGOScientific:2016aoc}%
  \BibitemOpen
  \bibfield  {author} {\bibinfo {author} {\bibfnamefont {B.~P.}\ \bibnamefont
  {Abbott}} \emph {et~al.} (\bibinfo {collaboration} {LIGO Scientific,
  Virgo}),\ }\href {https://doi.org/10.1103/PhysRevLett.116.061102} {\bibfield
  {journal} {\bibinfo  {journal} {Phys. Rev. Lett.}\ }\textbf {\bibinfo
  {volume} {116}},\ \bibinfo {pages} {061102} (\bibinfo {year} {2016})},\
  \Eprint {https://arxiv.org/abs/1602.03837} {arXiv:1602.03837 [gr-qc]}
  \BibitemShut {NoStop}%
\bibitem [{\citenamefont {Abbott}\ \emph {et~al.}(2017)\citenamefont {Abbott}
  \emph {et~al.}}]{LIGOScientific:2017vwq}%
  \BibitemOpen
  \bibfield  {author} {\bibinfo {author} {\bibfnamefont {B.~P.}\ \bibnamefont
  {Abbott}} \emph {et~al.} (\bibinfo {collaboration} {LIGO Scientific,
  Virgo}),\ }\href {https://doi.org/10.1103/PhysRevLett.119.161101} {\bibfield
  {journal} {\bibinfo  {journal} {Phys. Rev. Lett.}\ }\textbf {\bibinfo
  {volume} {119}},\ \bibinfo {pages} {161101} (\bibinfo {year} {2017})},\
  \Eprint {https://arxiv.org/abs/1710.05832} {arXiv:1710.05832 [gr-qc]}
  \BibitemShut {NoStop}%
\bibitem [{\citenamefont {Abbott}\ \emph {et~al.}(2021)\citenamefont {Abbott}
  \emph {et~al.}}]{LIGOScientific:2021djp}%
  \BibitemOpen
  \bibfield  {author} {\bibinfo {author} {\bibfnamefont {R.}~\bibnamefont
  {Abbott}} \emph {et~al.} (\bibinfo {collaboration} {LIGO Scientific, VIRGO,
  KAGRA}),\ }\href@noop {} {\  (\bibinfo {year} {2021})},\ \Eprint
  {https://arxiv.org/abs/2111.03606} {arXiv:2111.03606 [gr-qc]} \BibitemShut
  {NoStop}%
\bibitem [{\citenamefont {Acernese}\ \emph {et~al.}(2015)\citenamefont
  {Acernese} \emph {et~al.}}]{VIRGO:2014yos}%
  \BibitemOpen
  \bibfield  {author} {\bibinfo {author} {\bibfnamefont {F.}~\bibnamefont
  {Acernese}} \emph {et~al.} (\bibinfo {collaboration} {VIRGO}),\ }\href
  {https://doi.org/10.1088/0264-9381/32/2/024001} {\bibfield  {journal}
  {\bibinfo  {journal} {Class. Quant. Grav.}\ }\textbf {\bibinfo {volume}
  {32}},\ \bibinfo {pages} {024001} (\bibinfo {year} {2015})},\ \Eprint
  {https://arxiv.org/abs/1408.3978} {arXiv:1408.3978 [gr-qc]} \BibitemShut
  {NoStop}%
\bibitem [{\citenamefont {Akutsu}\ \emph {et~al.}(2021)\citenamefont {Akutsu}
  \emph {et~al.}}]{KAGRA:2020tym}%
  \BibitemOpen
  \bibfield  {author} {\bibinfo {author} {\bibfnamefont {T.}~\bibnamefont
  {Akutsu}} \emph {et~al.} (\bibinfo {collaboration} {KAGRA}),\ }\href
  {https://doi.org/10.1093/ptep/ptaa125} {\bibfield  {journal} {\bibinfo
  {journal} {PTEP}\ }\textbf {\bibinfo {volume} {2021}},\ \bibinfo {pages}
  {05A101} (\bibinfo {year} {2021})},\ \Eprint
  {https://arxiv.org/abs/2005.05574} {arXiv:2005.05574 [physics.ins-det]}
  \BibitemShut {NoStop}%
\bibitem [{\citenamefont {Bailes}\ \emph {et~al.}(2021)\citenamefont {Bailes}
  \emph {et~al.}}]{Bailes:2021tot}%
  \BibitemOpen
  \bibfield  {author} {\bibinfo {author} {\bibfnamefont {M.}~\bibnamefont
  {Bailes}} \emph {et~al.},\ }\href
  {https://doi.org/10.1038/s42254-021-00303-8} {\bibfield  {journal} {\bibinfo
  {journal} {Nature Rev. Phys.}\ }\textbf {\bibinfo {volume} {3}},\ \bibinfo
  {pages} {344} (\bibinfo {year} {2021})}\BibitemShut {NoStop}%
\bibitem [{\citenamefont {Schmidt}(2020)}]{Schmidt:2020ekt}%
  \BibitemOpen
  \bibfield  {author} {\bibinfo {author} {\bibfnamefont {P.}~\bibnamefont
  {Schmidt}},\ }\href {https://doi.org/10.3389/fspas.2020.00028} {\bibfield
  {journal} {\bibinfo  {journal} {Front. Astron. Space Sci.}\ }\textbf
  {\bibinfo {volume} {7}},\ \bibinfo {pages} {28} (\bibinfo {year}
  {2020})}\BibitemShut {NoStop}%
\bibitem [{\citenamefont {Barack}\ \emph {et~al.}(2019)\citenamefont {Barack}
  \emph {et~al.}}]{Barack:2018yly}%
  \BibitemOpen
  \bibfield  {author} {\bibinfo {author} {\bibfnamefont {L.}~\bibnamefont
  {Barack}} \emph {et~al.},\ }\href {https://doi.org/10.1088/1361-6382/ab0587}
  {\bibfield  {journal} {\bibinfo  {journal} {Class. Quant. Grav.}\ }\textbf
  {\bibinfo {volume} {36}},\ \bibinfo {pages} {143001} (\bibinfo {year}
  {2019})},\ \Eprint {https://arxiv.org/abs/1806.05195} {arXiv:1806.05195
  [gr-qc]} \BibitemShut {NoStop}%
\bibitem [{\citenamefont {Abbott}\ \emph {et~al.}(2019)\citenamefont {Abbott}
  \emph {et~al.}}]{LIGOScientific:2018jsj}%
  \BibitemOpen
  \bibfield  {author} {\bibinfo {author} {\bibfnamefont {B.~P.}\ \bibnamefont
  {Abbott}} \emph {et~al.} (\bibinfo {collaboration} {LIGO Scientific,
  Virgo}),\ }\href {https://doi.org/10.3847/2041-8213/ab3800} {\bibfield
  {journal} {\bibinfo  {journal} {Astrophys. J. Lett.}\ }\textbf {\bibinfo
  {volume} {882}},\ \bibinfo {pages} {L24} (\bibinfo {year} {2019})},\ \Eprint
  {https://arxiv.org/abs/1811.12940} {arXiv:1811.12940 [astro-ph.HE]}
  \BibitemShut {NoStop}%
\bibitem [{\citenamefont {Nishizawa}\ \emph {et~al.}(2016)\citenamefont
  {Nishizawa}, \citenamefont {Berti}, \citenamefont {Klein},\ and\
  \citenamefont {Sesana}}]{Nishizawa:2016jji}%
  \BibitemOpen
  \bibfield  {author} {\bibinfo {author} {\bibfnamefont {A.}~\bibnamefont
  {Nishizawa}}, \bibinfo {author} {\bibfnamefont {E.}~\bibnamefont {Berti}},
  \bibinfo {author} {\bibfnamefont {A.}~\bibnamefont {Klein}},\ and\ \bibinfo
  {author} {\bibfnamefont {A.}~\bibnamefont {Sesana}},\ }\href
  {https://doi.org/10.1103/PhysRevD.94.064020} {\bibfield  {journal} {\bibinfo
  {journal} {Phys. Rev. D}\ }\textbf {\bibinfo {volume} {94}},\ \bibinfo
  {pages} {064020} (\bibinfo {year} {2016})},\ \Eprint
  {https://arxiv.org/abs/1605.01341} {arXiv:1605.01341 [gr-qc]} \BibitemShut
  {NoStop}%
\bibitem [{\citenamefont {Barausse}\ \emph {et~al.}(2015)\citenamefont
  {Barausse}, \citenamefont {Bellovary}, \citenamefont {Berti}, \citenamefont
  {Holley-Bockelmann}, \citenamefont {Farris}, \citenamefont {Sathyaprakash},\
  and\ \citenamefont {Sesana}}]{Barausse:2014oca}%
  \BibitemOpen
  \bibfield  {author} {\bibinfo {author} {\bibfnamefont {E.}~\bibnamefont
  {Barausse}}, \bibinfo {author} {\bibfnamefont {J.}~\bibnamefont {Bellovary}},
  \bibinfo {author} {\bibfnamefont {E.}~\bibnamefont {Berti}}, \bibinfo
  {author} {\bibfnamefont {K.}~\bibnamefont {Holley-Bockelmann}}, \bibinfo
  {author} {\bibfnamefont {B.}~\bibnamefont {Farris}}, \bibinfo {author}
  {\bibfnamefont {B.}~\bibnamefont {Sathyaprakash}},\ and\ \bibinfo {author}
  {\bibfnamefont {A.}~\bibnamefont {Sesana}},\ }\href
  {https://doi.org/10.1088/1742-6596/610/1/012001} {\bibfield  {journal}
  {\bibinfo  {journal} {J. Phys. Conf. Ser.}\ }\textbf {\bibinfo {volume}
  {610}},\ \bibinfo {pages} {012001} (\bibinfo {year} {2015})},\ \Eprint
  {https://arxiv.org/abs/1410.2907} {arXiv:1410.2907 [astro-ph.HE]}
  \BibitemShut {NoStop}%
\bibitem [{\citenamefont {Kavanagh}\ \emph {et~al.}(2018)\citenamefont
  {Kavanagh}, \citenamefont {Gaggero},\ and\ \citenamefont
  {Bertone}}]{Kavanagh:2018ggo}%
  \BibitemOpen
  \bibfield  {author} {\bibinfo {author} {\bibfnamefont {B.~J.}\ \bibnamefont
  {Kavanagh}}, \bibinfo {author} {\bibfnamefont {D.}~\bibnamefont {Gaggero}},\
  and\ \bibinfo {author} {\bibfnamefont {G.}~\bibnamefont {Bertone}},\ }\href
  {https://doi.org/10.1103/PhysRevD.98.023536} {\bibfield  {journal} {\bibinfo
  {journal} {Phys. Rev. D}\ }\textbf {\bibinfo {volume} {98}},\ \bibinfo
  {pages} {023536} (\bibinfo {year} {2018})},\ \Eprint
  {https://arxiv.org/abs/1805.09034} {arXiv:1805.09034 [astro-ph.CO]}
  \BibitemShut {NoStop}%
\bibitem [{\citenamefont {Talbot}\ and\ \citenamefont
  {Thrane}(2017)}]{Talbot:2017yur}%
  \BibitemOpen
  \bibfield  {author} {\bibinfo {author} {\bibfnamefont {C.}~\bibnamefont
  {Talbot}}\ and\ \bibinfo {author} {\bibfnamefont {E.}~\bibnamefont
  {Thrane}},\ }\href {https://doi.org/10.1103/PhysRevD.96.023012} {\bibfield
  {journal} {\bibinfo  {journal} {Phys. Rev. D}\ }\textbf {\bibinfo {volume}
  {96}},\ \bibinfo {pages} {023012} (\bibinfo {year} {2017})},\ \Eprint
  {https://arxiv.org/abs/1704.08370} {arXiv:1704.08370 [astro-ph.HE]}
  \BibitemShut {NoStop}%
\bibitem [{\citenamefont {Mapelli}(2020)}]{Mapelli:2020vfa}%
  \BibitemOpen
  \bibfield  {author} {\bibinfo {author} {\bibfnamefont {M.}~\bibnamefont
  {Mapelli}},\ }\href {https://doi.org/10.3389/fspas.2020.00038} {\bibfield
  {journal} {\bibinfo  {journal} {Front. Astron. Space Sci.}\ }\textbf
  {\bibinfo {volume} {7}},\ \bibinfo {pages} {38} (\bibinfo {year}
  {2020})}\BibitemShut {NoStop}%
\bibitem [{\citenamefont {Breivik}\ \emph {et~al.}(2016)\citenamefont
  {Breivik}, \citenamefont {Rodriguez}, \citenamefont {Larson}, \citenamefont
  {Kalogera},\ and\ \citenamefont {Rasio}}]{Breivik:2016ddj}%
  \BibitemOpen
  \bibfield  {author} {\bibinfo {author} {\bibfnamefont {K.}~\bibnamefont
  {Breivik}}, \bibinfo {author} {\bibfnamefont {C.~L.}\ \bibnamefont
  {Rodriguez}}, \bibinfo {author} {\bibfnamefont {S.~L.}\ \bibnamefont
  {Larson}}, \bibinfo {author} {\bibfnamefont {V.}~\bibnamefont {Kalogera}},\
  and\ \bibinfo {author} {\bibfnamefont {F.~A.}\ \bibnamefont {Rasio}},\ }\href
  {https://doi.org/10.3847/2041-8205/830/1/L18} {\bibfield  {journal} {\bibinfo
   {journal} {Astrophys. J. Lett.}\ }\textbf {\bibinfo {volume} {830}},\
  \bibinfo {pages} {L18} (\bibinfo {year} {2016})},\ \Eprint
  {https://arxiv.org/abs/1606.09558} {arXiv:1606.09558 [astro-ph.GA]}
  \BibitemShut {NoStop}%
\bibitem [{\citenamefont {Fragione}\ and\ \citenamefont
  {Kocsis}(2018)}]{Fragione:2018vty}%
  \BibitemOpen
  \bibfield  {author} {\bibinfo {author} {\bibfnamefont {G.}~\bibnamefont
  {Fragione}}\ and\ \bibinfo {author} {\bibfnamefont {B.}~\bibnamefont
  {Kocsis}},\ }\href {https://doi.org/10.1103/PhysRevLett.121.161103}
  {\bibfield  {journal} {\bibinfo  {journal} {Phys. Rev. Lett.}\ }\textbf
  {\bibinfo {volume} {121}},\ \bibinfo {pages} {161103} (\bibinfo {year}
  {2018})},\ \Eprint {https://arxiv.org/abs/1806.02351} {arXiv:1806.02351
  [astro-ph.GA]} \BibitemShut {NoStop}%
\bibitem [{\citenamefont {Cardoso}\ and\ \citenamefont
  {Pani}(2019)}]{Cardoso:2019rvt}%
  \BibitemOpen
  \bibfield  {author} {\bibinfo {author} {\bibfnamefont {V.}~\bibnamefont
  {Cardoso}}\ and\ \bibinfo {author} {\bibfnamefont {P.}~\bibnamefont {Pani}},\
  }\href {https://doi.org/10.1007/s41114-019-0020-4} {\bibfield  {journal}
  {\bibinfo  {journal} {Living Rev. Rel.}\ }\textbf {\bibinfo {volume} {22}},\
  \bibinfo {pages} {4} (\bibinfo {year} {2019})},\ \Eprint
  {https://arxiv.org/abs/1904.05363} {arXiv:1904.05363 [gr-qc]} \BibitemShut
  {NoStop}%
\bibitem [{\citenamefont {Barausse}\ \emph {et~al.}(2014)\citenamefont
  {Barausse}, \citenamefont {Cardoso},\ and\ \citenamefont
  {Pani}}]{Barausse:2014tra}%
  \BibitemOpen
  \bibfield  {author} {\bibinfo {author} {\bibfnamefont {E.}~\bibnamefont
  {Barausse}}, \bibinfo {author} {\bibfnamefont {V.}~\bibnamefont {Cardoso}},\
  and\ \bibinfo {author} {\bibfnamefont {P.}~\bibnamefont {Pani}},\ }\href
  {https://doi.org/10.1103/PhysRevD.89.104059} {\bibfield  {journal} {\bibinfo
  {journal} {Phys. Rev. D}\ }\textbf {\bibinfo {volume} {89}},\ \bibinfo
  {pages} {104059} (\bibinfo {year} {2014})},\ \Eprint
  {https://arxiv.org/abs/1404.7149} {arXiv:1404.7149 [gr-qc]} \BibitemShut
  {NoStop}%
\bibitem [{\citenamefont {Cardoso}\ and\ \citenamefont
  {Maselli}(2020)}]{Cardoso:2019rou}%
  \BibitemOpen
  \bibfield  {author} {\bibinfo {author} {\bibfnamefont {V.}~\bibnamefont
  {Cardoso}}\ and\ \bibinfo {author} {\bibfnamefont {A.}~\bibnamefont
  {Maselli}},\ }\href {https://doi.org/10.1051/0004-6361/202037654} {\bibfield
  {journal} {\bibinfo  {journal} {Astron. Astrophys.}\ }\textbf {\bibinfo
  {volume} {644}},\ \bibinfo {pages} {A147} (\bibinfo {year} {2020})},\ \Eprint
  {https://arxiv.org/abs/1909.05870} {arXiv:1909.05870 [astro-ph.HE]}
  \BibitemShut {NoStop}%
\bibitem [{\citenamefont {Yunes}\ \emph {et~al.}(2011)\citenamefont {Yunes},
  \citenamefont {Kocsis}, \citenamefont {Loeb},\ and\ \citenamefont
  {Haiman}}]{Yunes:2011ws}%
  \BibitemOpen
  \bibfield  {author} {\bibinfo {author} {\bibfnamefont {N.}~\bibnamefont
  {Yunes}}, \bibinfo {author} {\bibfnamefont {B.}~\bibnamefont {Kocsis}},
  \bibinfo {author} {\bibfnamefont {A.}~\bibnamefont {Loeb}},\ and\ \bibinfo
  {author} {\bibfnamefont {Z.}~\bibnamefont {Haiman}},\ }\href
  {https://doi.org/10.1103/PhysRevLett.107.171103} {\bibfield  {journal}
  {\bibinfo  {journal} {Phys. Rev. Lett.}\ }\textbf {\bibinfo {volume} {107}},\
  \bibinfo {pages} {171103} (\bibinfo {year} {2011})},\ \Eprint
  {https://arxiv.org/abs/1103.4609} {arXiv:1103.4609 [astro-ph.CO]}
  \BibitemShut {NoStop}%
\bibitem [{\citenamefont {Kocsis}\ \emph {et~al.}(2011)\citenamefont {Kocsis},
  \citenamefont {Yunes},\ and\ \citenamefont {Loeb}}]{Kocsis:2011dr}%
  \BibitemOpen
  \bibfield  {author} {\bibinfo {author} {\bibfnamefont {B.}~\bibnamefont
  {Kocsis}}, \bibinfo {author} {\bibfnamefont {N.}~\bibnamefont {Yunes}},\ and\
  \bibinfo {author} {\bibfnamefont {A.}~\bibnamefont {Loeb}},\ }\href
  {https://doi.org/10.1103/PhysRevD.86.049907} {\bibfield  {journal} {\bibinfo
  {journal} {Phys. Rev. D}\ }\textbf {\bibinfo {volume} {84}},\ \bibinfo
  {pages} {024032} (\bibinfo {year} {2011})},\ \Eprint
  {https://arxiv.org/abs/1104.2322} {arXiv:1104.2322 [astro-ph.GA]}
  \BibitemShut {NoStop}%
\bibitem [{\citenamefont {Eda}\ \emph {et~al.}(2013)\citenamefont {Eda},
  \citenamefont {Itoh}, \citenamefont {Kuroyanagi},\ and\ \citenamefont
  {Silk}}]{Eda:2013gg}%
  \BibitemOpen
  \bibfield  {author} {\bibinfo {author} {\bibfnamefont {K.}~\bibnamefont
  {Eda}}, \bibinfo {author} {\bibfnamefont {Y.}~\bibnamefont {Itoh}}, \bibinfo
  {author} {\bibfnamefont {S.}~\bibnamefont {Kuroyanagi}},\ and\ \bibinfo
  {author} {\bibfnamefont {J.}~\bibnamefont {Silk}},\ }\href
  {https://doi.org/10.1103/PhysRevLett.110.221101} {\bibfield  {journal}
  {\bibinfo  {journal} {Phys. Rev. Lett.}\ }\textbf {\bibinfo {volume} {110}},\
  \bibinfo {pages} {221101} (\bibinfo {year} {2013})},\ \Eprint
  {https://arxiv.org/abs/1301.5971} {arXiv:1301.5971 [gr-qc]} \BibitemShut
  {NoStop}%
\bibitem [{\citenamefont {Macedo}\ \emph {et~al.}(2013)\citenamefont {Macedo},
  \citenamefont {Pani}, \citenamefont {Cardoso},\ and\ \citenamefont
  {Crispino}}]{Macedo:2013qea}%
  \BibitemOpen
  \bibfield  {author} {\bibinfo {author} {\bibfnamefont {C.~F.~B.}\
  \bibnamefont {Macedo}}, \bibinfo {author} {\bibfnamefont {P.}~\bibnamefont
  {Pani}}, \bibinfo {author} {\bibfnamefont {V.}~\bibnamefont {Cardoso}},\ and\
  \bibinfo {author} {\bibfnamefont {L.~C.~B.}\ \bibnamefont {Crispino}},\
  }\href {https://doi.org/10.1088/0004-637X/774/1/48} {\bibfield  {journal}
  {\bibinfo  {journal} {Astrophys. J.}\ }\textbf {\bibinfo {volume} {774}},\
  \bibinfo {pages} {48} (\bibinfo {year} {2013})},\ \Eprint
  {https://arxiv.org/abs/1302.2646} {arXiv:1302.2646 [gr-qc]} \BibitemShut
  {NoStop}%
\bibitem [{\citenamefont {Coogan}\ \emph {et~al.}(2022)\citenamefont {Coogan},
  \citenamefont {Bertone}, \citenamefont {Gaggero}, \citenamefont {Kavanagh},\
  and\ \citenamefont {Nichols}}]{Coogan:2021uqv}%
  \BibitemOpen
  \bibfield  {author} {\bibinfo {author} {\bibfnamefont {A.}~\bibnamefont
  {Coogan}}, \bibinfo {author} {\bibfnamefont {G.}~\bibnamefont {Bertone}},
  \bibinfo {author} {\bibfnamefont {D.}~\bibnamefont {Gaggero}}, \bibinfo
  {author} {\bibfnamefont {B.~J.}\ \bibnamefont {Kavanagh}},\ and\ \bibinfo
  {author} {\bibfnamefont {D.~A.}\ \bibnamefont {Nichols}},\ }\href
  {https://doi.org/10.1103/PhysRevD.105.043009} {\bibfield  {journal} {\bibinfo
   {journal} {Phys. Rev. D}\ }\textbf {\bibinfo {volume} {105}},\ \bibinfo
  {pages} {043009} (\bibinfo {year} {2022})},\ \Eprint
  {https://arxiv.org/abs/2108.04154} {arXiv:2108.04154 [gr-qc]} \BibitemShut
  {NoStop}%
\bibitem [{\citenamefont {Cardoso}\ and\ \citenamefont
  {Macedo}(2020)}]{Cardoso:2020lxx}%
  \BibitemOpen
  \bibfield  {author} {\bibinfo {author} {\bibfnamefont {V.}~\bibnamefont
  {Cardoso}}\ and\ \bibinfo {author} {\bibfnamefont {C.~F.~B.}\ \bibnamefont
  {Macedo}},\ }\href {https://doi.org/10.1093/mnras/staa2396} {\bibfield
  {journal} {\bibinfo  {journal} {Mon. Not. Roy. Astron. Soc.}\ }\textbf
  {\bibinfo {volume} {498}},\ \bibinfo {pages} {1963} (\bibinfo {year}
  {2020})},\ \Eprint {https://arxiv.org/abs/2008.01091} {arXiv:2008.01091
  [astro-ph.HE]} \BibitemShut {NoStop}%
\bibitem [{\citenamefont {Bertone}\ and\ \citenamefont
  {Tait}(2018)}]{Bertone:2018krk}%
  \BibitemOpen
  \bibfield  {author} {\bibinfo {author} {\bibfnamefont {G.}~\bibnamefont
  {Bertone}}\ and\ \bibinfo {author} {\bibfnamefont {T.}~\bibnamefont {Tait},
  \bibfnamefont {M.~P.}},\ }\href {https://doi.org/10.1038/s41586-018-0542-z}
  {\bibfield  {journal} {\bibinfo  {journal} {Nature}\ }\textbf {\bibinfo
  {volume} {562}},\ \bibinfo {pages} {51} (\bibinfo {year} {2018})},\ \Eprint
  {https://arxiv.org/abs/1810.01668} {arXiv:1810.01668 [astro-ph.CO]}
  \BibitemShut {NoStop}%
\bibitem [{\citenamefont {Cole}\ \emph {et~al.}(2022)\citenamefont {Cole},
  \citenamefont {Coogan}, \citenamefont {Kavanagh},\ and\ \citenamefont
  {Bertone}}]{Cole:2022ucw}%
  \BibitemOpen
  \bibfield  {author} {\bibinfo {author} {\bibfnamefont {P.~S.}\ \bibnamefont
  {Cole}}, \bibinfo {author} {\bibfnamefont {A.}~\bibnamefont {Coogan}},
  \bibinfo {author} {\bibfnamefont {B.~J.}\ \bibnamefont {Kavanagh}},\ and\
  \bibinfo {author} {\bibfnamefont {G.}~\bibnamefont {Bertone}},\ }\href@noop
  {} {\  (\bibinfo {year} {2022})},\ \Eprint {https://arxiv.org/abs/2207.07576}
  {arXiv:2207.07576 [astro-ph.CO]} \BibitemShut {NoStop}%
\bibitem [{\citenamefont {Kavanagh}\ \emph {et~al.}(2020)\citenamefont
  {Kavanagh}, \citenamefont {Nichols}, \citenamefont {Bertone},\ and\
  \citenamefont {Gaggero}}]{Kavanagh:2020cfn}%
  \BibitemOpen
  \bibfield  {author} {\bibinfo {author} {\bibfnamefont {B.~J.}\ \bibnamefont
  {Kavanagh}}, \bibinfo {author} {\bibfnamefont {D.~A.}\ \bibnamefont
  {Nichols}}, \bibinfo {author} {\bibfnamefont {G.}~\bibnamefont {Bertone}},\
  and\ \bibinfo {author} {\bibfnamefont {D.}~\bibnamefont {Gaggero}},\ }\href
  {https://doi.org/10.1103/PhysRevD.102.083006} {\bibfield  {journal} {\bibinfo
   {journal} {Phys. Rev. D}\ }\textbf {\bibinfo {volume} {102}},\ \bibinfo
  {pages} {083006} (\bibinfo {year} {2020})},\ \Eprint
  {https://arxiv.org/abs/2002.12811} {arXiv:2002.12811 [gr-qc]} \BibitemShut
  {NoStop}%
\bibitem [{\citenamefont {Hannuksela}\ \emph {et~al.}(2019)\citenamefont
  {Hannuksela}, \citenamefont {Wong}, \citenamefont {Brito}, \citenamefont
  {Berti},\ and\ \citenamefont {Li}}]{Hannuksela:2018izj}%
  \BibitemOpen
  \bibfield  {author} {\bibinfo {author} {\bibfnamefont {O.~A.}\ \bibnamefont
  {Hannuksela}}, \bibinfo {author} {\bibfnamefont {K.~W.~K.}\ \bibnamefont
  {Wong}}, \bibinfo {author} {\bibfnamefont {R.}~\bibnamefont {Brito}},
  \bibinfo {author} {\bibfnamefont {E.}~\bibnamefont {Berti}},\ and\ \bibinfo
  {author} {\bibfnamefont {T.~G.~F.}\ \bibnamefont {Li}},\ }\href
  {https://doi.org/10.1038/s41550-019-0712-4} {\bibfield  {journal} {\bibinfo
  {journal} {Nature Astron.}\ }\textbf {\bibinfo {volume} {3}},\ \bibinfo
  {pages} {447} (\bibinfo {year} {2019})},\ \Eprint
  {https://arxiv.org/abs/1804.09659} {arXiv:1804.09659 [astro-ph.HE]}
  \BibitemShut {NoStop}%
\bibitem [{\citenamefont {Baryakhtar}\ \emph {et~al.}(2022)\citenamefont
  {Baryakhtar} \emph {et~al.}}]{Baryakhtar:2022hbu}%
  \BibitemOpen
  \bibfield  {author} {\bibinfo {author} {\bibfnamefont {M.}~\bibnamefont
  {Baryakhtar}} \emph {et~al.},\ }in\ \href@noop {} {\emph {\bibinfo
  {booktitle} {{2022 Snowmass Summer Study}}}}\ (\bibinfo {year} {2022})\
  \Eprint {https://arxiv.org/abs/2203.07984} {arXiv:2203.07984 [hep-ph]}
  \BibitemShut {NoStop}%
\bibitem [{\citenamefont {Alves~Batista}\ \emph {et~al.}(2021)\citenamefont
  {Alves~Batista} \emph {et~al.}}]{AlvesBatista:2021gzc}%
  \BibitemOpen
  \bibfield  {author} {\bibinfo {author} {\bibfnamefont {R.}~\bibnamefont
  {Alves~Batista}} \emph {et~al.},\ }\href@noop {} {\  (\bibinfo {year}
  {2021})},\ \Eprint {https://arxiv.org/abs/2110.10074} {arXiv:2110.10074
  [astro-ph.HE]} \BibitemShut {NoStop}%
\bibitem [{\citenamefont {Zwick}\ \emph {et~al.}(2022)\citenamefont {Zwick},
  \citenamefont {Derdzinski}, \citenamefont {Garg}, \citenamefont {Capelo},\
  and\ \citenamefont {Mayer}}]{Zwick:2021dlg}%
  \BibitemOpen
  \bibfield  {author} {\bibinfo {author} {\bibfnamefont {L.}~\bibnamefont
  {Zwick}}, \bibinfo {author} {\bibfnamefont {A.}~\bibnamefont {Derdzinski}},
  \bibinfo {author} {\bibfnamefont {M.}~\bibnamefont {Garg}}, \bibinfo {author}
  {\bibfnamefont {P.~R.}\ \bibnamefont {Capelo}},\ and\ \bibinfo {author}
  {\bibfnamefont {L.}~\bibnamefont {Mayer}},\ }\href
  {https://doi.org/10.1093/mnras/stac299} {\bibfield  {journal} {\bibinfo
  {journal} {Mon. Not. Roy. Astron. Soc.}\ }\textbf {\bibinfo {volume} {511}},\
  \bibinfo {pages} {6143} (\bibinfo {year} {2022})},\ \Eprint
  {https://arxiv.org/abs/2110.09097} {arXiv:2110.09097 [astro-ph.HE]}
  \BibitemShut {NoStop}%
\bibitem [{\citenamefont {Cardoso}\ and\ \citenamefont
  {Duque}(2020)}]{Cardoso:2019upw}%
  \BibitemOpen
  \bibfield  {author} {\bibinfo {author} {\bibfnamefont {V.}~\bibnamefont
  {Cardoso}}\ and\ \bibinfo {author} {\bibfnamefont {F.}~\bibnamefont
  {Duque}},\ }\href {https://doi.org/10.1103/PhysRevD.101.064028} {\bibfield
  {journal} {\bibinfo  {journal} {Phys. Rev. D}\ }\textbf {\bibinfo {volume}
  {101}},\ \bibinfo {pages} {064028} (\bibinfo {year} {2020})},\ \Eprint
  {https://arxiv.org/abs/1912.07616} {arXiv:1912.07616 [gr-qc]} \BibitemShut
  {NoStop}%
\bibitem [{\citenamefont {Pato}\ \emph {et~al.}(2015)\citenamefont {Pato},
  \citenamefont {Iocco},\ and\ \citenamefont {Bertone}}]{Pato:2015dua}%
  \BibitemOpen
  \bibfield  {author} {\bibinfo {author} {\bibfnamefont {M.}~\bibnamefont
  {Pato}}, \bibinfo {author} {\bibfnamefont {F.}~\bibnamefont {Iocco}},\ and\
  \bibinfo {author} {\bibfnamefont {G.}~\bibnamefont {Bertone}},\ }\href
  {https://doi.org/10.1088/1475-7516/2015/12/001} {\bibfield  {journal}
  {\bibinfo  {journal} {JCAP}\ }\textbf {\bibinfo {volume} {12}},\ \bibinfo
  {pages} {001}},\ \Eprint {https://arxiv.org/abs/1504.06324} {arXiv:1504.06324
  [astro-ph.GA]} \BibitemShut {NoStop}%
\bibitem [{\citenamefont {Nesti}\ and\ \citenamefont
  {Salucci}(2013)}]{Nesti:2013uwa}%
  \BibitemOpen
  \bibfield  {author} {\bibinfo {author} {\bibfnamefont {F.}~\bibnamefont
  {Nesti}}\ and\ \bibinfo {author} {\bibfnamefont {P.}~\bibnamefont
  {Salucci}},\ }\href {https://doi.org/10.1088/1475-7516/2013/07/016}
  {\bibfield  {journal} {\bibinfo  {journal} {JCAP}\ }\textbf {\bibinfo
  {volume} {07}},\ \bibinfo {pages} {016}},\ \Eprint
  {https://arxiv.org/abs/1304.5127} {arXiv:1304.5127 [astro-ph.GA]}
  \BibitemShut {NoStop}%
\bibitem [{\citenamefont {Li}\ \emph {et~al.}(2020)\citenamefont {Li},
  \citenamefont {Shen},\ and\ \citenamefont {Schive}}]{Li:2020qva}%
  \BibitemOpen
  \bibfield  {author} {\bibinfo {author} {\bibfnamefont {Z.}~\bibnamefont
  {Li}}, \bibinfo {author} {\bibfnamefont {J.}~\bibnamefont {Shen}},\ and\
  \bibinfo {author} {\bibfnamefont {H.-Y.}\ \bibnamefont {Schive}}\ }\href
  {https://doi.org/10.3847/1538-4357/ab6598} {10.3847/1538-4357/ab6598}
  (\bibinfo {year} {2020}),\ \Eprint {https://arxiv.org/abs/2001.00318}
  {arXiv:2001.00318 [astro-ph.GA]} \BibitemShut {NoStop}%
\bibitem [{\citenamefont {De~Martino}\ \emph {et~al.}(2020)\citenamefont
  {De~Martino}, \citenamefont {Broadhurst}, \citenamefont {Tye}, \citenamefont
  {Chiueh},\ and\ \citenamefont {Schive}}]{DeMartino:2018zkx}%
  \BibitemOpen
  \bibfield  {author} {\bibinfo {author} {\bibfnamefont {I.}~\bibnamefont
  {De~Martino}}, \bibinfo {author} {\bibfnamefont {T.}~\bibnamefont
  {Broadhurst}}, \bibinfo {author} {\bibfnamefont {S.~H.~H.}\ \bibnamefont
  {Tye}}, \bibinfo {author} {\bibfnamefont {T.}~\bibnamefont {Chiueh}},\ and\
  \bibinfo {author} {\bibfnamefont {H.-Y.}\ \bibnamefont {Schive}},\ }\href
  {https://doi.org/10.1016/j.dark.2020.100503} {\bibfield  {journal} {\bibinfo
  {journal} {Phys. Dark Univ.}\ }\textbf {\bibinfo {volume} {28}},\ \bibinfo
  {pages} {100503} (\bibinfo {year} {2020})},\ \Eprint
  {https://arxiv.org/abs/1807.08153} {arXiv:1807.08153 [astro-ph.GA]}
  \BibitemShut {NoStop}%
\bibitem [{\citenamefont {Ablimit}\ \emph {et~al.}(2020)\citenamefont
  {Ablimit}, \citenamefont {Zhao}, \citenamefont {Flynn},\ and\ \citenamefont
  {Bird}}]{Ablimit:2020gxw}%
  \BibitemOpen
  \bibfield  {author} {\bibinfo {author} {\bibfnamefont {I.}~\bibnamefont
  {Ablimit}}, \bibinfo {author} {\bibfnamefont {G.}~\bibnamefont {Zhao}},
  \bibinfo {author} {\bibfnamefont {C.}~\bibnamefont {Flynn}},\ and\ \bibinfo
  {author} {\bibfnamefont {S.~A.}\ \bibnamefont {Bird}},\ }\href
  {https://doi.org/10.3847/2041-8213/ab8d45} {\bibfield  {journal} {\bibinfo
  {journal} {Astrophys. J.}\ }\textbf {\bibinfo {volume} {895}},\ \bibinfo
  {pages} {L12} (\bibinfo {year} {2020})},\ \Eprint
  {https://arxiv.org/abs/2004.13768} {arXiv:2004.13768 [astro-ph.GA]}
  \BibitemShut {NoStop}%
\bibitem [{\citenamefont {Gondolo}\ and\ \citenamefont
  {Silk}(1999)}]{Gondolo:1999ef}%
  \BibitemOpen
  \bibfield  {author} {\bibinfo {author} {\bibfnamefont {P.}~\bibnamefont
  {Gondolo}}\ and\ \bibinfo {author} {\bibfnamefont {J.}~\bibnamefont {Silk}},\
  }\href {https://doi.org/10.1103/PhysRevLett.83.1719} {\bibfield  {journal}
  {\bibinfo  {journal} {Phys. Rev. Lett.}\ }\textbf {\bibinfo {volume} {83}},\
  \bibinfo {pages} {1719} (\bibinfo {year} {1999})},\ \Eprint
  {https://arxiv.org/abs/astro-ph/9906391} {arXiv:astro-ph/9906391}
  \BibitemShut {NoStop}%
\bibitem [{\citenamefont {Ferrer}\ \emph {et~al.}(2017)\citenamefont {Ferrer},
  \citenamefont {da~Rosa},\ and\ \citenamefont {Will}}]{Ferrer:2017xwm}%
  \BibitemOpen
  \bibfield  {author} {\bibinfo {author} {\bibfnamefont {F.}~\bibnamefont
  {Ferrer}}, \bibinfo {author} {\bibfnamefont {A.~M.}\ \bibnamefont
  {da~Rosa}},\ and\ \bibinfo {author} {\bibfnamefont {C.~M.}\ \bibnamefont
  {Will}},\ }\href {https://doi.org/10.1103/PhysRevD.96.083014} {\bibfield
  {journal} {\bibinfo  {journal} {Phys. Rev. D}\ }\textbf {\bibinfo {volume}
  {96}},\ \bibinfo {pages} {083014} (\bibinfo {year} {2017})},\ \Eprint
  {https://arxiv.org/abs/1707.06302} {arXiv:1707.06302 [astro-ph.CO]}
  \BibitemShut {NoStop}%
\bibitem [{\citenamefont {Bertone}(2007)}]{Bertone:2007zz}%
  \BibitemOpen
  \bibfield  {author} {\bibinfo {author} {\bibfnamefont {G.}~\bibnamefont
  {Bertone}},\ }\href@noop {} {\bibfield  {journal} {\bibinfo  {journal}
  {Frascati Phys. Ser.}\ }\textbf {\bibinfo {volume} {45}},\ \bibinfo {pages}
  {313} (\bibinfo {year} {2007})}\BibitemShut {NoStop}%
\bibitem [{\citenamefont {Speeney}\ \emph {et~al.}(2022)\citenamefont
  {Speeney}, \citenamefont {Antonelli}, \citenamefont {Baibhav},\ and\
  \citenamefont {Berti}}]{Speeney:2022ryg}%
  \BibitemOpen
  \bibfield  {author} {\bibinfo {author} {\bibfnamefont {N.}~\bibnamefont
  {Speeney}}, \bibinfo {author} {\bibfnamefont {A.}~\bibnamefont {Antonelli}},
  \bibinfo {author} {\bibfnamefont {V.}~\bibnamefont {Baibhav}},\ and\ \bibinfo
  {author} {\bibfnamefont {E.}~\bibnamefont {Berti}},\ }\href
  {https://doi.org/10.1103/PhysRevD.106.044027} {\bibfield  {journal} {\bibinfo
   {journal} {Phys. Rev. D}\ }\textbf {\bibinfo {volume} {106}},\ \bibinfo
  {pages} {044027} (\bibinfo {year} {2022})},\ \Eprint
  {https://arxiv.org/abs/2204.12508} {arXiv:2204.12508 [gr-qc]} \BibitemShut
  {NoStop}%
\bibitem [{\citenamefont {Lacroix}(2018)}]{Lacroix:2018zmg}%
  \BibitemOpen
  \bibfield  {author} {\bibinfo {author} {\bibfnamefont {T.}~\bibnamefont
  {Lacroix}},\ }\href {https://doi.org/10.1051/0004-6361/201832652} {\bibfield
  {journal} {\bibinfo  {journal} {Astron. Astrophys.}\ }\textbf {\bibinfo
  {volume} {619}},\ \bibinfo {pages} {A46} (\bibinfo {year} {2018})},\ \Eprint
  {https://arxiv.org/abs/1801.01308} {arXiv:1801.01308 [astro-ph.GA]}
  \BibitemShut {NoStop}%
\bibitem [{\citenamefont {de~Jong}\ \emph {et~al.}(2022)\citenamefont
  {de~Jong}, \citenamefont {Aurrekoetxea},\ and\ \citenamefont
  {Lim}}]{deJong:2021bbo}%
  \BibitemOpen
  \bibfield  {author} {\bibinfo {author} {\bibfnamefont {E.}~\bibnamefont
  {de~Jong}}, \bibinfo {author} {\bibfnamefont {J.~C.}\ \bibnamefont
  {Aurrekoetxea}},\ and\ \bibinfo {author} {\bibfnamefont {E.~A.}\ \bibnamefont
  {Lim}},\ }\href {https://doi.org/10.1088/1475-7516/2022/03/029} {\bibfield
  {journal} {\bibinfo  {journal} {JCAP}\ }\textbf {\bibinfo {volume}
  {03}}\bibfield  {number} {\bibinfo  {number} { (03)},\ \bibinfo {pages}
  {029}},\ }\Eprint {https://arxiv.org/abs/2109.04896} {arXiv:2109.04896
  [astro-ph.CO]} \BibitemShut {NoStop}%
\bibitem [{\citenamefont {Ure\~na L\'opez}(2019)}]{Urena-Lopez:2019kud}%
  \BibitemOpen
  \bibfield  {author} {\bibinfo {author} {\bibfnamefont {L.~A.}\ \bibnamefont
  {Ure\~na L\'opez}},\ }\href {https://doi.org/10.3389/fspas.2019.00047}
  {\bibfield  {journal} {\bibinfo  {journal} {Front. Astron. Space Sci.}\
  }\textbf {\bibinfo {volume} {6}},\ \bibinfo {pages} {47} (\bibinfo {year}
  {2019})}\BibitemShut {NoStop}%
\bibitem [{\citenamefont {Hui}\ \emph {et~al.}(2017)\citenamefont {Hui},
  \citenamefont {Ostriker}, \citenamefont {Tremaine},\ and\ \citenamefont
  {Witten}}]{Hui:2016ltb}%
  \BibitemOpen
  \bibfield  {author} {\bibinfo {author} {\bibfnamefont {L.}~\bibnamefont
  {Hui}}, \bibinfo {author} {\bibfnamefont {J.~P.}\ \bibnamefont {Ostriker}},
  \bibinfo {author} {\bibfnamefont {S.}~\bibnamefont {Tremaine}},\ and\
  \bibinfo {author} {\bibfnamefont {E.}~\bibnamefont {Witten}},\ }\href
  {https://doi.org/10.1103/PhysRevD.95.043541} {\bibfield  {journal} {\bibinfo
  {journal} {Phys. Rev. D}\ }\textbf {\bibinfo {volume} {95}},\ \bibinfo
  {pages} {043541} (\bibinfo {year} {2017})},\ \Eprint
  {https://arxiv.org/abs/1610.08297} {arXiv:1610.08297 [astro-ph.CO]}
  \BibitemShut {NoStop}%
\bibitem [{\citenamefont {Hui}(2021)}]{Hui:2021tkt}%
  \BibitemOpen
  \bibfield  {author} {\bibinfo {author} {\bibfnamefont {L.}~\bibnamefont
  {Hui}},\ }\href {https://doi.org/10.1146/annurev-astro-120920-010024}
  {\bibfield  {journal} {\bibinfo  {journal} {Ann. Rev. Astron. Astrophys.}\
  }\textbf {\bibinfo {volume} {59}},\ \bibinfo {pages} {247} (\bibinfo {year}
  {2021})},\ \Eprint {https://arxiv.org/abs/2101.11735} {arXiv:2101.11735
  [astro-ph.CO]} \BibitemShut {NoStop}%
\bibitem [{\citenamefont {Ferreira}(2021)}]{Ferreira:2020fam}%
  \BibitemOpen
  \bibfield  {author} {\bibinfo {author} {\bibfnamefont {E.~G.~M.}\
  \bibnamefont {Ferreira}},\ }\href
  {https://doi.org/10.1007/s00159-021-00135-6} {\bibfield  {journal} {\bibinfo
  {journal} {Astron. Astrophys. Rev.}\ }\textbf {\bibinfo {volume} {29}},\
  \bibinfo {pages} {7} (\bibinfo {year} {2021})},\ \Eprint
  {https://arxiv.org/abs/2005.03254} {arXiv:2005.03254 [astro-ph.CO]}
  \BibitemShut {NoStop}%
\bibitem [{\citenamefont {Hu}\ \emph {et~al.}(2000)\citenamefont {Hu},
  \citenamefont {Barkana},\ and\ \citenamefont {Gruzinov}}]{Hu:2000ke}%
  \BibitemOpen
  \bibfield  {author} {\bibinfo {author} {\bibfnamefont {W.}~\bibnamefont
  {Hu}}, \bibinfo {author} {\bibfnamefont {R.}~\bibnamefont {Barkana}},\ and\
  \bibinfo {author} {\bibfnamefont {A.}~\bibnamefont {Gruzinov}},\ }\href
  {https://doi.org/10.1103/PhysRevLett.85.1158} {\bibfield  {journal} {\bibinfo
   {journal} {Phys. Rev. Lett.}\ }\textbf {\bibinfo {volume} {85}},\ \bibinfo
  {pages} {1158} (\bibinfo {year} {2000})},\ \Eprint
  {https://arxiv.org/abs/astro-ph/0003365} {arXiv:astro-ph/0003365}
  \BibitemShut {NoStop}%
\bibitem [{\citenamefont {Niemeyer}(2019)}]{Niemeyer:2019aqm}%
  \BibitemOpen
  \bibfield  {author} {\bibinfo {author} {\bibfnamefont {J.~C.}\ \bibnamefont
  {Niemeyer}}\ }\href {https://doi.org/10.1016/j.ppnp.2020.103787}
  {10.1016/j.ppnp.2020.103787} (\bibinfo {year} {2019}),\ \Eprint
  {https://arxiv.org/abs/1912.07064} {arXiv:1912.07064 [astro-ph.CO]}
  \BibitemShut {NoStop}%
\bibitem [{\citenamefont {Sikivie}(2011)}]{Sikivie:2010bq}%
  \BibitemOpen
  \bibfield  {author} {\bibinfo {author} {\bibfnamefont {P.}~\bibnamefont
  {Sikivie}},\ }\href {https://doi.org/10.1016/j.physletb.2010.11.027}
  {\bibfield  {journal} {\bibinfo  {journal} {Phys. Lett. B}\ }\textbf
  {\bibinfo {volume} {695}},\ \bibinfo {pages} {22} (\bibinfo {year} {2011})},\
  \Eprint {https://arxiv.org/abs/1003.2426} {arXiv:1003.2426 [astro-ph.GA]}
  \BibitemShut {NoStop}%
\bibitem [{\citenamefont {Hui}\ \emph {et~al.}(2021)\citenamefont {Hui},
  \citenamefont {Joyce}, \citenamefont {Landry},\ and\ \citenamefont
  {Li}}]{Hui:2020hbq}%
  \BibitemOpen
  \bibfield  {author} {\bibinfo {author} {\bibfnamefont {L.}~\bibnamefont
  {Hui}}, \bibinfo {author} {\bibfnamefont {A.}~\bibnamefont {Joyce}}, \bibinfo
  {author} {\bibfnamefont {M.~J.}\ \bibnamefont {Landry}},\ and\ \bibinfo
  {author} {\bibfnamefont {X.}~\bibnamefont {Li}},\ }\href
  {https://doi.org/10.1088/1475-7516/2021/01/011} {\bibfield  {journal}
  {\bibinfo  {journal} {JCAP}\ }\textbf {\bibinfo {volume} {01}},\ \bibinfo
  {pages} {011}},\ \Eprint {https://arxiv.org/abs/2004.01188} {arXiv:2004.01188
  [astro-ph.CO]} \BibitemShut {NoStop}%
\bibitem [{\citenamefont {Arvanitaki}\ \emph {et~al.}(2010)\citenamefont
  {Arvanitaki}, \citenamefont {Dimopoulos}, \citenamefont {Dubovsky},
  \citenamefont {Kaloper},\ and\ \citenamefont
  {March-Russell}}]{Arvanitaki:2009fg}%
  \BibitemOpen
  \bibfield  {author} {\bibinfo {author} {\bibfnamefont {A.}~\bibnamefont
  {Arvanitaki}}, \bibinfo {author} {\bibfnamefont {S.}~\bibnamefont
  {Dimopoulos}}, \bibinfo {author} {\bibfnamefont {S.}~\bibnamefont
  {Dubovsky}}, \bibinfo {author} {\bibfnamefont {N.}~\bibnamefont {Kaloper}},\
  and\ \bibinfo {author} {\bibfnamefont {J.}~\bibnamefont {March-Russell}},\
  }\href {https://doi.org/10.1103/PhysRevD.81.123530} {\bibfield  {journal}
  {\bibinfo  {journal} {Phys. Rev. D}\ }\textbf {\bibinfo {volume} {81}},\
  \bibinfo {pages} {123530} (\bibinfo {year} {2010})},\ \Eprint
  {https://arxiv.org/abs/0905.4720} {arXiv:0905.4720 [hep-th]} \BibitemShut
  {NoStop}%
\bibitem [{\citenamefont {Deng}\ \emph {et~al.}(2018)\citenamefont {Deng},
  \citenamefont {Hertzberg}, \citenamefont {Namjoo},\ and\ \citenamefont
  {Masoumi}}]{Deng:2018jjz}%
  \BibitemOpen
  \bibfield  {author} {\bibinfo {author} {\bibfnamefont {H.}~\bibnamefont
  {Deng}}, \bibinfo {author} {\bibfnamefont {M.~P.}\ \bibnamefont {Hertzberg}},
  \bibinfo {author} {\bibfnamefont {M.~H.}\ \bibnamefont {Namjoo}},\ and\
  \bibinfo {author} {\bibfnamefont {A.}~\bibnamefont {Masoumi}},\ }\href
  {https://doi.org/10.1103/PhysRevD.98.023513} {\bibfield  {journal} {\bibinfo
  {journal} {Phys. Rev. D}\ }\textbf {\bibinfo {volume} {98}},\ \bibinfo
  {pages} {023513} (\bibinfo {year} {2018})},\ \Eprint
  {https://arxiv.org/abs/1804.05921} {arXiv:1804.05921 [astro-ph.CO]}
  \BibitemShut {NoStop}%
\bibitem [{\citenamefont {Schive}\ \emph {et~al.}(2014)\citenamefont {Schive},
  \citenamefont {Chiueh},\ and\ \citenamefont {Broadhurst}}]{Schive:2014dra}%
  \BibitemOpen
  \bibfield  {author} {\bibinfo {author} {\bibfnamefont {H.-Y.}\ \bibnamefont
  {Schive}}, \bibinfo {author} {\bibfnamefont {T.}~\bibnamefont {Chiueh}},\
  and\ \bibinfo {author} {\bibfnamefont {T.}~\bibnamefont {Broadhurst}},\
  }\href {https://doi.org/10.1038/nphys2996} {\bibfield  {journal} {\bibinfo
  {journal} {Nature Phys.}\ }\textbf {\bibinfo {volume} {10}},\ \bibinfo
  {pages} {496} (\bibinfo {year} {2014})},\ \Eprint
  {https://arxiv.org/abs/1406.6586} {arXiv:1406.6586 [astro-ph.GA]}
  \BibitemShut {NoStop}%
\bibitem [{\citenamefont {Grin}\ \emph {et~al.}(2019)\citenamefont {Grin},
  \citenamefont {Amin}, \citenamefont {Gluscevic}, \citenamefont {Hlǒzek},
  \citenamefont {Marsh}, \citenamefont {Poulin}, \citenamefont
  {Prescod-Weinstein},\ and\ \citenamefont {Smith}}]{Grin:2019mub}%
  \BibitemOpen
  \bibfield  {author} {\bibinfo {author} {\bibfnamefont {D.}~\bibnamefont
  {Grin}}, \bibinfo {author} {\bibfnamefont {M.~A.}\ \bibnamefont {Amin}},
  \bibinfo {author} {\bibfnamefont {V.}~\bibnamefont {Gluscevic}}, \bibinfo
  {author} {\bibfnamefont {R.}~\bibnamefont {Hlǒzek}}, \bibinfo {author}
  {\bibfnamefont {D.~J.~E.}\ \bibnamefont {Marsh}}, \bibinfo {author}
  {\bibfnamefont {V.}~\bibnamefont {Poulin}}, \bibinfo {author} {\bibfnamefont
  {C.}~\bibnamefont {Prescod-Weinstein}},\ and\ \bibinfo {author}
  {\bibfnamefont {T.~L.}\ \bibnamefont {Smith}},\ }\href@noop {} {\  (\bibinfo
  {year} {2019})},\ \Eprint {https://arxiv.org/abs/1904.09003}
  {arXiv:1904.09003 [astro-ph.CO]} \BibitemShut {NoStop}%
\bibitem [{\citenamefont {Stott}(2020)}]{Stott:2020gjj}%
  \BibitemOpen
  \bibfield  {author} {\bibinfo {author} {\bibfnamefont {M.~J.}\ \bibnamefont
  {Stott}},\ }\href@noop {} {\  (\bibinfo {year} {2020})},\ \Eprint
  {https://arxiv.org/abs/2009.07206} {arXiv:2009.07206 [hep-ph]} \BibitemShut
  {NoStop}%
\bibitem [{\citenamefont {El-Zant}\ \emph {et~al.}(2020)\citenamefont
  {El-Zant}, \citenamefont {Roupas},\ and\ \citenamefont
  {Silk}}]{El-Zant:2020god}%
  \BibitemOpen
  \bibfield  {author} {\bibinfo {author} {\bibfnamefont {A.}~\bibnamefont
  {El-Zant}}, \bibinfo {author} {\bibfnamefont {Z.}~\bibnamefont {Roupas}},\
  and\ \bibinfo {author} {\bibfnamefont {J.}~\bibnamefont {Silk}},\ }\href
  {https://doi.org/10.1093/mnras/staa2972} {\bibfield  {journal} {\bibinfo
  {journal} {Mon. Not. Roy. Astron. Soc.}\ }\textbf {\bibinfo {volume} {499}},\
  \bibinfo {pages} {2575} (\bibinfo {year} {2020})},\ \Eprint
  {https://arxiv.org/abs/2009.10167} {arXiv:2009.10167 [astro-ph.GA]}
  \BibitemShut {NoStop}%
\bibitem [{\citenamefont {Marsh}(2016)}]{Marsh:2015xka}%
  \BibitemOpen
  \bibfield  {author} {\bibinfo {author} {\bibfnamefont {D.~J.~E.}\
  \bibnamefont {Marsh}},\ }\href
  {https://doi.org/10.1016/j.physrep.2016.06.005} {\bibfield  {journal}
  {\bibinfo  {journal} {Phys. Rept.}\ }\textbf {\bibinfo {volume} {643}},\
  \bibinfo {pages} {1} (\bibinfo {year} {2016})},\ \Eprint
  {https://arxiv.org/abs/1510.07633} {arXiv:1510.07633 [astro-ph.CO]}
  \BibitemShut {NoStop}%
\bibitem [{\citenamefont {Brito}\ \emph {et~al.}(2022)\citenamefont {Brito},
  \citenamefont {Chakrabarti}, \citenamefont {Clesse}, \citenamefont {Dvorkin},
  \citenamefont {Garcia-Bellido}, \citenamefont {Meyers}, \citenamefont {Ng},
  \citenamefont {Miller}, \citenamefont {Shandera},\ and\ \citenamefont
  {Sun}}]{Brito:2022lmd}%
  \BibitemOpen
  \bibfield  {author} {\bibinfo {author} {\bibfnamefont {R.}~\bibnamefont
  {Brito}}, \bibinfo {author} {\bibfnamefont {S.}~\bibnamefont {Chakrabarti}},
  \bibinfo {author} {\bibfnamefont {S.}~\bibnamefont {Clesse}}, \bibinfo
  {author} {\bibfnamefont {C.}~\bibnamefont {Dvorkin}}, \bibinfo {author}
  {\bibfnamefont {J.}~\bibnamefont {Garcia-Bellido}}, \bibinfo {author}
  {\bibfnamefont {J.}~\bibnamefont {Meyers}}, \bibinfo {author} {\bibfnamefont
  {K.~K.~Y.}\ \bibnamefont {Ng}}, \bibinfo {author} {\bibfnamefont {A.~L.}\
  \bibnamefont {Miller}}, \bibinfo {author} {\bibfnamefont {S.}~\bibnamefont
  {Shandera}},\ and\ \bibinfo {author} {\bibfnamefont {L.}~\bibnamefont
  {Sun}},\ }\href@noop {} {\  (\bibinfo {year} {2022})},\ \Eprint
  {https://arxiv.org/abs/2203.15954} {arXiv:2203.15954 [hep-ph]} \BibitemShut
  {NoStop}%
\bibitem [{\citenamefont {Hui}\ \emph {et~al.}(2022)\citenamefont {Hui},
  \citenamefont {Law}, \citenamefont {Santoni}, \citenamefont {Sun},
  \citenamefont {Tomaselli},\ and\ \citenamefont {Trincherini}}]{Hui:2022sri}%
  \BibitemOpen
  \bibfield  {author} {\bibinfo {author} {\bibfnamefont {L.}~\bibnamefont
  {Hui}}, \bibinfo {author} {\bibfnamefont {Y.~T.~A.}\ \bibnamefont {Law}},
  \bibinfo {author} {\bibfnamefont {L.}~\bibnamefont {Santoni}}, \bibinfo
  {author} {\bibfnamefont {G.}~\bibnamefont {Sun}}, \bibinfo {author}
  {\bibfnamefont {G.~M.}\ \bibnamefont {Tomaselli}},\ and\ \bibinfo {author}
  {\bibfnamefont {E.}~\bibnamefont {Trincherini}},\ }\href@noop {} {\
  (\bibinfo {year} {2022})},\ \Eprint {https://arxiv.org/abs/2208.06408}
  {arXiv:2208.06408 [gr-qc]} \BibitemShut {NoStop}%
\bibitem [{\citenamefont {Traykova}\ \emph {et~al.}(2021)\citenamefont
  {Traykova}, \citenamefont {Clough}, \citenamefont {Helfer}, \citenamefont
  {Berti}, \citenamefont {Ferreira},\ and\ \citenamefont
  {Hui}}]{Traykova:2021dua}%
  \BibitemOpen
  \bibfield  {author} {\bibinfo {author} {\bibfnamefont {D.}~\bibnamefont
  {Traykova}}, \bibinfo {author} {\bibfnamefont {K.}~\bibnamefont {Clough}},
  \bibinfo {author} {\bibfnamefont {T.}~\bibnamefont {Helfer}}, \bibinfo
  {author} {\bibfnamefont {E.}~\bibnamefont {Berti}}, \bibinfo {author}
  {\bibfnamefont {P.~G.}\ \bibnamefont {Ferreira}},\ and\ \bibinfo {author}
  {\bibfnamefont {L.}~\bibnamefont {Hui}},\ }\href
  {https://doi.org/10.1103/PhysRevD.104.103014} {\bibfield  {journal} {\bibinfo
   {journal} {Phys. Rev. D}\ }\textbf {\bibinfo {volume} {104}},\ \bibinfo
  {pages} {103014} (\bibinfo {year} {2021})},\ \Eprint
  {https://arxiv.org/abs/2106.08280} {arXiv:2106.08280 [gr-qc]} \BibitemShut
  {NoStop}%
\bibitem [{\citenamefont {Brax}\ \emph {et~al.}(2020)\citenamefont {Brax},
  \citenamefont {Valageas},\ and\ \citenamefont {Cembranos}}]{Brax:2019npi}%
  \BibitemOpen
  \bibfield  {author} {\bibinfo {author} {\bibfnamefont {P.}~\bibnamefont
  {Brax}}, \bibinfo {author} {\bibfnamefont {P.}~\bibnamefont {Valageas}},\
  and\ \bibinfo {author} {\bibfnamefont {J.~A.~R.}\ \bibnamefont {Cembranos}},\
  }\href {https://doi.org/10.1103/PhysRevD.101.023521} {\bibfield  {journal}
  {\bibinfo  {journal} {Phys. Rev. D}\ }\textbf {\bibinfo {volume} {101}},\
  \bibinfo {pages} {023521} (\bibinfo {year} {2020})},\ \Eprint
  {https://arxiv.org/abs/1909.02614} {arXiv:1909.02614 [astro-ph.CO]}
  \BibitemShut {NoStop}%
\bibitem [{\citenamefont {Brito}\ \emph
  {et~al.}(2015{\natexlab{a}})\citenamefont {Brito}, \citenamefont {Cardoso},\
  and\ \citenamefont {Okawa}}]{Brito:2015yga}%
  \BibitemOpen
  \bibfield  {author} {\bibinfo {author} {\bibfnamefont {R.}~\bibnamefont
  {Brito}}, \bibinfo {author} {\bibfnamefont {V.}~\bibnamefont {Cardoso}},\
  and\ \bibinfo {author} {\bibfnamefont {H.}~\bibnamefont {Okawa}},\ }\href
  {https://doi.org/10.1103/PhysRevLett.115.111301} {\bibfield  {journal}
  {\bibinfo  {journal} {Phys. Rev. Lett.}\ }\textbf {\bibinfo {volume} {115}},\
  \bibinfo {pages} {111301} (\bibinfo {year} {2015}{\natexlab{a}})},\ \Eprint
  {https://arxiv.org/abs/1508.04773} {arXiv:1508.04773 [gr-qc]} \BibitemShut
  {NoStop}%
\bibitem [{\citenamefont {Marsh}\ and\ \citenamefont
  {Pop}(2015)}]{Marsh:2015wka}%
  \BibitemOpen
  \bibfield  {author} {\bibinfo {author} {\bibfnamefont {D.~J.~E.}\
  \bibnamefont {Marsh}}\ and\ \bibinfo {author} {\bibfnamefont {A.-R.}\
  \bibnamefont {Pop}},\ }\href {https://doi.org/10.1093/mnras/stv1050}
  {\bibfield  {journal} {\bibinfo  {journal} {Mon. Not. Roy. Astron. Soc.}\
  }\textbf {\bibinfo {volume} {451}},\ \bibinfo {pages} {2479} (\bibinfo {year}
  {2015})},\ \Eprint {https://arxiv.org/abs/1502.03456} {arXiv:1502.03456
  [astro-ph.CO]} \BibitemShut {NoStop}%
\bibitem [{\citenamefont {Khlopov}\ \emph {et~al.}(1985)\citenamefont
  {Khlopov}, \citenamefont {Malomed},\ and\ \citenamefont
  {Zeldovich}}]{Khlopov:1985jw}%
  \BibitemOpen
  \bibfield  {author} {\bibinfo {author} {\bibfnamefont {M.}~\bibnamefont
  {Khlopov}}, \bibinfo {author} {\bibfnamefont {B.~A.}\ \bibnamefont
  {Malomed}},\ and\ \bibinfo {author} {\bibfnamefont {I.~B.}\ \bibnamefont
  {Zeldovich}},\ }\href@noop {} {\bibfield  {journal} {\bibinfo  {journal}
  {Mon. Not. Roy. Astron. Soc.}\ }\textbf {\bibinfo {volume} {215}},\ \bibinfo
  {pages} {575} (\bibinfo {year} {1985})}\BibitemShut {NoStop}%
\bibitem [{\citenamefont {Clough}\ \emph {et~al.}(2019)\citenamefont {Clough},
  \citenamefont {Ferreira},\ and\ \citenamefont {Lagos}}]{Clough:2019jpm}%
  \BibitemOpen
  \bibfield  {author} {\bibinfo {author} {\bibfnamefont {K.}~\bibnamefont
  {Clough}}, \bibinfo {author} {\bibfnamefont {P.~G.}\ \bibnamefont
  {Ferreira}},\ and\ \bibinfo {author} {\bibfnamefont {M.}~\bibnamefont
  {Lagos}},\ }\href {https://doi.org/10.1103/PhysRevD.100.063014} {\bibfield
  {journal} {\bibinfo  {journal} {Phys. Rev. D}\ }\textbf {\bibinfo {volume}
  {100}},\ \bibinfo {pages} {063014} (\bibinfo {year} {2019})},\ \Eprint
  {https://arxiv.org/abs/1904.12783} {arXiv:1904.12783 [gr-qc]} \BibitemShut
  {NoStop}%
\bibitem [{\citenamefont {Bamber}\ \emph
  {et~al.}(2021{\natexlab{a}})\citenamefont {Bamber}, \citenamefont {Clough},
  \citenamefont {Ferreira}, \citenamefont {Hui},\ and\ \citenamefont
  {Lagos}}]{Bamber:2020bpu}%
  \BibitemOpen
  \bibfield  {author} {\bibinfo {author} {\bibfnamefont {J.}~\bibnamefont
  {Bamber}}, \bibinfo {author} {\bibfnamefont {K.}~\bibnamefont {Clough}},
  \bibinfo {author} {\bibfnamefont {P.~G.}\ \bibnamefont {Ferreira}}, \bibinfo
  {author} {\bibfnamefont {L.}~\bibnamefont {Hui}},\ and\ \bibinfo {author}
  {\bibfnamefont {M.}~\bibnamefont {Lagos}},\ }\href
  {https://doi.org/10.1103/PhysRevD.103.044059} {\bibfield  {journal} {\bibinfo
   {journal} {Phys. Rev. D}\ }\textbf {\bibinfo {volume} {103}},\ \bibinfo
  {pages} {044059} (\bibinfo {year} {2021}{\natexlab{a}})},\ \Eprint
  {https://arxiv.org/abs/2011.07870} {arXiv:2011.07870 [gr-qc]} \BibitemShut
  {NoStop}%
\bibitem [{\citenamefont {Hui}\ \emph {et~al.}(2019)\citenamefont {Hui},
  \citenamefont {Kabat}, \citenamefont {Li}, \citenamefont {Santoni},\ and\
  \citenamefont {Wong}}]{Hui:2019aqm}%
  \BibitemOpen
  \bibfield  {author} {\bibinfo {author} {\bibfnamefont {L.}~\bibnamefont
  {Hui}}, \bibinfo {author} {\bibfnamefont {D.}~\bibnamefont {Kabat}}, \bibinfo
  {author} {\bibfnamefont {X.}~\bibnamefont {Li}}, \bibinfo {author}
  {\bibfnamefont {L.}~\bibnamefont {Santoni}},\ and\ \bibinfo {author}
  {\bibfnamefont {S.~S.~C.}\ \bibnamefont {Wong}},\ }\href
  {https://doi.org/10.1088/1475-7516/2019/06/038} {\bibfield  {journal}
  {\bibinfo  {journal} {JCAP}\ }\textbf {\bibinfo {volume} {06}},\ \bibinfo
  {pages} {038}},\ \Eprint {https://arxiv.org/abs/1904.12803} {arXiv:1904.12803
  [gr-qc]} \BibitemShut {NoStop}%
\bibitem [{\citenamefont {Vieira}\ \emph {et~al.}(2014)\citenamefont {Vieira},
  \citenamefont {Bezerra},\ and\ \citenamefont {Muniz}}]{Vieira:2014waa}%
  \BibitemOpen
  \bibfield  {author} {\bibinfo {author} {\bibfnamefont {H.~S.}\ \bibnamefont
  {Vieira}}, \bibinfo {author} {\bibfnamefont {V.~B.}\ \bibnamefont
  {Bezerra}},\ and\ \bibinfo {author} {\bibfnamefont {C.~R.}\ \bibnamefont
  {Muniz}},\ }\href {https://doi.org/10.1016/j.aop.2014.07.011} {\bibfield
  {journal} {\bibinfo  {journal} {Annals Phys.}\ }\textbf {\bibinfo {volume}
  {350}},\ \bibinfo {pages} {14} (\bibinfo {year} {2014})},\ \Eprint
  {https://arxiv.org/abs/1401.5397} {arXiv:1401.5397 [gr-qc]} \BibitemShut
  {NoStop}%
\bibitem [{\citenamefont {Santos}\ and\ \citenamefont
  {Herdeiro}(2020)}]{Santos:2020sut}%
  \BibitemOpen
  \bibfield  {author} {\bibinfo {author} {\bibfnamefont {N.~M.}\ \bibnamefont
  {Santos}}\ and\ \bibinfo {author} {\bibfnamefont {C.~A.~R.}\ \bibnamefont
  {Herdeiro}},\ }\href {https://doi.org/10.1142/S0218271820410138} {\bibfield
  {journal} {\bibinfo  {journal} {Int. J. Mod. Phys. D}\ }\textbf {\bibinfo
  {volume} {29}},\ \bibinfo {pages} {2041013} (\bibinfo {year} {2020})},\
  \Eprint {https://arxiv.org/abs/2005.07201} {arXiv:2005.07201 [gr-qc]}
  \BibitemShut {NoStop}%
\bibitem [{\citenamefont {Detweiler}(1980)}]{Detweiler:1980uk}%
  \BibitemOpen
  \bibfield  {author} {\bibinfo {author} {\bibfnamefont {S.~L.}\ \bibnamefont
  {Detweiler}},\ }\href {https://doi.org/10.1103/PhysRevD.22.2323} {\bibfield
  {journal} {\bibinfo  {journal} {Phys. Rev. D}\ }\textbf {\bibinfo {volume}
  {22}},\ \bibinfo {pages} {2323} (\bibinfo {year} {1980})}\BibitemShut
  {NoStop}%
\bibitem [{\citenamefont {Baumann}\ \emph
  {et~al.}(2019{\natexlab{a}})\citenamefont {Baumann}, \citenamefont {Chia},
  \citenamefont {Stout},\ and\ \citenamefont {ter Haar}}]{Baumann:2019eav}%
  \BibitemOpen
  \bibfield  {author} {\bibinfo {author} {\bibfnamefont {D.}~\bibnamefont
  {Baumann}}, \bibinfo {author} {\bibfnamefont {H.~S.}\ \bibnamefont {Chia}},
  \bibinfo {author} {\bibfnamefont {J.}~\bibnamefont {Stout}},\ and\ \bibinfo
  {author} {\bibfnamefont {L.}~\bibnamefont {ter Haar}},\ }\href
  {https://doi.org/10.1088/1475-7516/2019/12/006} {\bibfield  {journal}
  {\bibinfo  {journal} {JCAP}\ }\textbf {\bibinfo {volume} {12}},\ \bibinfo
  {pages} {006}},\ \Eprint {https://arxiv.org/abs/1908.10370} {arXiv:1908.10370
  [gr-qc]} \BibitemShut {NoStop}%
\bibitem [{\citenamefont {Sanchis-Gual}\ \emph {et~al.}(2020)\citenamefont
  {Sanchis-Gual}, \citenamefont {Zilh\~ao}, \citenamefont {Herdeiro},
  \citenamefont {Di~Giovanni}, \citenamefont {Font},\ and\ \citenamefont
  {Radu}}]{Sanchis-Gual:2020mzb}%
  \BibitemOpen
  \bibfield  {author} {\bibinfo {author} {\bibfnamefont {N.}~\bibnamefont
  {Sanchis-Gual}}, \bibinfo {author} {\bibfnamefont {M.}~\bibnamefont
  {Zilh\~ao}}, \bibinfo {author} {\bibfnamefont {C.}~\bibnamefont {Herdeiro}},
  \bibinfo {author} {\bibfnamefont {F.}~\bibnamefont {Di~Giovanni}}, \bibinfo
  {author} {\bibfnamefont {J.~A.}\ \bibnamefont {Font}},\ and\ \bibinfo
  {author} {\bibfnamefont {E.}~\bibnamefont {Radu}},\ }\href
  {https://doi.org/10.1103/PhysRevD.102.101504} {\bibfield  {journal} {\bibinfo
   {journal} {Phys. Rev. D}\ }\textbf {\bibinfo {volume} {102}},\ \bibinfo
  {pages} {101504(R)} (\bibinfo {year} {2020})},\ \Eprint
  {https://arxiv.org/abs/2007.11584} {arXiv:2007.11584 [gr-qc]} \BibitemShut
  {NoStop}%
\bibitem [{\citenamefont {Clough}\ \emph {et~al.}(2018)\citenamefont {Clough},
  \citenamefont {Dietrich},\ and\ \citenamefont {Niemeyer}}]{Clough:2018exo}%
  \BibitemOpen
  \bibfield  {author} {\bibinfo {author} {\bibfnamefont {K.}~\bibnamefont
  {Clough}}, \bibinfo {author} {\bibfnamefont {T.}~\bibnamefont {Dietrich}},\
  and\ \bibinfo {author} {\bibfnamefont {J.~C.}\ \bibnamefont {Niemeyer}},\
  }\href {https://doi.org/10.1103/PhysRevD.98.083020} {\bibfield  {journal}
  {\bibinfo  {journal} {Phys. Rev. D}\ }\textbf {\bibinfo {volume} {98}},\
  \bibinfo {pages} {083020} (\bibinfo {year} {2018})},\ \Eprint
  {https://arxiv.org/abs/1808.04668} {arXiv:1808.04668 [gr-qc]} \BibitemShut
  {NoStop}%
\bibitem [{\citenamefont {Sanchis-Gual}\ \emph {et~al.}(2016)\citenamefont
  {Sanchis-Gual}, \citenamefont {Degollado}, \citenamefont {Izquierdo},
  \citenamefont {Font},\ and\ \citenamefont {Montero}}]{Sanchis-Gual:2016jst}%
  \BibitemOpen
  \bibfield  {author} {\bibinfo {author} {\bibfnamefont {N.}~\bibnamefont
  {Sanchis-Gual}}, \bibinfo {author} {\bibfnamefont {J.~C.}\ \bibnamefont
  {Degollado}}, \bibinfo {author} {\bibfnamefont {P.}~\bibnamefont
  {Izquierdo}}, \bibinfo {author} {\bibfnamefont {J.~A.}\ \bibnamefont
  {Font}},\ and\ \bibinfo {author} {\bibfnamefont {P.~J.}\ \bibnamefont
  {Montero}},\ }\href {https://doi.org/10.1103/PhysRevD.94.043004} {\bibfield
  {journal} {\bibinfo  {journal} {Phys. Rev. D}\ }\textbf {\bibinfo {volume}
  {94}},\ \bibinfo {pages} {043004} (\bibinfo {year} {2016})},\ \Eprint
  {https://arxiv.org/abs/1606.05146} {arXiv:1606.05146 [gr-qc]} \BibitemShut
  {NoStop}%
\bibitem [{\citenamefont {Brito}\ \emph
  {et~al.}(2015{\natexlab{b}})\citenamefont {Brito}, \citenamefont {Cardoso},\
  and\ \citenamefont {Pani}}]{Brito:2015oca}%
  \BibitemOpen
  \bibfield  {author} {\bibinfo {author} {\bibfnamefont {R.}~\bibnamefont
  {Brito}}, \bibinfo {author} {\bibfnamefont {V.}~\bibnamefont {Cardoso}},\
  and\ \bibinfo {author} {\bibfnamefont {P.}~\bibnamefont {Pani}},\ }\href
  {https://doi.org/10.1007\/978-3-319-19000-6} {\bibfield  {journal} {\bibinfo
  {journal} {Lect. Notes Phys.}\ }\textbf {\bibinfo {volume} {906}},\ \bibinfo
  {pages} {pp.1} (\bibinfo {year} {2015}{\natexlab{b}})},\ \Eprint
  {https://arxiv.org/abs/1501.06570} {arXiv:1501.06570 [gr-qc]} \BibitemShut
  {NoStop}%
\bibitem [{\citenamefont {East}\ and\ \citenamefont
  {Pretorius}(2017)}]{East:2017ovw}%
  \BibitemOpen
  \bibfield  {author} {\bibinfo {author} {\bibfnamefont {W.~E.}\ \bibnamefont
  {East}}\ and\ \bibinfo {author} {\bibfnamefont {F.}~\bibnamefont
  {Pretorius}},\ }\href {https://doi.org/10.1103/PhysRevLett.119.041101}
  {\bibfield  {journal} {\bibinfo  {journal} {Phys. Rev. Lett.}\ }\textbf
  {\bibinfo {volume} {119}},\ \bibinfo {pages} {041101} (\bibinfo {year}
  {2017})},\ \Eprint {https://arxiv.org/abs/1704.04791} {arXiv:1704.04791
  [gr-qc]} \BibitemShut {NoStop}%
\bibitem [{\citenamefont {Ikeda}\ \emph {et~al.}(2021)\citenamefont {Ikeda},
  \citenamefont {Bernard}, \citenamefont {Cardoso},\ and\ \citenamefont
  {Zilh\~ao}}]{Ikeda:2020xvt}%
  \BibitemOpen
  \bibfield  {author} {\bibinfo {author} {\bibfnamefont {T.}~\bibnamefont
  {Ikeda}}, \bibinfo {author} {\bibfnamefont {L.}~\bibnamefont {Bernard}},
  \bibinfo {author} {\bibfnamefont {V.}~\bibnamefont {Cardoso}},\ and\ \bibinfo
  {author} {\bibfnamefont {M.}~\bibnamefont {Zilh\~ao}},\ }\href
  {https://doi.org/10.1103/PhysRevD.103.024020} {\bibfield  {journal} {\bibinfo
   {journal} {Phys. Rev. D}\ }\textbf {\bibinfo {volume} {103}},\ \bibinfo
  {pages} {024020} (\bibinfo {year} {2021})},\ \Eprint
  {https://arxiv.org/abs/2010.00008} {arXiv:2010.00008 [gr-qc]} \BibitemShut
  {NoStop}%
\bibitem [{\citenamefont {Day}\ and\ \citenamefont
  {McDonald}(2019)}]{Day:2019bbh}%
  \BibitemOpen
  \bibfield  {author} {\bibinfo {author} {\bibfnamefont {F.~V.}\ \bibnamefont
  {Day}}\ and\ \bibinfo {author} {\bibfnamefont {J.~I.}\ \bibnamefont
  {McDonald}},\ }\href {https://doi.org/10.1088/1475-7516/2019/10/051}
  {\bibfield  {journal} {\bibinfo  {journal} {JCAP}\ }\textbf {\bibinfo
  {volume} {10}},\ \bibinfo {pages} {051}},\ \Eprint
  {https://arxiv.org/abs/1904.08341} {arXiv:1904.08341 [hep-ph]} \BibitemShut
  {NoStop}%
\bibitem [{\citenamefont {Cardoso}\ \emph {et~al.}(2015)\citenamefont
  {Cardoso}, \citenamefont {Brito},\ and\ \citenamefont
  {Rosa}}]{Cardoso:2015zqa}%
  \BibitemOpen
  \bibfield  {author} {\bibinfo {author} {\bibfnamefont {V.}~\bibnamefont
  {Cardoso}}, \bibinfo {author} {\bibfnamefont {R.}~\bibnamefont {Brito}},\
  and\ \bibinfo {author} {\bibfnamefont {J.~L.}\ \bibnamefont {Rosa}},\ }\href
  {https://doi.org/10.1103/PhysRevD.91.124026} {\bibfield  {journal} {\bibinfo
  {journal} {Phys. Rev. D}\ }\textbf {\bibinfo {volume} {91}},\ \bibinfo
  {pages} {124026} (\bibinfo {year} {2015})},\ \Eprint
  {https://arxiv.org/abs/1505.05509} {arXiv:1505.05509 [gr-qc]} \BibitemShut
  {NoStop}%
\bibitem [{\citenamefont {Richartz}\ and\ \citenamefont
  {Saa}(2013)}]{Richartz:2013unq}%
  \BibitemOpen
  \bibfield  {author} {\bibinfo {author} {\bibfnamefont {M.}~\bibnamefont
  {Richartz}}\ and\ \bibinfo {author} {\bibfnamefont {A.}~\bibnamefont {Saa}},\
  }\href {https://doi.org/10.1103/PhysRevD.88.044008} {\bibfield  {journal}
  {\bibinfo  {journal} {Phys. Rev. D}\ }\textbf {\bibinfo {volume} {88}},\
  \bibinfo {pages} {044008} (\bibinfo {year} {2013})},\ \Eprint
  {https://arxiv.org/abs/1306.3137} {arXiv:1306.3137 [gr-qc]} \BibitemShut
  {NoStop}%
\bibitem [{\citenamefont {Cardoso}\ \emph {et~al.}(2017)\citenamefont
  {Cardoso}, \citenamefont {Pani},\ and\ \citenamefont {Yu}}]{Cardoso:2017kgn}%
  \BibitemOpen
  \bibfield  {author} {\bibinfo {author} {\bibfnamefont {V.}~\bibnamefont
  {Cardoso}}, \bibinfo {author} {\bibfnamefont {P.}~\bibnamefont {Pani}},\ and\
  \bibinfo {author} {\bibfnamefont {T.-T.}\ \bibnamefont {Yu}},\ }\href
  {https://doi.org/10.1103/PhysRevD.95.124056} {\bibfield  {journal} {\bibinfo
  {journal} {Phys. Rev. D}\ }\textbf {\bibinfo {volume} {95}},\ \bibinfo
  {pages} {124056} (\bibinfo {year} {2017})},\ \Eprint
  {https://arxiv.org/abs/1704.06151} {arXiv:1704.06151 [gr-qc]} \BibitemShut
  {NoStop}%
\bibitem [{\citenamefont {Merritt}\ and\ \citenamefont
  {Milosavljevic}(2002)}]{Merritt:2002jz}%
  \BibitemOpen
  \bibfield  {author} {\bibinfo {author} {\bibfnamefont {D.}~\bibnamefont
  {Merritt}}\ and\ \bibinfo {author} {\bibfnamefont {M.}~\bibnamefont
  {Milosavljevic}},\ }in\ \href@noop {} {\emph {\bibinfo {booktitle} {{4th
  International Heidelberg Conference on Dark Matter in Astro and Particle
  Physics}}}}\ (\bibinfo {year} {2002})\ pp.\ \bibinfo {pages} {79--89},\
  \Eprint {https://arxiv.org/abs/astro-ph/0205140} {arXiv:astro-ph/0205140}
  \BibitemShut {NoStop}%
\bibitem [{\citenamefont {Bertone}\ and\ \citenamefont
  {Merritt}(2005)}]{Bertone:2005hw}%
  \BibitemOpen
  \bibfield  {author} {\bibinfo {author} {\bibfnamefont {G.}~\bibnamefont
  {Bertone}}\ and\ \bibinfo {author} {\bibfnamefont {D.}~\bibnamefont
  {Merritt}},\ }\href {https://doi.org/10.1103/PhysRevD.72.103502} {\bibfield
  {journal} {\bibinfo  {journal} {Phys. Rev. D}\ }\textbf {\bibinfo {volume}
  {72}},\ \bibinfo {pages} {103502} (\bibinfo {year} {2005})},\ \Eprint
  {https://arxiv.org/abs/astro-ph/0501555} {arXiv:astro-ph/0501555}
  \BibitemShut {NoStop}%
\bibitem [{\citenamefont {Barausse}\ \emph {et~al.}(2020)\citenamefont
  {Barausse} \emph {et~al.}}]{Barausse:2020rsu}%
  \BibitemOpen
  \bibfield  {author} {\bibinfo {author} {\bibfnamefont {E.}~\bibnamefont
  {Barausse}} \emph {et~al.},\ }\href
  {https://doi.org/10.1007/s10714-020-02691-1} {\bibfield  {journal} {\bibinfo
  {journal} {Gen. Rel. Grav.}\ }\textbf {\bibinfo {volume} {52}},\ \bibinfo
  {pages} {81} (\bibinfo {year} {2020})},\ \Eprint
  {https://arxiv.org/abs/2001.09793} {arXiv:2001.09793 [gr-qc]} \BibitemShut
  {NoStop}%
\bibitem [{\citenamefont {Maselli}\ \emph {et~al.}(2022)\citenamefont
  {Maselli}, \citenamefont {Franchini}, \citenamefont {Gualtieri},
  \citenamefont {Sotiriou}, \citenamefont {Barsanti},\ and\ \citenamefont
  {Pani}}]{Maselli:2021men}%
  \BibitemOpen
  \bibfield  {author} {\bibinfo {author} {\bibfnamefont {A.}~\bibnamefont
  {Maselli}}, \bibinfo {author} {\bibfnamefont {N.}~\bibnamefont {Franchini}},
  \bibinfo {author} {\bibfnamefont {L.}~\bibnamefont {Gualtieri}}, \bibinfo
  {author} {\bibfnamefont {T.~P.}\ \bibnamefont {Sotiriou}}, \bibinfo {author}
  {\bibfnamefont {S.}~\bibnamefont {Barsanti}},\ and\ \bibinfo {author}
  {\bibfnamefont {P.}~\bibnamefont {Pani}},\ }\href
  {https://doi.org/10.1038/s41550-021-01589-5} {\bibfield  {journal} {\bibinfo
  {journal} {Nature Astron.}\ }\textbf {\bibinfo {volume} {6}},\ \bibinfo
  {pages} {464} (\bibinfo {year} {2022})},\ \Eprint
  {https://arxiv.org/abs/2106.11325} {arXiv:2106.11325 [gr-qc]} \BibitemShut
  {NoStop}%
\bibitem [{\citenamefont {Xie}\ and\ \citenamefont
  {Huang}(2022)}]{Xie:2022uvp}%
  \BibitemOpen
  \bibfield  {author} {\bibinfo {author} {\bibfnamefont {N.}~\bibnamefont
  {Xie}}\ and\ \bibinfo {author} {\bibfnamefont {F.~P.}\ \bibnamefont
  {Huang}},\ }\href@noop {} {\  (\bibinfo {year} {2022})},\ \Eprint
  {https://arxiv.org/abs/2207.11145} {arXiv:2207.11145 [hep-ph]} \BibitemShut
  {NoStop}%
\bibitem [{\citenamefont {Hannuksela}\ \emph {et~al.}(2020)\citenamefont
  {Hannuksela}, \citenamefont {Ng},\ and\ \citenamefont
  {Li}}]{Hannuksela:2019vip}%
  \BibitemOpen
  \bibfield  {author} {\bibinfo {author} {\bibfnamefont {O.~A.}\ \bibnamefont
  {Hannuksela}}, \bibinfo {author} {\bibfnamefont {K.~C.~Y.}\ \bibnamefont
  {Ng}},\ and\ \bibinfo {author} {\bibfnamefont {T.~G.~F.}\ \bibnamefont
  {Li}},\ }\href {https://doi.org/10.1103/PhysRevD.102.103022} {\bibfield
  {journal} {\bibinfo  {journal} {Phys. Rev. D}\ }\textbf {\bibinfo {volume}
  {102}},\ \bibinfo {pages} {103022} (\bibinfo {year} {2020})},\ \Eprint
  {https://arxiv.org/abs/1906.11845} {arXiv:1906.11845 [astro-ph.CO]}
  \BibitemShut {NoStop}%
\bibitem [{\citenamefont {Amaro-Seoane}(2018)}]{Amaro-Seoane:2012lgq}%
  \BibitemOpen
  \bibfield  {author} {\bibinfo {author} {\bibfnamefont {P.}~\bibnamefont
  {Amaro-Seoane}},\ }\href {https://doi.org/10.1007/s41114-018-0013-8}
  {\bibfield  {journal} {\bibinfo  {journal} {Living Rev. Rel.}\ }\textbf
  {\bibinfo {volume} {21}},\ \bibinfo {pages} {4} (\bibinfo {year} {2018})},\
  \Eprint {https://arxiv.org/abs/1205.5240} {arXiv:1205.5240 [astro-ph.CO]}
  \BibitemShut {NoStop}%
\bibitem [{\citenamefont {Polcar}\ \emph {et~al.}(2022)\citenamefont {Polcar},
  \citenamefont {Lukes-Gerakopoulos},\ and\ \citenamefont
  {Witzany}}]{Polcar:2022bwv}%
  \BibitemOpen
  \bibfield  {author} {\bibinfo {author} {\bibfnamefont {L.}~\bibnamefont
  {Polcar}}, \bibinfo {author} {\bibfnamefont {G.}~\bibnamefont
  {Lukes-Gerakopoulos}},\ and\ \bibinfo {author} {\bibfnamefont
  {V.}~\bibnamefont {Witzany}},\ }\href
  {https://doi.org/10.1103/PhysRevD.106.044069} {\bibfield  {journal} {\bibinfo
   {journal} {Phys. Rev. D}\ }\textbf {\bibinfo {volume} {106}},\ \bibinfo
  {pages} {044069} (\bibinfo {year} {2022})},\ \Eprint
  {https://arxiv.org/abs/2205.08516} {arXiv:2205.08516 [gr-qc]} \BibitemShut
  {NoStop}%
\bibitem [{\citenamefont {Amaro-Seoane}\ \emph {et~al.}(2007)\citenamefont
  {Amaro-Seoane}, \citenamefont {Gair}, \citenamefont {Freitag}, \citenamefont
  {Coleman~Miller}, \citenamefont {Mandel}, \citenamefont {Cutler},\ and\
  \citenamefont {Babak}}]{Amaro-Seoane:2007osp}%
  \BibitemOpen
  \bibfield  {author} {\bibinfo {author} {\bibfnamefont {P.}~\bibnamefont
  {Amaro-Seoane}}, \bibinfo {author} {\bibfnamefont {J.~R.}\ \bibnamefont
  {Gair}}, \bibinfo {author} {\bibfnamefont {M.}~\bibnamefont {Freitag}},
  \bibinfo {author} {\bibfnamefont {M.}~\bibnamefont {Coleman~Miller}},
  \bibinfo {author} {\bibfnamefont {I.}~\bibnamefont {Mandel}}, \bibinfo
  {author} {\bibfnamefont {C.~J.}\ \bibnamefont {Cutler}},\ and\ \bibinfo
  {author} {\bibfnamefont {S.}~\bibnamefont {Babak}},\ }\href
  {https://doi.org/10.1088/0264-9381/24/17/R01} {\bibfield  {journal} {\bibinfo
   {journal} {Class. Quant. Grav.}\ }\textbf {\bibinfo {volume} {24}},\
  \bibinfo {pages} {R113} (\bibinfo {year} {2007})},\ \Eprint
  {https://arxiv.org/abs/astro-ph/0703495} {arXiv:astro-ph/0703495}
  \BibitemShut {NoStop}%
\bibitem [{\citenamefont {Baumann}\ \emph
  {et~al.}(2022{\natexlab{a}})\citenamefont {Baumann}, \citenamefont {Bertone},
  \citenamefont {Stout},\ and\ \citenamefont {Tomaselli}}]{Baumann:2022pkl}%
  \BibitemOpen
  \bibfield  {author} {\bibinfo {author} {\bibfnamefont {D.}~\bibnamefont
  {Baumann}}, \bibinfo {author} {\bibfnamefont {G.}~\bibnamefont {Bertone}},
  \bibinfo {author} {\bibfnamefont {J.}~\bibnamefont {Stout}},\ and\ \bibinfo
  {author} {\bibfnamefont {G.~M.}\ \bibnamefont {Tomaselli}},\ }\href
  {https://doi.org/10.1103/PhysRevLett.128.221102} {\bibfield  {journal}
  {\bibinfo  {journal} {Phys. Rev. Lett.}\ }\textbf {\bibinfo {volume} {128}},\
  \bibinfo {pages} {221102} (\bibinfo {year} {2022}{\natexlab{a}})},\ \Eprint
  {https://arxiv.org/abs/2206.01212} {arXiv:2206.01212 [gr-qc]} \BibitemShut
  {NoStop}%
\bibitem [{\citenamefont {Cardoso}\ \emph
  {et~al.}(2022{\natexlab{a}})\citenamefont {Cardoso}, \citenamefont
  {Destounis}, \citenamefont {Duque}, \citenamefont {Panosso~Macedo},\ and\
  \citenamefont {Maselli}}]{Cardoso:2022whc}%
  \BibitemOpen
  \bibfield  {author} {\bibinfo {author} {\bibfnamefont {V.}~\bibnamefont
  {Cardoso}}, \bibinfo {author} {\bibfnamefont {K.}~\bibnamefont {Destounis}},
  \bibinfo {author} {\bibfnamefont {F.}~\bibnamefont {Duque}}, \bibinfo
  {author} {\bibfnamefont {R.}~\bibnamefont {Panosso~Macedo}},\ and\ \bibinfo
  {author} {\bibfnamefont {A.}~\bibnamefont {Maselli}},\ }\href@noop {} {\
  (\bibinfo {year} {2022}{\natexlab{a}})},\ \Eprint
  {https://arxiv.org/abs/2210.01133} {arXiv:2210.01133 [gr-qc]} \BibitemShut
  {NoStop}%
\bibitem [{\citenamefont {Li}\ \emph {et~al.}(2022)\citenamefont {Li},
  \citenamefont {Tang},\ and\ \citenamefont {Wu}}]{Li:2021pxf}%
  \BibitemOpen
  \bibfield  {author} {\bibinfo {author} {\bibfnamefont {G.-L.}\ \bibnamefont
  {Li}}, \bibinfo {author} {\bibfnamefont {Y.}~\bibnamefont {Tang}},\ and\
  \bibinfo {author} {\bibfnamefont {Y.-L.}\ \bibnamefont {Wu}},\ }\href
  {https://doi.org/10.1007/s11433-022-1930-9} {\bibfield  {journal} {\bibinfo
  {journal} {Sci. China Phys. Mech. Astron.}\ }\textbf {\bibinfo {volume}
  {65}},\ \bibinfo {pages} {100412} (\bibinfo {year} {2022})},\ \Eprint
  {https://arxiv.org/abs/2112.14041} {arXiv:2112.14041 [astro-ph.CO]}
  \BibitemShut {NoStop}%
\bibitem [{\citenamefont {Yue}\ and\ \citenamefont {Cao}(2019)}]{Yue:2019ozq}%
  \BibitemOpen
  \bibfield  {author} {\bibinfo {author} {\bibfnamefont {X.-J.}\ \bibnamefont
  {Yue}}\ and\ \bibinfo {author} {\bibfnamefont {Z.}~\bibnamefont {Cao}},\
  }\href {https://doi.org/10.1103/PhysRevD.100.043013} {\bibfield  {journal}
  {\bibinfo  {journal} {Phys. Rev. D}\ }\textbf {\bibinfo {volume} {100}},\
  \bibinfo {pages} {043013} (\bibinfo {year} {2019})},\ \Eprint
  {https://arxiv.org/abs/1908.10241} {arXiv:1908.10241 [astro-ph.HE]}
  \BibitemShut {NoStop}%
\bibitem [{\citenamefont {Yue}\ and\ \citenamefont {Han}(2018)}]{Yue:2017iwc}%
  \BibitemOpen
  \bibfield  {author} {\bibinfo {author} {\bibfnamefont {X.-J.}\ \bibnamefont
  {Yue}}\ and\ \bibinfo {author} {\bibfnamefont {W.-B.}\ \bibnamefont {Han}},\
  }\href {https://doi.org/10.1103/PhysRevD.97.064003} {\bibfield  {journal}
  {\bibinfo  {journal} {Phys. Rev. D}\ }\textbf {\bibinfo {volume} {97}},\
  \bibinfo {pages} {064003} (\bibinfo {year} {2018})},\ \Eprint
  {https://arxiv.org/abs/1711.09706} {arXiv:1711.09706 [gr-qc]} \BibitemShut
  {NoStop}%
\bibitem [{\citenamefont {Baumann}\ \emph
  {et~al.}(2022{\natexlab{b}})\citenamefont {Baumann}, \citenamefont {Bertone},
  \citenamefont {Stout},\ and\ \citenamefont {Tomaselli}}]{Baumann:2021fkf}%
  \BibitemOpen
  \bibfield  {author} {\bibinfo {author} {\bibfnamefont {D.}~\bibnamefont
  {Baumann}}, \bibinfo {author} {\bibfnamefont {G.}~\bibnamefont {Bertone}},
  \bibinfo {author} {\bibfnamefont {J.}~\bibnamefont {Stout}},\ and\ \bibinfo
  {author} {\bibfnamefont {G.~M.}\ \bibnamefont {Tomaselli}},\ }\href
  {https://doi.org/10.1103/PhysRevD.105.115036} {\bibfield  {journal} {\bibinfo
   {journal} {Phys. Rev. D}\ }\textbf {\bibinfo {volume} {105}},\ \bibinfo
  {pages} {115036} (\bibinfo {year} {2022}{\natexlab{b}})},\ \Eprint
  {https://arxiv.org/abs/2112.14777} {arXiv:2112.14777 [gr-qc]} \BibitemShut
  {NoStop}%
\bibitem [{\citenamefont {Baumann}\ \emph
  {et~al.}(2019{\natexlab{b}})\citenamefont {Baumann}, \citenamefont {Chia},\
  and\ \citenamefont {Porto}}]{Baumann:2018vus}%
  \BibitemOpen
  \bibfield  {author} {\bibinfo {author} {\bibfnamefont {D.}~\bibnamefont
  {Baumann}}, \bibinfo {author} {\bibfnamefont {H.~S.}\ \bibnamefont {Chia}},\
  and\ \bibinfo {author} {\bibfnamefont {R.~A.}\ \bibnamefont {Porto}},\ }\href
  {https://doi.org/10.1103/PhysRevD.99.044001} {\bibfield  {journal} {\bibinfo
  {journal} {Phys. Rev. D}\ }\textbf {\bibinfo {volume} {99}},\ \bibinfo
  {pages} {044001} (\bibinfo {year} {2019}{\natexlab{b}})},\ \Eprint
  {https://arxiv.org/abs/1804.03208} {arXiv:1804.03208 [gr-qc]} \BibitemShut
  {NoStop}%
\bibitem [{\citenamefont {Poddar}\ \emph {et~al.}(2020)\citenamefont {Poddar},
  \citenamefont {Mohanty},\ and\ \citenamefont {Jana}}]{KumarPoddar:2019jxe}%
  \BibitemOpen
  \bibfield  {author} {\bibinfo {author} {\bibfnamefont {T.~K.}\ \bibnamefont
  {Poddar}}, \bibinfo {author} {\bibfnamefont {S.}~\bibnamefont {Mohanty}},\
  and\ \bibinfo {author} {\bibfnamefont {S.}~\bibnamefont {Jana}},\ }\href
  {https://doi.org/10.1103/PhysRevD.101.083007} {\bibfield  {journal} {\bibinfo
   {journal} {Phys. Rev. D}\ }\textbf {\bibinfo {volume} {101}},\ \bibinfo
  {pages} {083007} (\bibinfo {year} {2020})},\ \Eprint
  {https://arxiv.org/abs/1906.00666} {arXiv:1906.00666 [hep-ph]} \BibitemShut
  {NoStop}%
\bibitem [{\citenamefont {Zhang}\ and\ \citenamefont
  {Yang}(2020)}]{Zhang:2019eid}%
  \BibitemOpen
  \bibfield  {author} {\bibinfo {author} {\bibfnamefont {J.}~\bibnamefont
  {Zhang}}\ and\ \bibinfo {author} {\bibfnamefont {H.}~\bibnamefont {Yang}},\
  }\href {https://doi.org/10.1103/PhysRevD.101.043020} {\bibfield  {journal}
  {\bibinfo  {journal} {Phys. Rev. D}\ }\textbf {\bibinfo {volume} {101}},\
  \bibinfo {pages} {043020} (\bibinfo {year} {2020})},\ \Eprint
  {https://arxiv.org/abs/1907.13582} {arXiv:1907.13582 [gr-qc]} \BibitemShut
  {NoStop}%
\bibitem [{\citenamefont {Barsanti}\ \emph {et~al.}(2022)\citenamefont
  {Barsanti}, \citenamefont {Franchini}, \citenamefont {Gualtieri},
  \citenamefont {Maselli},\ and\ \citenamefont {Sotiriou}}]{Barsanti:2022ana}%
  \BibitemOpen
  \bibfield  {author} {\bibinfo {author} {\bibfnamefont {S.}~\bibnamefont
  {Barsanti}}, \bibinfo {author} {\bibfnamefont {N.}~\bibnamefont {Franchini}},
  \bibinfo {author} {\bibfnamefont {L.}~\bibnamefont {Gualtieri}}, \bibinfo
  {author} {\bibfnamefont {A.}~\bibnamefont {Maselli}},\ and\ \bibinfo {author}
  {\bibfnamefont {T.~P.}\ \bibnamefont {Sotiriou}},\ }\href
  {https://doi.org/10.1103/PhysRevD.106.044029} {\bibfield  {journal} {\bibinfo
   {journal} {Phys. Rev. D}\ }\textbf {\bibinfo {volume} {106}},\ \bibinfo
  {pages} {044029} (\bibinfo {year} {2022})},\ \Eprint
  {https://arxiv.org/abs/2203.05003} {arXiv:2203.05003 [gr-qc]} \BibitemShut
  {NoStop}%
\bibitem [{\citenamefont {Ferreira}\ \emph {et~al.}(2017)\citenamefont
  {Ferreira}, \citenamefont {Macedo},\ and\ \citenamefont
  {Cardoso}}]{Ferreira:2017pth}%
  \BibitemOpen
  \bibfield  {author} {\bibinfo {author} {\bibfnamefont {M.~C.}\ \bibnamefont
  {Ferreira}}, \bibinfo {author} {\bibfnamefont {C.~F.~B.}\ \bibnamefont
  {Macedo}},\ and\ \bibinfo {author} {\bibfnamefont {V.}~\bibnamefont
  {Cardoso}},\ }\href {https://doi.org/10.1103/PhysRevD.96.083017} {\bibfield
  {journal} {\bibinfo  {journal} {Phys. Rev. D}\ }\textbf {\bibinfo {volume}
  {96}},\ \bibinfo {pages} {083017} (\bibinfo {year} {2017})},\ \Eprint
  {https://arxiv.org/abs/1710.00830} {arXiv:1710.00830 [gr-qc]} \BibitemShut
  {NoStop}%
\bibitem [{\citenamefont {Edwards}\ \emph {et~al.}(2020)\citenamefont
  {Edwards}, \citenamefont {Chianese}, \citenamefont {Kavanagh}, \citenamefont
  {Nissanke},\ and\ \citenamefont {Weniger}}]{Edwards:2019tzf}%
  \BibitemOpen
  \bibfield  {author} {\bibinfo {author} {\bibfnamefont {T.~D.~P.}\
  \bibnamefont {Edwards}}, \bibinfo {author} {\bibfnamefont {M.}~\bibnamefont
  {Chianese}}, \bibinfo {author} {\bibfnamefont {B.~J.}\ \bibnamefont
  {Kavanagh}}, \bibinfo {author} {\bibfnamefont {S.~M.}\ \bibnamefont
  {Nissanke}},\ and\ \bibinfo {author} {\bibfnamefont {C.}~\bibnamefont
  {Weniger}},\ }\href {https://doi.org/10.1103/PhysRevLett.124.161101}
  {\bibfield  {journal} {\bibinfo  {journal} {Phys. Rev. Lett.}\ }\textbf
  {\bibinfo {volume} {124}},\ \bibinfo {pages} {161101} (\bibinfo {year}
  {2020})},\ \Eprint {https://arxiv.org/abs/1905.04686} {arXiv:1905.04686
  [hep-ph]} \BibitemShut {NoStop}%
\bibitem [{\citenamefont {Yuan}\ \emph {et~al.}(2021)\citenamefont {Yuan},
  \citenamefont {Xia}, \citenamefont {Tang}, \citenamefont {Zhao},
  \citenamefont {Cai}, \citenamefont {Chen}, \citenamefont {Shu},\ and\
  \citenamefont {Yuan}}]{Yuan:2020xui}%
  \BibitemOpen
  \bibfield  {author} {\bibinfo {author} {\bibfnamefont {G.-W.}\ \bibnamefont
  {Yuan}}, \bibinfo {author} {\bibfnamefont {Z.}~\bibnamefont {Xia}}, \bibinfo
  {author} {\bibfnamefont {C.}~\bibnamefont {Tang}}, \bibinfo {author}
  {\bibfnamefont {Y.}~\bibnamefont {Zhao}}, \bibinfo {author} {\bibfnamefont
  {Y.-F.}\ \bibnamefont {Cai}}, \bibinfo {author} {\bibfnamefont
  {Y.}~\bibnamefont {Chen}}, \bibinfo {author} {\bibfnamefont {J.}~\bibnamefont
  {Shu}},\ and\ \bibinfo {author} {\bibfnamefont {Q.}~\bibnamefont {Yuan}},\
  }\href {https://doi.org/10.1088/1475-7516/2021/03/018} {\bibfield  {journal}
  {\bibinfo  {journal} {Journal of Cosmology and Astroparticle Physics}\
  }\textbf {\bibinfo {volume} {2021}}\bibfield  {number} {\bibinfo  {number} {
  (03)},\ \bibinfo {pages} {018}},\ }\Eprint {https://arxiv.org/abs/2008.13662}
  {arXiv:2008.13662 [astro-ph.HE]} \BibitemShut {NoStop}%
\bibitem [{\citenamefont {Workman}(2022)}]{Workman:2022ynf}%
  \BibitemOpen
  \bibfield  {author} {\bibinfo {author} {\bibfnamefont {R.~L.}\ \bibnamefont
  {Workman}} (\bibinfo {collaboration} {Particle Data Group}),\ }\href
  {https://doi.org/10.1093/ptep/ptac097} {\bibfield  {journal} {\bibinfo
  {journal} {PTEP}\ }\textbf {\bibinfo {volume} {2022}},\ \bibinfo {pages}
  {083C01} (\bibinfo {year} {2022})}\BibitemShut {NoStop}%
\bibitem [{\citenamefont {Kouvaris}\ \emph {et~al.}(2022)\citenamefont
  {Kouvaris}, \citenamefont {Liu},\ and\ \citenamefont
  {Lyu}}]{Kouvaris:2022guf}%
  \BibitemOpen
  \bibfield  {author} {\bibinfo {author} {\bibfnamefont {C.}~\bibnamefont
  {Kouvaris}}, \bibinfo {author} {\bibfnamefont {T.}~\bibnamefont {Liu}},\ and\
  \bibinfo {author} {\bibfnamefont {K.-F.}\ \bibnamefont {Lyu}},\ }\href@noop
  {} {\  (\bibinfo {year} {2022})},\ \Eprint {https://arxiv.org/abs/2202.11096}
  {arXiv:2202.11096 [astro-ph.HE]} \BibitemShut {NoStop}%
\bibitem [{\citenamefont {Aurrekoetxea}\ \emph {et~al.}(2022)\citenamefont
  {Aurrekoetxea}, \citenamefont {Clough},\ and\ \citenamefont
  {Lim}}]{Aurrekoetxea:2022mpw}%
  \BibitemOpen
  \bibfield  {author} {\bibinfo {author} {\bibfnamefont {J.~C.}\ \bibnamefont
  {Aurrekoetxea}}, \bibinfo {author} {\bibfnamefont {K.}~\bibnamefont
  {Clough}},\ and\ \bibinfo {author} {\bibfnamefont {E.~A.}\ \bibnamefont
  {Lim}},\ }\href@noop {} {\  (\bibinfo {year} {2022})},\ \Eprint
  {https://arxiv.org/abs/2207.03125} {arXiv:2207.03125 [gr-qc]} \BibitemShut
  {NoStop}%
\bibitem [{\citenamefont {Wong}(2019)}]{Wong:2019kru}%
  \BibitemOpen
  \bibfield  {author} {\bibinfo {author} {\bibfnamefont {L.~K.}\ \bibnamefont
  {Wong}},\ }\href {https://doi.org/10.1103/PhysRevD.100.044051} {\bibfield
  {journal} {\bibinfo  {journal} {Phys. Rev. D}\ }\textbf {\bibinfo {volume}
  {100}},\ \bibinfo {pages} {044051} (\bibinfo {year} {2019})},\ \Eprint
  {https://arxiv.org/abs/1905.08543} {arXiv:1905.08543 [hep-th]} \BibitemShut
  {NoStop}%
\bibitem [{\citenamefont {Wong}(2020)}]{Wong:2020qom}%
  \BibitemOpen
  \bibfield  {author} {\bibinfo {author} {\bibfnamefont {L.~K.}\ \bibnamefont
  {Wong}},\ }\href {https://doi.org/10.1103/PhysRevD.101.124049} {\bibfield
  {journal} {\bibinfo  {journal} {Phys. Rev. D}\ }\textbf {\bibinfo {volume}
  {101}},\ \bibinfo {pages} {124049} (\bibinfo {year} {2020})},\ \Eprint
  {https://arxiv.org/abs/2004.03570} {arXiv:2004.03570 [hep-th]} \BibitemShut
  {NoStop}%
\bibitem [{\citenamefont {Khmelnitsky}\ and\ \citenamefont
  {Rubakov}(2014)}]{Khmelnitsky:2013lxt}%
  \BibitemOpen
  \bibfield  {author} {\bibinfo {author} {\bibfnamefont {A.}~\bibnamefont
  {Khmelnitsky}}\ and\ \bibinfo {author} {\bibfnamefont {V.}~\bibnamefont
  {Rubakov}},\ }\href {https://doi.org/10.1088/1475-7516/2014/02/019}
  {\bibfield  {journal} {\bibinfo  {journal} {JCAP}\ }\textbf {\bibinfo
  {volume} {02}},\ \bibinfo {pages} {019}},\ \Eprint
  {https://arxiv.org/abs/1309.5888} {arXiv:1309.5888 [astro-ph.CO]}
  \BibitemShut {NoStop}%
\bibitem [{\citenamefont {Blas}\ \emph {et~al.}(2017)\citenamefont {Blas},
  \citenamefont {Nacir},\ and\ \citenamefont {Sibiryakov}}]{Blas:2016ddr}%
  \BibitemOpen
  \bibfield  {author} {\bibinfo {author} {\bibfnamefont {D.}~\bibnamefont
  {Blas}}, \bibinfo {author} {\bibfnamefont {D.~L.}\ \bibnamefont {Nacir}},\
  and\ \bibinfo {author} {\bibfnamefont {S.}~\bibnamefont {Sibiryakov}},\
  }\href {https://doi.org/10.1103/PhysRevLett.118.261102} {\bibfield  {journal}
  {\bibinfo  {journal} {Phys. Rev. Lett.}\ }\textbf {\bibinfo {volume} {118}},\
  \bibinfo {pages} {261102} (\bibinfo {year} {2017})},\ \Eprint
  {https://arxiv.org/abs/1612.06789} {arXiv:1612.06789 [hep-ph]} \BibitemShut
  {NoStop}%
\bibitem [{\citenamefont {Blas}\ \emph {et~al.}(2020)\citenamefont {Blas},
  \citenamefont {L\'opez~Nacir},\ and\ \citenamefont
  {Sibiryakov}}]{Blas:2019hxz}%
  \BibitemOpen
  \bibfield  {author} {\bibinfo {author} {\bibfnamefont {D.}~\bibnamefont
  {Blas}}, \bibinfo {author} {\bibfnamefont {D.}~\bibnamefont
  {L\'opez~Nacir}},\ and\ \bibinfo {author} {\bibfnamefont {S.}~\bibnamefont
  {Sibiryakov}},\ }\href {https://doi.org/10.1103/PhysRevD.101.063016}
  {\bibfield  {journal} {\bibinfo  {journal} {Phys. Rev. D}\ }\textbf {\bibinfo
  {volume} {101}},\ \bibinfo {pages} {063016} (\bibinfo {year} {2020})},\
  \Eprint {https://arxiv.org/abs/1910.08544} {arXiv:1910.08544 [gr-qc]}
  \BibitemShut {NoStop}%
\bibitem [{\citenamefont {Ribeiro}\ \emph {et~al.}(2022)\citenamefont
  {Ribeiro}, \citenamefont {Zilh\~ao},\ and\ \citenamefont
  {Cardoso}}]{Ribeiro:2022ohq}%
  \BibitemOpen
  \bibfield  {author} {\bibinfo {author} {\bibfnamefont {D.~C.}\ \bibnamefont
  {Ribeiro}}, \bibinfo {author} {\bibfnamefont {M.}~\bibnamefont {Zilh\~ao}},\
  and\ \bibinfo {author} {\bibfnamefont {V.}~\bibnamefont {Cardoso}},\ }\href
  {https://doi.org/10.1103/PhysRevD.105.084004} {\bibfield  {journal} {\bibinfo
   {journal} {Phys. Rev. D}\ }\textbf {\bibinfo {volume} {105}},\ \bibinfo
  {pages} {084004} (\bibinfo {year} {2022})},\ \Eprint
  {https://arxiv.org/abs/2201.13407} {arXiv:2201.13407 [gr-qc]} \BibitemShut
  {NoStop}%
\bibitem [{\citenamefont {Annulli}\ \emph {et~al.}(2020)\citenamefont
  {Annulli}, \citenamefont {Cardoso},\ and\ \citenamefont
  {Vicente}}]{Annulli:2020lyc}%
  \BibitemOpen
  \bibfield  {author} {\bibinfo {author} {\bibfnamefont {L.}~\bibnamefont
  {Annulli}}, \bibinfo {author} {\bibfnamefont {V.}~\bibnamefont {Cardoso}},\
  and\ \bibinfo {author} {\bibfnamefont {R.}~\bibnamefont {Vicente}},\ }\href
  {https://doi.org/10.1103/PhysRevD.102.063022} {\bibfield  {journal} {\bibinfo
   {journal} {Phys. Rev. D}\ }\textbf {\bibinfo {volume} {102}},\ \bibinfo
  {pages} {063022} (\bibinfo {year} {2020})},\ \Eprint
  {https://arxiv.org/abs/2009.00012} {arXiv:2009.00012 [gr-qc]} \BibitemShut
  {NoStop}%
\bibitem [{\citenamefont {Liu}\ and\ \citenamefont {Lyu}(2021)}]{Liu:2021llm}%
  \BibitemOpen
  \bibfield  {author} {\bibinfo {author} {\bibfnamefont {T.}~\bibnamefont
  {Liu}}\ and\ \bibinfo {author} {\bibfnamefont {K.-F.}\ \bibnamefont {Lyu}},\
  }\href@noop {} {\  (\bibinfo {year} {2021})},\ \Eprint
  {https://arxiv.org/abs/2107.09971} {arXiv:2107.09971 [astro-ph.HE]}
  \BibitemShut {NoStop}%
\bibitem [{\citenamefont {Kavic}\ \emph {et~al.}(2020)\citenamefont {Kavic},
  \citenamefont {Liebling}, \citenamefont {Lippert},\ and\ \citenamefont
  {Simonetti}}]{Kavic:2019cgk}%
  \BibitemOpen
  \bibfield  {author} {\bibinfo {author} {\bibfnamefont {M.}~\bibnamefont
  {Kavic}}, \bibinfo {author} {\bibfnamefont {S.~L.}\ \bibnamefont {Liebling}},
  \bibinfo {author} {\bibfnamefont {M.}~\bibnamefont {Lippert}},\ and\ \bibinfo
  {author} {\bibfnamefont {J.~H.}\ \bibnamefont {Simonetti}},\ }\href
  {https://doi.org/10.1088/1475-7516/2020/08/005} {\bibfield  {journal}
  {\bibinfo  {journal} {JCAP}\ }\textbf {\bibinfo {volume} {08}},\ \bibinfo
  {pages} {005}},\ \Eprint {https://arxiv.org/abs/1910.06977} {arXiv:1910.06977
  [astro-ph.HE]} \BibitemShut {NoStop}%
\bibitem [{\citenamefont {Bo\v{s}kovi\'c}\ \emph {et~al.}(2018)\citenamefont
  {Bo\v{s}kovi\'c}, \citenamefont {Duque}, \citenamefont {Ferreira},
  \citenamefont {Miguel},\ and\ \citenamefont {Cardoso}}]{Boskovic:2018rub}%
  \BibitemOpen
  \bibfield  {author} {\bibinfo {author} {\bibfnamefont {M.}~\bibnamefont
  {Bo\v{s}kovi\'c}}, \bibinfo {author} {\bibfnamefont {F.}~\bibnamefont
  {Duque}}, \bibinfo {author} {\bibfnamefont {M.~C.}\ \bibnamefont {Ferreira}},
  \bibinfo {author} {\bibfnamefont {F.~S.}\ \bibnamefont {Miguel}},\ and\
  \bibinfo {author} {\bibfnamefont {V.}~\bibnamefont {Cardoso}},\ }\href
  {https://doi.org/10.1103/PhysRevD.98.024037} {\bibfield  {journal} {\bibinfo
  {journal} {Phys. Rev. D}\ }\textbf {\bibinfo {volume} {98}},\ \bibinfo
  {pages} {024037} (\bibinfo {year} {2018})},\ \Eprint
  {https://arxiv.org/abs/1806.07331} {arXiv:1806.07331 [gr-qc]} \BibitemShut
  {NoStop}%
\bibitem [{\citenamefont {Dave}\ and\ \citenamefont
  {Digal}(2022)}]{Dave:2021lcv}%
  \BibitemOpen
  \bibfield  {author} {\bibinfo {author} {\bibfnamefont {S.~S.}\ \bibnamefont
  {Dave}}\ and\ \bibinfo {author} {\bibfnamefont {S.}~\bibnamefont {Digal}},\
  }\href {https://doi.org/10.1103/PhysRevD.105.024039} {\bibfield  {journal}
  {\bibinfo  {journal} {Phys. Rev. D}\ }\textbf {\bibinfo {volume} {105}},\
  \bibinfo {pages} {024039} (\bibinfo {year} {2022})},\ \Eprint
  {https://arxiv.org/abs/2106.05812} {arXiv:2106.05812 [gr-qc]} \BibitemShut
  {NoStop}%
\bibitem [{\citenamefont {Berti}\ \emph {et~al.}(2019)\citenamefont {Berti},
  \citenamefont {Brito}, \citenamefont {Macedo}, \citenamefont {Raposo},\ and\
  \citenamefont {Rosa}}]{Berti:2019wnn}%
  \BibitemOpen
  \bibfield  {author} {\bibinfo {author} {\bibfnamefont {E.}~\bibnamefont
  {Berti}}, \bibinfo {author} {\bibfnamefont {R.}~\bibnamefont {Brito}},
  \bibinfo {author} {\bibfnamefont {C.~F.~B.}\ \bibnamefont {Macedo}}, \bibinfo
  {author} {\bibfnamefont {G.}~\bibnamefont {Raposo}},\ and\ \bibinfo {author}
  {\bibfnamefont {J.~L.}\ \bibnamefont {Rosa}},\ }\href
  {https://doi.org/10.1103/PhysRevD.99.104039} {\bibfield  {journal} {\bibinfo
  {journal} {Phys. Rev. D}\ }\textbf {\bibinfo {volume} {99}},\ \bibinfo
  {pages} {104039} (\bibinfo {year} {2019})},\ \Eprint
  {https://arxiv.org/abs/1904.03131} {arXiv:1904.03131 [gr-qc]} \BibitemShut
  {NoStop}%
\bibitem [{\citenamefont {Horbatsch}\ and\ \citenamefont
  {Burgess}(2012)}]{Horbatsch:2011ye}%
  \BibitemOpen
  \bibfield  {author} {\bibinfo {author} {\bibfnamefont {M.~W.}\ \bibnamefont
  {Horbatsch}}\ and\ \bibinfo {author} {\bibfnamefont {C.~P.}\ \bibnamefont
  {Burgess}},\ }\href {https://doi.org/10.1088/1475-7516/2012/05/010}
  {\bibfield  {journal} {\bibinfo  {journal} {JCAP}\ }\textbf {\bibinfo
  {volume} {05}},\ \bibinfo {pages} {010}},\ \Eprint
  {https://arxiv.org/abs/1111.4009} {arXiv:1111.4009 [gr-qc]} \BibitemShut
  {NoStop}%
\bibitem [{\citenamefont {Brito}\ \emph {et~al.}(2017)\citenamefont {Brito},
  \citenamefont {Ghosh}, \citenamefont {Barausse}, \citenamefont {Berti},
  \citenamefont {Cardoso}, \citenamefont {Dvorkin}, \citenamefont {Klein},\
  and\ \citenamefont {Pani}}]{Brito:2017zvb}%
  \BibitemOpen
  \bibfield  {author} {\bibinfo {author} {\bibfnamefont {R.}~\bibnamefont
  {Brito}}, \bibinfo {author} {\bibfnamefont {S.}~\bibnamefont {Ghosh}},
  \bibinfo {author} {\bibfnamefont {E.}~\bibnamefont {Barausse}}, \bibinfo
  {author} {\bibfnamefont {E.}~\bibnamefont {Berti}}, \bibinfo {author}
  {\bibfnamefont {V.}~\bibnamefont {Cardoso}}, \bibinfo {author} {\bibfnamefont
  {I.}~\bibnamefont {Dvorkin}}, \bibinfo {author} {\bibfnamefont
  {A.}~\bibnamefont {Klein}},\ and\ \bibinfo {author} {\bibfnamefont
  {P.}~\bibnamefont {Pani}},\ }\href
  {https://doi.org/10.1103/PhysRevD.96.064050} {\bibfield  {journal} {\bibinfo
  {journal} {Phys. Rev. D}\ }\textbf {\bibinfo {volume} {96}},\ \bibinfo
  {pages} {064050} (\bibinfo {year} {2017})},\ \Eprint
  {https://arxiv.org/abs/1706.06311} {arXiv:1706.06311 [gr-qc]} \BibitemShut
  {NoStop}%
\bibitem [{\citenamefont {Takahashi}\ and\ \citenamefont
  {Tanaka}(2021)}]{Takahashi:2021eso}%
  \BibitemOpen
  \bibfield  {author} {\bibinfo {author} {\bibfnamefont {T.}~\bibnamefont
  {Takahashi}}\ and\ \bibinfo {author} {\bibfnamefont {T.}~\bibnamefont
  {Tanaka}},\ }\href {https://doi.org/10.1088/1475-7516/2021/10/031} {\bibfield
   {journal} {\bibinfo  {journal} {JCAP}\ }\textbf {\bibinfo {volume} {10}},\
  \bibinfo {pages} {031}},\ \Eprint {https://arxiv.org/abs/2106.08836}
  {arXiv:2106.08836 [gr-qc]} \BibitemShut {NoStop}%
\bibitem [{\citenamefont {Takahashi}\ \emph {et~al.}(2022)\citenamefont
  {Takahashi}, \citenamefont {Omiya},\ and\ \citenamefont
  {Tanaka}}]{Takahashi:2021yhy}%
  \BibitemOpen
  \bibfield  {author} {\bibinfo {author} {\bibfnamefont {T.}~\bibnamefont
  {Takahashi}}, \bibinfo {author} {\bibfnamefont {H.}~\bibnamefont {Omiya}},\
  and\ \bibinfo {author} {\bibfnamefont {T.}~\bibnamefont {Tanaka}},\ }\href
  {https://doi.org/10.1093/ptep/ptac044} {\bibfield  {journal} {\bibinfo
  {journal} {PTEP}\ }\textbf {\bibinfo {volume} {2022}},\ \bibinfo {pages}
  {043E01} (\bibinfo {year} {2022})},\ \Eprint
  {https://arxiv.org/abs/2112.05774} {arXiv:2112.05774 [gr-qc]} \BibitemShut
  {NoStop}%
\bibitem [{\citenamefont {Su}\ \emph {et~al.}(2021)\citenamefont {Su},
  \citenamefont {Xianyu},\ and\ \citenamefont {Zhang}}]{Su:2021dwz}%
  \BibitemOpen
  \bibfield  {author} {\bibinfo {author} {\bibfnamefont {B.}~\bibnamefont
  {Su}}, \bibinfo {author} {\bibfnamefont {Z.-Z.}\ \bibnamefont {Xianyu}},\
  and\ \bibinfo {author} {\bibfnamefont {X.}~\bibnamefont {Zhang}},\ }\href
  {https://doi.org/10.3847/1538-4357/ac2d91} {\bibfield  {journal} {\bibinfo
  {journal} {Astrophys. J.}\ }\textbf {\bibinfo {volume} {923}},\ \bibinfo
  {pages} {114} (\bibinfo {year} {2021})},\ \Eprint
  {https://arxiv.org/abs/2107.13527} {arXiv:2107.13527 [gr-qc]} \BibitemShut
  {NoStop}%
\bibitem [{\citenamefont {Ficarra}(2021)}]{Ficarra:2021qeh}%
  \BibitemOpen
  \bibfield  {author} {\bibinfo {author} {\bibfnamefont {G.}~\bibnamefont
  {Ficarra}},\ }in\ \href@noop {} {\emph {\bibinfo {booktitle} {{55th
  Rencontres de Moriond on Gravitation}}}}\ (\bibinfo {year} {2021})\ \Eprint
  {https://arxiv.org/abs/2105.05918} {arXiv:2105.05918 [gr-qc]} \BibitemShut
  {NoStop}%
\bibitem [{\citenamefont {Rozner}\ \emph {et~al.}(2020)\citenamefont {Rozner},
  \citenamefont {Grishin}, \citenamefont {Ginat}, \citenamefont {Igoshev},\
  and\ \citenamefont {Desjacques}}]{Rozner:2019gba}%
  \BibitemOpen
  \bibfield  {author} {\bibinfo {author} {\bibfnamefont {M.}~\bibnamefont
  {Rozner}}, \bibinfo {author} {\bibfnamefont {E.}~\bibnamefont {Grishin}},
  \bibinfo {author} {\bibfnamefont {Y.~B.}\ \bibnamefont {Ginat}}, \bibinfo
  {author} {\bibfnamefont {A.~P.}\ \bibnamefont {Igoshev}},\ and\ \bibinfo
  {author} {\bibfnamefont {V.}~\bibnamefont {Desjacques}},\ }\href
  {https://doi.org/10.1088/1475-7516/2020/03/061} {\bibfield  {journal}
  {\bibinfo  {journal} {JCAP}\ }\textbf {\bibinfo {volume} {03}},\ \bibinfo
  {pages} {061}},\ \Eprint {https://arxiv.org/abs/1904.01958} {arXiv:1904.01958
  [astro-ph.CO]} \BibitemShut {NoStop}%
\bibitem [{\citenamefont {Cardoso}\ \emph {et~al.}(2020)\citenamefont
  {Cardoso}, \citenamefont {Duque},\ and\ \citenamefont
  {Ikeda}}]{Cardoso:2020hca}%
  \BibitemOpen
  \bibfield  {author} {\bibinfo {author} {\bibfnamefont {V.}~\bibnamefont
  {Cardoso}}, \bibinfo {author} {\bibfnamefont {F.}~\bibnamefont {Duque}},\
  and\ \bibinfo {author} {\bibfnamefont {T.}~\bibnamefont {Ikeda}},\ }\href
  {https://doi.org/10.1103/PhysRevD.101.064054} {\bibfield  {journal} {\bibinfo
   {journal} {Phys. Rev. D}\ }\textbf {\bibinfo {volume} {101}},\ \bibinfo
  {pages} {064054} (\bibinfo {year} {2020})},\ \Eprint
  {https://arxiv.org/abs/2001.01729} {arXiv:2001.01729 [gr-qc]} \BibitemShut
  {NoStop}%
\bibitem [{\citenamefont {Leung}\ \emph {et~al.}(1997)\citenamefont {Leung},
  \citenamefont {Liu}, \citenamefont {Suen}, \citenamefont {Tam},\ and\
  \citenamefont {Young}}]{Leung:1997was}%
  \BibitemOpen
  \bibfield  {author} {\bibinfo {author} {\bibfnamefont {P.~T.}\ \bibnamefont
  {Leung}}, \bibinfo {author} {\bibfnamefont {Y.~T.}\ \bibnamefont {Liu}},
  \bibinfo {author} {\bibfnamefont {W.~M.}\ \bibnamefont {Suen}}, \bibinfo
  {author} {\bibfnamefont {C.~Y.}\ \bibnamefont {Tam}},\ and\ \bibinfo {author}
  {\bibfnamefont {K.}~\bibnamefont {Young}},\ }\href
  {https://doi.org/10.1103/PhysRevLett.78.2894} {\bibfield  {journal} {\bibinfo
   {journal} {Phys. Rev. Lett.}\ }\textbf {\bibinfo {volume} {78}},\ \bibinfo
  {pages} {2894} (\bibinfo {year} {1997})},\ \Eprint
  {https://arxiv.org/abs/gr-qc/9903031} {arXiv:gr-qc/9903031} \BibitemShut
  {NoStop}%
\bibitem [{\citenamefont {Medved}\ \emph {et~al.}(2004)\citenamefont {Medved},
  \citenamefont {Martin},\ and\ \citenamefont {Visser}}]{Medved:2003pr}%
  \BibitemOpen
  \bibfield  {author} {\bibinfo {author} {\bibfnamefont {A.~J.~M.}\
  \bibnamefont {Medved}}, \bibinfo {author} {\bibfnamefont {D.}~\bibnamefont
  {Martin}},\ and\ \bibinfo {author} {\bibfnamefont {M.}~\bibnamefont
  {Visser}},\ }\href {https://doi.org/10.1088/0264-9381/21/9/013} {\bibfield
  {journal} {\bibinfo  {journal} {Class. Quant. Grav.}\ }\textbf {\bibinfo
  {volume} {21}},\ \bibinfo {pages} {2393} (\bibinfo {year} {2004})},\ \Eprint
  {https://arxiv.org/abs/gr-qc/0310097} {arXiv:gr-qc/0310097} \BibitemShut
  {NoStop}%
\bibitem [{\citenamefont {Bamber}\ \emph
  {et~al.}(2021{\natexlab{b}})\citenamefont {Bamber}, \citenamefont
  {Tattersall}, \citenamefont {Clough},\ and\ \citenamefont
  {Ferreira}}]{Bamber:2021knr}%
  \BibitemOpen
  \bibfield  {author} {\bibinfo {author} {\bibfnamefont {J.}~\bibnamefont
  {Bamber}}, \bibinfo {author} {\bibfnamefont {O.~J.}\ \bibnamefont
  {Tattersall}}, \bibinfo {author} {\bibfnamefont {K.}~\bibnamefont {Clough}},\
  and\ \bibinfo {author} {\bibfnamefont {P.~G.}\ \bibnamefont {Ferreira}},\
  }\href {https://doi.org/10.1103/PhysRevD.103.124013} {\bibfield  {journal}
  {\bibinfo  {journal} {Phys. Rev. D}\ }\textbf {\bibinfo {volume} {103}},\
  \bibinfo {pages} {124013} (\bibinfo {year} {2021}{\natexlab{b}})},\ \Eprint
  {https://arxiv.org/abs/2103.00026} {arXiv:2103.00026 [gr-qc]} \BibitemShut
  {NoStop}%
\bibitem [{\citenamefont {Zhang}\ \emph
  {et~al.}(2022{\natexlab{a}})\citenamefont {Zhang}, \citenamefont {Zhu},
  \citenamefont {Fang},\ and\ \citenamefont {Wang}}]{Zhang:2022roh}%
  \BibitemOpen
  \bibfield  {author} {\bibinfo {author} {\bibfnamefont {C.}~\bibnamefont
  {Zhang}}, \bibinfo {author} {\bibfnamefont {T.}~\bibnamefont {Zhu}}, \bibinfo
  {author} {\bibfnamefont {X.}~\bibnamefont {Fang}},\ and\ \bibinfo {author}
  {\bibfnamefont {A.}~\bibnamefont {Wang}},\ }\href
  {https://doi.org/10.1016/j.dark.2022.101078} {\bibfield  {journal} {\bibinfo
  {journal} {Phys. Dark Univ.}\ }\textbf {\bibinfo {volume} {37}},\ \bibinfo
  {pages} {101078} (\bibinfo {year} {2022}{\natexlab{a}})},\ \Eprint
  {https://arxiv.org/abs/2201.11352} {arXiv:2201.11352 [gr-qc]} \BibitemShut
  {NoStop}%
\bibitem [{\citenamefont {Zhang}\ \emph {et~al.}(2021)\citenamefont {Zhang},
  \citenamefont {Zhu},\ and\ \citenamefont {Wang}}]{Zhang:2021bdr}%
  \BibitemOpen
  \bibfield  {author} {\bibinfo {author} {\bibfnamefont {C.}~\bibnamefont
  {Zhang}}, \bibinfo {author} {\bibfnamefont {T.}~\bibnamefont {Zhu}},\ and\
  \bibinfo {author} {\bibfnamefont {A.}~\bibnamefont {Wang}},\ }\href
  {https://doi.org/10.1103/PhysRevD.104.124082} {\bibfield  {journal} {\bibinfo
   {journal} {Phys. Rev. D}\ }\textbf {\bibinfo {volume} {104}},\ \bibinfo
  {pages} {124082} (\bibinfo {year} {2021})},\ \Eprint
  {https://arxiv.org/abs/2111.04966} {arXiv:2111.04966 [gr-qc]} \BibitemShut
  {NoStop}%
\bibitem [{\citenamefont {Yang}\ \emph {et~al.}(2018)\citenamefont {Yang},
  \citenamefont {Ji}, \citenamefont {Hu}, \citenamefont {Cao},\ and\
  \citenamefont {Cai}}]{Yang:2017lpm}%
  \BibitemOpen
  \bibfield  {author} {\bibinfo {author} {\bibfnamefont {Q.}~\bibnamefont
  {Yang}}, \bibinfo {author} {\bibfnamefont {L.-W.}\ \bibnamefont {Ji}},
  \bibinfo {author} {\bibfnamefont {B.}~\bibnamefont {Hu}}, \bibinfo {author}
  {\bibfnamefont {Z.-J.}\ \bibnamefont {Cao}},\ and\ \bibinfo {author}
  {\bibfnamefont {R.-G.}\ \bibnamefont {Cai}},\ }\href
  {https://doi.org/10.1088/1674-4527/18/6/65} {\bibfield  {journal} {\bibinfo
  {journal} {Res. Astron. Astrophys.}\ }\textbf {\bibinfo {volume} {18}},\
  \bibinfo {pages} {065} (\bibinfo {year} {2018})},\ \Eprint
  {https://arxiv.org/abs/1706.00678} {arXiv:1706.00678 [gr-qc]} \BibitemShut
  {NoStop}%
\bibitem [{\citenamefont {Choudhary}\ \emph {et~al.}(2021)\citenamefont
  {Choudhary}, \citenamefont {Sanchis-Gual}, \citenamefont {Gupta},
  \citenamefont {Degollado}, \citenamefont {Bose},\ and\ \citenamefont
  {Font}}]{Choudhary:2020pxy}%
  \BibitemOpen
  \bibfield  {author} {\bibinfo {author} {\bibfnamefont {S.}~\bibnamefont
  {Choudhary}}, \bibinfo {author} {\bibfnamefont {N.}~\bibnamefont
  {Sanchis-Gual}}, \bibinfo {author} {\bibfnamefont {A.}~\bibnamefont {Gupta}},
  \bibinfo {author} {\bibfnamefont {J.~C.}\ \bibnamefont {Degollado}}, \bibinfo
  {author} {\bibfnamefont {S.}~\bibnamefont {Bose}},\ and\ \bibinfo {author}
  {\bibfnamefont {J.~A.}\ \bibnamefont {Font}},\ }\href
  {https://doi.org/10.1103/PhysRevD.103.044032} {\bibfield  {journal} {\bibinfo
   {journal} {Phys. Rev. D}\ }\textbf {\bibinfo {volume} {103}},\ \bibinfo
  {pages} {044032} (\bibinfo {year} {2021})},\ \Eprint
  {https://arxiv.org/abs/2010.00935} {arXiv:2010.00935 [gr-qc]} \BibitemShut
  {NoStop}%
\bibitem [{\citenamefont {Zhang}\ \emph
  {et~al.}(2022{\natexlab{b}})\citenamefont {Zhang}, \citenamefont
  {Gracia-Linares}, \citenamefont {Laguna}, \citenamefont {Shoemaker},\ and\
  \citenamefont {Liu}}]{Zhang:2022rex}%
  \BibitemOpen
  \bibfield  {author} {\bibinfo {author} {\bibfnamefont {Y.-P.}\ \bibnamefont
  {Zhang}}, \bibinfo {author} {\bibfnamefont {M.}~\bibnamefont
  {Gracia-Linares}}, \bibinfo {author} {\bibfnamefont {P.}~\bibnamefont
  {Laguna}}, \bibinfo {author} {\bibfnamefont {D.}~\bibnamefont {Shoemaker}},\
  and\ \bibinfo {author} {\bibfnamefont {Y.-X.}\ \bibnamefont {Liu}},\
  }\href@noop {} {\  (\bibinfo {year} {2022}{\natexlab{b}})},\ \Eprint
  {https://arxiv.org/abs/2209.11814} {arXiv:2209.11814 [gr-qc]} \BibitemShut
  {NoStop}%
\bibitem [{\citenamefont {Healy}\ \emph {et~al.}(2012)\citenamefont {Healy},
  \citenamefont {Bode}, \citenamefont {Haas}, \citenamefont {Pazos},
  \citenamefont {Laguna}, \citenamefont {Shoemaker},\ and\ \citenamefont
  {Yunes}}]{Healy:2011ef}%
  \BibitemOpen
  \bibfield  {author} {\bibinfo {author} {\bibfnamefont {J.}~\bibnamefont
  {Healy}}, \bibinfo {author} {\bibfnamefont {T.}~\bibnamefont {Bode}},
  \bibinfo {author} {\bibfnamefont {R.}~\bibnamefont {Haas}}, \bibinfo {author}
  {\bibfnamefont {E.}~\bibnamefont {Pazos}}, \bibinfo {author} {\bibfnamefont
  {P.}~\bibnamefont {Laguna}}, \bibinfo {author} {\bibfnamefont
  {D.}~\bibnamefont {Shoemaker}},\ and\ \bibinfo {author} {\bibfnamefont
  {N.}~\bibnamefont {Yunes}},\ }\href
  {https://doi.org/10.1088/0264-9381/29/23/232002} {\bibfield  {journal}
  {\bibinfo  {journal} {Class. Quant. Grav.}\ }\textbf {\bibinfo {volume}
  {29}},\ \bibinfo {pages} {232002} (\bibinfo {year} {2012})},\ \Eprint
  {https://arxiv.org/abs/1112.3928} {arXiv:1112.3928 [gr-qc]} \BibitemShut
  {NoStop}%
\bibitem [{\citenamefont {Berti}\ \emph {et~al.}(2013)\citenamefont {Berti},
  \citenamefont {Cardoso}, \citenamefont {Gualtieri}, \citenamefont
  {Horbatsch},\ and\ \citenamefont {Sperhake}}]{Berti:2013gfa}%
  \BibitemOpen
  \bibfield  {author} {\bibinfo {author} {\bibfnamefont {E.}~\bibnamefont
  {Berti}}, \bibinfo {author} {\bibfnamefont {V.}~\bibnamefont {Cardoso}},
  \bibinfo {author} {\bibfnamefont {L.}~\bibnamefont {Gualtieri}}, \bibinfo
  {author} {\bibfnamefont {M.}~\bibnamefont {Horbatsch}},\ and\ \bibinfo
  {author} {\bibfnamefont {U.}~\bibnamefont {Sperhake}},\ }\href
  {https://doi.org/10.1103/PhysRevD.87.124020} {\bibfield  {journal} {\bibinfo
  {journal} {Phys. Rev. D}\ }\textbf {\bibinfo {volume} {87}},\ \bibinfo
  {pages} {124020} (\bibinfo {year} {2013})},\ \Eprint
  {https://arxiv.org/abs/1304.2836} {arXiv:1304.2836 [gr-qc]} \BibitemShut
  {NoStop}%
\bibitem [{\citenamefont {Cao}\ \emph {et~al.}(2013)\citenamefont {Cao},
  \citenamefont {Galaviz},\ and\ \citenamefont {Li}}]{Cao:2013osa}%
  \BibitemOpen
  \bibfield  {author} {\bibinfo {author} {\bibfnamefont {Z.}~\bibnamefont
  {Cao}}, \bibinfo {author} {\bibfnamefont {P.}~\bibnamefont {Galaviz}},\ and\
  \bibinfo {author} {\bibfnamefont {L.-F.}\ \bibnamefont {Li}},\ }\href
  {https://doi.org/10.1103/PhysRevD.87.104029} {\bibfield  {journal} {\bibinfo
  {journal} {Phys. Rev. D}\ }\textbf {\bibinfo {volume} {87}},\ \bibinfo
  {pages} {104029} (\bibinfo {year} {2013})},\ \Eprint
  {https://arxiv.org/abs/1608.07816} {arXiv:1608.07816 [gr-qc]} \BibitemShut
  {NoStop}%
\bibitem [{\citenamefont {Witek}\ \emph {et~al.}(2019)\citenamefont {Witek},
  \citenamefont {Gualtieri}, \citenamefont {Pani},\ and\ \citenamefont
  {Sotiriou}}]{Witek:2018dmd}%
  \BibitemOpen
  \bibfield  {author} {\bibinfo {author} {\bibfnamefont {H.}~\bibnamefont
  {Witek}}, \bibinfo {author} {\bibfnamefont {L.}~\bibnamefont {Gualtieri}},
  \bibinfo {author} {\bibfnamefont {P.}~\bibnamefont {Pani}},\ and\ \bibinfo
  {author} {\bibfnamefont {T.~P.}\ \bibnamefont {Sotiriou}},\ }\href
  {https://doi.org/10.1103/PhysRevD.99.064035} {\bibfield  {journal} {\bibinfo
  {journal} {Phys. Rev. D}\ }\textbf {\bibinfo {volume} {99}},\ \bibinfo
  {pages} {064035} (\bibinfo {year} {2019})},\ \Eprint
  {https://arxiv.org/abs/1810.05177} {arXiv:1810.05177 [gr-qc]} \BibitemShut
  {NoStop}%
\bibitem [{\citenamefont {Okounkova}(2020)}]{Okounkova:2020rqw}%
  \BibitemOpen
  \bibfield  {author} {\bibinfo {author} {\bibfnamefont {M.}~\bibnamefont
  {Okounkova}},\ }\href {https://doi.org/10.1103/PhysRevD.102.084046}
  {\bibfield  {journal} {\bibinfo  {journal} {Phys. Rev. D}\ }\textbf {\bibinfo
  {volume} {102}},\ \bibinfo {pages} {084046} (\bibinfo {year} {2020})},\
  \Eprint {https://arxiv.org/abs/2001.03571} {arXiv:2001.03571 [gr-qc]}
  \BibitemShut {NoStop}%
\bibitem [{\citenamefont {Okounkova}\ \emph {et~al.}(2022)\citenamefont
  {Okounkova}, \citenamefont {Isi}, \citenamefont {Chatziioannou},\ and\
  \citenamefont {Farr}}]{Okounkova:2022grv}%
  \BibitemOpen
  \bibfield  {author} {\bibinfo {author} {\bibfnamefont {M.}~\bibnamefont
  {Okounkova}}, \bibinfo {author} {\bibfnamefont {M.}~\bibnamefont {Isi}},
  \bibinfo {author} {\bibfnamefont {K.}~\bibnamefont {Chatziioannou}},\ and\
  \bibinfo {author} {\bibfnamefont {W.~M.}\ \bibnamefont {Farr}},\ }\href@noop
  {} {\  (\bibinfo {year} {2022})},\ \Eprint {https://arxiv.org/abs/2208.02805}
  {arXiv:2208.02805 [gr-qc]} \BibitemShut {NoStop}%
\bibitem [{\citenamefont {Hortacsu}(2012)}]{Hortacsu:2011rr}%
  \BibitemOpen
  \bibfield  {author} {\bibinfo {author} {\bibfnamefont {M.}~\bibnamefont
  {Hortacsu}},\ }\href {https://doi.org/10.1142/9789814417532_0002} {\bibfield
  {journal} {\bibinfo  {journal} {Adv. High Energy Phys}\ ,\ \bibinfo {pages}
  {23}} (\bibinfo {year} {2012})},\ \Eprint {https://arxiv.org/abs/1101.0471}
  {arXiv:1101.0471 [math-ph]} \BibitemShut {NoStop}%
\bibitem [{\citenamefont {Clough}(2021)}]{Clough:2021qlv}%
  \BibitemOpen
  \bibfield  {author} {\bibinfo {author} {\bibfnamefont {K.}~\bibnamefont
  {Clough}},\ }\href {https://doi.org/10.1088/1361-6382/ac10ee} {\bibfield
  {journal} {\bibinfo  {journal} {Class. Quant. Grav.}\ }\textbf {\bibinfo
  {volume} {38}},\ \bibinfo {pages} {167001} (\bibinfo {year} {2021})},\
  \Eprint {https://arxiv.org/abs/2104.13420} {arXiv:2104.13420 [gr-qc]}
  \BibitemShut {NoStop}%
\bibitem [{\citenamefont {Croft}(2022)}]{Croft:2022gks}%
  \BibitemOpen
  \bibfield  {author} {\bibinfo {author} {\bibfnamefont {R.}~\bibnamefont
  {Croft}},\ }\href@noop {} {\  (\bibinfo {year} {2022})},\ \Eprint
  {https://arxiv.org/abs/2203.13845} {arXiv:2203.13845 [gr-qc]} \BibitemShut
  {NoStop}%
\bibitem [{\citenamefont {Peters}(1964)}]{Peters:1964zz}%
  \BibitemOpen
  \bibfield  {author} {\bibinfo {author} {\bibfnamefont {P.~C.}\ \bibnamefont
  {Peters}},\ }\href {https://doi.org/10.1103/PhysRev.136.B1224} {\bibfield
  {journal} {\bibinfo  {journal} {Phys. Rev.}\ }\textbf {\bibinfo {volume}
  {136}},\ \bibinfo {pages} {B1224} (\bibinfo {year} {1964})}\BibitemShut
  {NoStop}%
\bibitem [{\citenamefont {Krolak}\ and\ \citenamefont
  {Schutz}(1987)}]{Krolak:1987ofj}%
  \BibitemOpen
  \bibfield  {author} {\bibinfo {author} {\bibfnamefont {A.}~\bibnamefont
  {Krolak}}\ and\ \bibinfo {author} {\bibfnamefont {B.~F.}\ \bibnamefont
  {Schutz}},\ }\href {https://doi.org/10.1007/BF00759095} {\bibfield  {journal}
  {\bibinfo  {journal} {Gen. Rel. Grav.}\ }\textbf {\bibinfo {volume} {19}},\
  \bibinfo {pages} {1163} (\bibinfo {year} {1987})}\BibitemShut {NoStop}%
\bibitem [{\citenamefont {Cardoso}\ \emph {et~al.}(2021)\citenamefont
  {Cardoso}, \citenamefont {Macedo},\ and\ \citenamefont
  {Vicente}}]{Cardoso:2020iji}%
  \BibitemOpen
  \bibfield  {author} {\bibinfo {author} {\bibfnamefont {V.}~\bibnamefont
  {Cardoso}}, \bibinfo {author} {\bibfnamefont {C.~F.~B.}\ \bibnamefont
  {Macedo}},\ and\ \bibinfo {author} {\bibfnamefont {R.}~\bibnamefont
  {Vicente}},\ }\href {https://doi.org/10.1103/PhysRevD.103.023015} {\bibfield
  {journal} {\bibinfo  {journal} {Phys. Rev. D}\ }\textbf {\bibinfo {volume}
  {103}},\ \bibinfo {pages} {023015} (\bibinfo {year} {2021})},\ \Eprint
  {https://arxiv.org/abs/2010.15151} {arXiv:2010.15151 [gr-qc]} \BibitemShut
  {NoStop}%
\bibitem [{\citenamefont {Will}(2019)}]{Will:2019lfe}%
  \BibitemOpen
  \bibfield  {author} {\bibinfo {author} {\bibfnamefont {C.~M.}\ \bibnamefont
  {Will}},\ }\href {https://doi.org/10.1088/1361-6382/ab41d0} {\bibfield
  {journal} {\bibinfo  {journal} {Class. Quant. Grav.}\ }\textbf {\bibinfo
  {volume} {36}},\ \bibinfo {pages} {195013} (\bibinfo {year} {2019})},\
  \Eprint {https://arxiv.org/abs/1906.08064} {arXiv:1906.08064 [gr-qc]}
  \BibitemShut {NoStop}%
\bibitem [{\citenamefont {Becker}\ \emph {et~al.}(2022)\citenamefont {Becker},
  \citenamefont {Sagunski}, \citenamefont {Prinz},\ and\ \citenamefont
  {Rastgoo}}]{Becker:2021ivq}%
  \BibitemOpen
  \bibfield  {author} {\bibinfo {author} {\bibfnamefont {N.}~\bibnamefont
  {Becker}}, \bibinfo {author} {\bibfnamefont {L.}~\bibnamefont {Sagunski}},
  \bibinfo {author} {\bibfnamefont {L.}~\bibnamefont {Prinz}},\ and\ \bibinfo
  {author} {\bibfnamefont {S.}~\bibnamefont {Rastgoo}},\ }\href
  {https://doi.org/10.1103/PhysRevD.105.063029} {\bibfield  {journal} {\bibinfo
   {journal} {Phys. Rev. D}\ }\textbf {\bibinfo {volume} {105}},\ \bibinfo
  {pages} {063029} (\bibinfo {year} {2022})},\ \Eprint
  {https://arxiv.org/abs/2112.09586} {arXiv:2112.09586 [gr-qc]} \BibitemShut
  {NoStop}%
\bibitem [{\citenamefont {Cardoso}\ \emph
  {et~al.}(2022{\natexlab{b}})\citenamefont {Cardoso}, \citenamefont {Ikeda},
  \citenamefont {Vicente},\ and\ \citenamefont {Zilh\~ao}}]{Cardoso:2022nzc}%
  \BibitemOpen
  \bibfield  {author} {\bibinfo {author} {\bibfnamefont {V.}~\bibnamefont
  {Cardoso}}, \bibinfo {author} {\bibfnamefont {T.}~\bibnamefont {Ikeda}},
  \bibinfo {author} {\bibfnamefont {R.}~\bibnamefont {Vicente}},\ and\ \bibinfo
  {author} {\bibfnamefont {M.}~\bibnamefont {Zilh\~ao}},\ }\href@noop {} {\
  (\bibinfo {year} {2022}{\natexlab{b}})},\ \Eprint
  {https://arxiv.org/abs/2207.09469} {arXiv:2207.09469 [gr-qc]} \BibitemShut
  {NoStop}%
\bibitem [{\citenamefont {Alcubierre}(2008)}]{AlcubierreMiguel2008It3N}%
  \BibitemOpen
  \bibfield  {author} {\bibinfo {author} {\bibfnamefont {M.}~\bibnamefont
  {Alcubierre}},\ }\href@noop {} {\emph {\bibinfo {title} {Introduction to 3+1
  Numerical Relativity}}},\ \bibinfo {series} {International Series of
  Monographs on Physics}, Vol.\ \bibinfo {volume} {140}\ (\bibinfo  {publisher}
  {Oxford University Press},\ \bibinfo {address} {Oxford},\ \bibinfo {year}
  {2008})\ pp.\ \bibinfo {pages} {1--458}\BibitemShut {NoStop}%
\bibitem [{\citenamefont {Brandt}\ and\ \citenamefont
  {Seidel}(1995)}]{Brandt:1994ee}%
  \BibitemOpen
  \bibfield  {author} {\bibinfo {author} {\bibfnamefont {S.~R.}\ \bibnamefont
  {Brandt}}\ and\ \bibinfo {author} {\bibfnamefont {E.}~\bibnamefont
  {Seidel}},\ }\href {https://doi.org/10.1103/PhysRevD.52.856} {\bibfield
  {journal} {\bibinfo  {journal} {Phys. Rev. D}\ }\textbf {\bibinfo {volume}
  {52}},\ \bibinfo {pages} {856} (\bibinfo {year} {1995})},\ \Eprint
  {https://arxiv.org/abs/gr-qc/9412072} {arXiv:gr-qc/9412072} \BibitemShut
  {NoStop}%
\bibitem [{\citenamefont {Brandt}\ and\ \citenamefont
  {Seidel}(1996)}]{Brandt:1996si}%
  \BibitemOpen
  \bibfield  {author} {\bibinfo {author} {\bibfnamefont {S.~R.}\ \bibnamefont
  {Brandt}}\ and\ \bibinfo {author} {\bibfnamefont {E.}~\bibnamefont
  {Seidel}},\ }\href {https://doi.org/10.1103/PhysRevD.54.1403} {\bibfield
  {journal} {\bibinfo  {journal} {Phys. Rev. D}\ }\textbf {\bibinfo {volume}
  {54}},\ \bibinfo {pages} {1403} (\bibinfo {year} {1996})},\ \Eprint
  {https://arxiv.org/abs/gr-qc/9601010} {arXiv:gr-qc/9601010} \BibitemShut
  {NoStop}%
\bibitem [{\citenamefont {Okawa}\ \emph {et~al.}(2014)\citenamefont {Okawa},
  \citenamefont {Witek},\ and\ \citenamefont {Cardoso}}]{Okawa:2014nda}%
  \BibitemOpen
  \bibfield  {author} {\bibinfo {author} {\bibfnamefont {H.}~\bibnamefont
  {Okawa}}, \bibinfo {author} {\bibfnamefont {H.}~\bibnamefont {Witek}},\ and\
  \bibinfo {author} {\bibfnamefont {V.}~\bibnamefont {Cardoso}},\ }\href
  {https://doi.org/10.1103/PhysRevD.89.104032} {\bibfield  {journal} {\bibinfo
  {journal} {Phys. Rev. D}\ }\textbf {\bibinfo {volume} {89}},\ \bibinfo
  {pages} {104032} (\bibinfo {year} {2014})},\ \Eprint
  {https://arxiv.org/abs/1401.1548} {arXiv:1401.1548 [gr-qc]} \BibitemShut
  {NoStop}%
\bibitem [{\citenamefont {Dennison}\ \emph {et~al.}(2006)\citenamefont
  {Dennison}, \citenamefont {Baumgarte},\ and\ \citenamefont
  {Pfeiffer}}]{Dennison:2006nq}%
  \BibitemOpen
  \bibfield  {author} {\bibinfo {author} {\bibfnamefont {K.~A.}\ \bibnamefont
  {Dennison}}, \bibinfo {author} {\bibfnamefont {T.~W.}\ \bibnamefont
  {Baumgarte}},\ and\ \bibinfo {author} {\bibfnamefont {H.~P.}\ \bibnamefont
  {Pfeiffer}},\ }\href {https://doi.org/10.1103/PhysRevD.74.064016} {\bibfield
  {journal} {\bibinfo  {journal} {Phys. Rev. D}\ }\textbf {\bibinfo {volume}
  {74}},\ \bibinfo {pages} {064016} (\bibinfo {year} {2006})},\ \Eprint
  {https://arxiv.org/abs/gr-qc/0606037} {arXiv:gr-qc/0606037} \BibitemShut
  {NoStop}%
\bibitem [{\citenamefont {Cook}(2000)}]{Cook:2000vr}%
  \BibitemOpen
  \bibfield  {author} {\bibinfo {author} {\bibfnamefont {G.~B.}\ \bibnamefont
  {Cook}},\ }\href {https://doi.org/10.12942/lrr-2000-5} {\bibfield  {journal}
  {\bibinfo  {journal} {Living Rev. Rel.}\ }\textbf {\bibinfo {volume} {3}},\
  \bibinfo {pages} {5} (\bibinfo {year} {2000})},\ \Eprint
  {https://arxiv.org/abs/gr-qc/0007085} {arXiv:gr-qc/0007085} \BibitemShut
  {NoStop}%
\bibitem [{\citenamefont {Bowen}(1979)}]{Bowen:1979}%
  \BibitemOpen
  \bibfield  {author} {\bibinfo {author} {\bibfnamefont {J.~M.}\ \bibnamefont
  {Bowen}},\ }\href@noop {} {\bibfield  {journal} {\bibinfo  {journal} {General
  Relativity and Gravitation}\ }\textbf {\bibinfo {volume} {11}},\ \bibinfo
  {pages} {227} (\bibinfo {year} {1979})}\BibitemShut {NoStop}%
\bibitem [{\citenamefont {Bowen}\ and\ \citenamefont
  {York}(1980)}]{Bowen:1980yu}%
  \BibitemOpen
  \bibfield  {author} {\bibinfo {author} {\bibfnamefont {J.~M.}\ \bibnamefont
  {Bowen}}\ and\ \bibinfo {author} {\bibfnamefont {J.~W.}\ \bibnamefont {York},
  \bibfnamefont {Jr.}},\ }\href {https://doi.org/10.1103/PhysRevD.21.2047}
  {\bibfield  {journal} {\bibinfo  {journal} {Phys. Rev. D}\ }\textbf {\bibinfo
  {volume} {21}},\ \bibinfo {pages} {2047} (\bibinfo {year}
  {1980})}\BibitemShut {NoStop}%
\bibitem [{\citenamefont {York}(1978)}]{York:1978gql}%
  \BibitemOpen
  \bibfield  {author} {\bibinfo {author} {\bibfnamefont {J.~W.}\ \bibnamefont
  {York}, \bibfnamefont {Jr.}},\ }in\ \href@noop {} {\emph {\bibinfo
  {booktitle} {{Workshop on Sources of Gravitational Radiation}}}}\ (\bibinfo
  {year} {1978})\ pp.\ \bibinfo {pages} {83--126}\BibitemShut {NoStop}%
\bibitem [{\citenamefont {Radia}\ \emph {et~al.}(2022)\citenamefont {Radia},
  \citenamefont {Sperhake}, \citenamefont {Drew}, \citenamefont {Clough},
  \citenamefont {Figueras}, \citenamefont {Lim}, \citenamefont {Ripley},
  \citenamefont {Aurrekoetxea}, \citenamefont {Fran\c{c}a},\ and\ \citenamefont
  {Helfer}}]{Radia:2021smk}%
  \BibitemOpen
  \bibfield  {author} {\bibinfo {author} {\bibfnamefont {M.}~\bibnamefont
  {Radia}}, \bibinfo {author} {\bibfnamefont {U.}~\bibnamefont {Sperhake}},
  \bibinfo {author} {\bibfnamefont {A.}~\bibnamefont {Drew}}, \bibinfo {author}
  {\bibfnamefont {K.}~\bibnamefont {Clough}}, \bibinfo {author} {\bibfnamefont
  {P.}~\bibnamefont {Figueras}}, \bibinfo {author} {\bibfnamefont {E.~A.}\
  \bibnamefont {Lim}}, \bibinfo {author} {\bibfnamefont {J.~L.}\ \bibnamefont
  {Ripley}}, \bibinfo {author} {\bibfnamefont {J.~C.}\ \bibnamefont
  {Aurrekoetxea}}, \bibinfo {author} {\bibfnamefont {T.}~\bibnamefont
  {Fran\c{c}a}},\ and\ \bibinfo {author} {\bibfnamefont {T.}~\bibnamefont
  {Helfer}},\ }\href {https://doi.org/10.1088/1361-6382/ac6fa9} {\bibfield
  {journal} {\bibinfo  {journal} {Class. Quant. Grav.}\ }\textbf {\bibinfo
  {volume} {39}},\ \bibinfo {pages} {135006} (\bibinfo {year} {2022})},\
  \Eprint {https://arxiv.org/abs/2112.10567} {arXiv:2112.10567 [gr-qc]}
  \BibitemShut {NoStop}%
\bibitem [{\citenamefont {Liu}\ \emph {et~al.}(2009)\citenamefont {Liu},
  \citenamefont {Etienne},\ and\ \citenamefont {Shapiro}}]{Liu:2009al}%
  \BibitemOpen
  \bibfield  {author} {\bibinfo {author} {\bibfnamefont {Y.~T.}\ \bibnamefont
  {Liu}}, \bibinfo {author} {\bibfnamefont {Z.~B.}\ \bibnamefont {Etienne}},\
  and\ \bibinfo {author} {\bibfnamefont {S.~L.}\ \bibnamefont {Shapiro}},\
  }\href {https://doi.org/10.1103/PhysRevD.80.121503} {\bibfield  {journal}
  {\bibinfo  {journal} {Phys. Rev. D}\ }\textbf {\bibinfo {volume} {80}},\
  \bibinfo {pages} {121503(R)} (\bibinfo {year} {2009})},\ \Eprint
  {https://arxiv.org/abs/1001.4077} {arXiv:1001.4077 [gr-qc]} \BibitemShut
  {NoStop}%
\bibitem [{\citenamefont {Hannam}\ \emph {et~al.}(2007)\citenamefont {Hannam},
  \citenamefont {Husa}, \citenamefont {Bruegmann}, \citenamefont {Gonzalez},
  \citenamefont {Sperhake},\ and\ \citenamefont {Murchadha}}]{Hannam:2006xw}%
  \BibitemOpen
  \bibfield  {author} {\bibinfo {author} {\bibfnamefont {M.}~\bibnamefont
  {Hannam}}, \bibinfo {author} {\bibfnamefont {S.}~\bibnamefont {Husa}},
  \bibinfo {author} {\bibfnamefont {B.}~\bibnamefont {Bruegmann}}, \bibinfo
  {author} {\bibfnamefont {J.~A.}\ \bibnamefont {Gonzalez}}, \bibinfo {author}
  {\bibfnamefont {U.}~\bibnamefont {Sperhake}},\ and\ \bibinfo {author}
  {\bibfnamefont {N.~O.}\ \bibnamefont {Murchadha}},\ }\href
  {https://doi.org/10.1088/1742-6596/66/1/012047} {\bibfield  {journal}
  {\bibinfo  {journal} {J. Phys. Conf. Ser.}\ }\textbf {\bibinfo {volume}
  {66}},\ \bibinfo {pages} {012047} (\bibinfo {year} {2007})},\ \Eprint
  {https://arxiv.org/abs/gr-qc/0612097} {arXiv:gr-qc/0612097} \BibitemShut
  {NoStop}%
\bibitem [{\citenamefont {Pfeiffer}\ \emph {et~al.}(2007)\citenamefont
  {Pfeiffer}, \citenamefont {Brown}, \citenamefont {Kidder}, \citenamefont
  {Lindblom}, \citenamefont {Lovelace},\ and\ \citenamefont
  {Scheel}}]{Pfeiffer:2007yz}%
  \BibitemOpen
  \bibfield  {author} {\bibinfo {author} {\bibfnamefont {H.~P.}\ \bibnamefont
  {Pfeiffer}}, \bibinfo {author} {\bibfnamefont {D.~A.}\ \bibnamefont {Brown}},
  \bibinfo {author} {\bibfnamefont {L.~E.}\ \bibnamefont {Kidder}}, \bibinfo
  {author} {\bibfnamefont {L.}~\bibnamefont {Lindblom}}, \bibinfo {author}
  {\bibfnamefont {G.}~\bibnamefont {Lovelace}},\ and\ \bibinfo {author}
  {\bibfnamefont {M.~A.}\ \bibnamefont {Scheel}},\ }\href
  {https://doi.org/10.1088/0264-9381/24/12/S06} {\bibfield  {journal} {\bibinfo
   {journal} {Class. Quant. Grav.}\ }\textbf {\bibinfo {volume} {24}},\
  \bibinfo {pages} {S59} (\bibinfo {year} {2007})},\ \Eprint
  {https://arxiv.org/abs/gr-qc/0702106} {arXiv:gr-qc/0702106} \BibitemShut
  {NoStop}%
\bibitem [{\citenamefont {Habib}\ \emph {et~al.}(2021)\citenamefont {Habib},
  \citenamefont {Ramos-Buades}, \citenamefont {Huerta}, \citenamefont {Husa},
  \citenamefont {Haas},\ and\ \citenamefont {Etienne}}]{Habib:2020dba}%
  \BibitemOpen
  \bibfield  {author} {\bibinfo {author} {\bibfnamefont {S.}~\bibnamefont
  {Habib}}, \bibinfo {author} {\bibfnamefont {A.}~\bibnamefont {Ramos-Buades}},
  \bibinfo {author} {\bibfnamefont {E.~A.}\ \bibnamefont {Huerta}}, \bibinfo
  {author} {\bibfnamefont {S.}~\bibnamefont {Husa}}, \bibinfo {author}
  {\bibfnamefont {R.}~\bibnamefont {Haas}},\ and\ \bibinfo {author}
  {\bibfnamefont {Z.}~\bibnamefont {Etienne}},\ }\href
  {https://doi.org/10.1088/1361-6382/abe691} {\bibfield  {journal} {\bibinfo
  {journal} {Class. Quant. Grav.}\ }\textbf {\bibinfo {volume} {38}},\ \bibinfo
  {pages} {125007} (\bibinfo {year} {2021})},\ \Eprint
  {https://arxiv.org/abs/2011.08878} {arXiv:2011.08878 [gr-qc]} \BibitemShut
  {NoStop}%
\bibitem [{\citenamefont {Buonanno}\ \emph {et~al.}(2011)\citenamefont
  {Buonanno}, \citenamefont {Kidder}, \citenamefont {Mroue}, \citenamefont
  {Pfeiffer},\ and\ \citenamefont {Taracchini}}]{Buonanno:2010yk}%
  \BibitemOpen
  \bibfield  {author} {\bibinfo {author} {\bibfnamefont {A.}~\bibnamefont
  {Buonanno}}, \bibinfo {author} {\bibfnamefont {L.~E.}\ \bibnamefont
  {Kidder}}, \bibinfo {author} {\bibfnamefont {A.~H.}\ \bibnamefont {Mroue}},
  \bibinfo {author} {\bibfnamefont {H.~P.}\ \bibnamefont {Pfeiffer}},\ and\
  \bibinfo {author} {\bibfnamefont {A.}~\bibnamefont {Taracchini}},\ }\href
  {https://doi.org/10.1103/PhysRevD.83.104034} {\bibfield  {journal} {\bibinfo
  {journal} {Phys. Rev. D}\ }\textbf {\bibinfo {volume} {83}},\ \bibinfo
  {pages} {104034} (\bibinfo {year} {2011})},\ \Eprint
  {https://arxiv.org/abs/1012.1549} {arXiv:1012.1549 [gr-qc]} \BibitemShut
  {NoStop}%
\bibitem [{\citenamefont {Ramos-Buades}\ \emph {et~al.}(2019)\citenamefont
  {Ramos-Buades}, \citenamefont {Husa},\ and\ \citenamefont
  {Pratten}}]{Ramos-Buades:2018azo}%
  \BibitemOpen
  \bibfield  {author} {\bibinfo {author} {\bibfnamefont {A.}~\bibnamefont
  {Ramos-Buades}}, \bibinfo {author} {\bibfnamefont {S.}~\bibnamefont {Husa}},\
  and\ \bibinfo {author} {\bibfnamefont {G.}~\bibnamefont {Pratten}},\ }\href
  {https://doi.org/10.1103/PhysRevD.99.023003} {\bibfield  {journal} {\bibinfo
  {journal} {Phys. Rev. D}\ }\textbf {\bibinfo {volume} {99}},\ \bibinfo
  {pages} {023003} (\bibinfo {year} {2019})},\ \Eprint
  {https://arxiv.org/abs/1810.00036} {arXiv:1810.00036 [gr-qc]} \BibitemShut
  {NoStop}%
\bibitem [{\citenamefont {Blanchet}(2006)}]{Blanchet:2006zz}%
  \BibitemOpen
  \bibfield  {author} {\bibinfo {author} {\bibfnamefont {L.}~\bibnamefont
  {Blanchet}},\ }\href@noop {} {\bibfield  {journal} {\bibinfo  {journal}
  {Living Rev. Rel.}\ }\textbf {\bibinfo {volume} {9}},\ \bibinfo {pages} {4}
  (\bibinfo {year} {2006})}\BibitemShut {NoStop}%
\bibitem [{\citenamefont {Bernard}(2018)}]{Bernard:2018hta}%
  \BibitemOpen
  \bibfield  {author} {\bibinfo {author} {\bibfnamefont {L.}~\bibnamefont
  {Bernard}},\ }\href {https://doi.org/10.1103/PhysRevD.98.044004} {\bibfield
  {journal} {\bibinfo  {journal} {Phys. Rev. D}\ }\textbf {\bibinfo {volume}
  {98}},\ \bibinfo {pages} {044004} (\bibinfo {year} {2018})},\ \Eprint
  {https://arxiv.org/abs/1802.10201} {arXiv:1802.10201 [gr-qc]} \BibitemShut
  {NoStop}%
\bibitem [{\citenamefont {Juli\'e}\ and\ \citenamefont
  {Berti}(2019)}]{Julie:2019sab}%
  \BibitemOpen
  \bibfield  {author} {\bibinfo {author} {\bibfnamefont {F.-L.}\ \bibnamefont
  {Juli\'e}}\ and\ \bibinfo {author} {\bibfnamefont {E.}~\bibnamefont
  {Berti}},\ }\href {https://doi.org/10.1103/PhysRevD.100.104061} {\bibfield
  {journal} {\bibinfo  {journal} {Phys. Rev. D}\ }\textbf {\bibinfo {volume}
  {100}},\ \bibinfo {pages} {104061} (\bibinfo {year} {2019})},\ \Eprint
  {https://arxiv.org/abs/1909.05258} {arXiv:1909.05258 [gr-qc]} \BibitemShut
  {NoStop}%
\bibitem [{\citenamefont {Yagi}\ \emph {et~al.}(2012)\citenamefont {Yagi},
  \citenamefont {Stein}, \citenamefont {Yunes},\ and\ \citenamefont
  {Tanaka}}]{Yagi:2011xp}%
  \BibitemOpen
  \bibfield  {author} {\bibinfo {author} {\bibfnamefont {K.}~\bibnamefont
  {Yagi}}, \bibinfo {author} {\bibfnamefont {L.~C.}\ \bibnamefont {Stein}},
  \bibinfo {author} {\bibfnamefont {N.}~\bibnamefont {Yunes}},\ and\ \bibinfo
  {author} {\bibfnamefont {T.}~\bibnamefont {Tanaka}},\ }\href
  {https://doi.org/10.1103/PhysRevD.85.064022} {\bibfield  {journal} {\bibinfo
  {journal} {Phys. Rev. D}\ }\textbf {\bibinfo {volume} {85}},\ \bibinfo
  {pages} {064022} (\bibinfo {year} {2012})},\ \bibinfo {note} {[Erratum:
  Phys.Rev.D 93, 029902 (2016)]},\ \Eprint {https://arxiv.org/abs/1110.5950}
  {arXiv:1110.5950 [gr-qc]} \BibitemShut {NoStop}%
\bibitem [{\citenamefont {Shiralilou}\ \emph {et~al.}(2021)\citenamefont
  {Shiralilou}, \citenamefont {Hinderer}, \citenamefont {Nissanke},
  \citenamefont {Ortiz},\ and\ \citenamefont {Witek}}]{Shiralilou:2020gah}%
  \BibitemOpen
  \bibfield  {author} {\bibinfo {author} {\bibfnamefont {B.}~\bibnamefont
  {Shiralilou}}, \bibinfo {author} {\bibfnamefont {T.}~\bibnamefont
  {Hinderer}}, \bibinfo {author} {\bibfnamefont {S.~M.}\ \bibnamefont
  {Nissanke}}, \bibinfo {author} {\bibfnamefont {N.}~\bibnamefont {Ortiz}},\
  and\ \bibinfo {author} {\bibfnamefont {H.}~\bibnamefont {Witek}},\ }\href
  {https://doi.org/10.1103/PhysRevD.103.L121503} {\bibfield  {journal}
  {\bibinfo  {journal} {Phys. Rev. D}\ }\textbf {\bibinfo {volume} {103}},\
  \bibinfo {pages} {L121503} (\bibinfo {year} {2021})},\ \Eprint
  {https://arxiv.org/abs/2012.09162} {arXiv:2012.09162 [gr-qc]} \BibitemShut
  {NoStop}%
\bibitem [{\citenamefont {Shiralilou}\ \emph {et~al.}(2022)\citenamefont
  {Shiralilou}, \citenamefont {Hinderer}, \citenamefont {Nissanke},
  \citenamefont {Ortiz},\ and\ \citenamefont {Witek}}]{Shiralilou:2021mfl}%
  \BibitemOpen
  \bibfield  {author} {\bibinfo {author} {\bibfnamefont {B.}~\bibnamefont
  {Shiralilou}}, \bibinfo {author} {\bibfnamefont {T.}~\bibnamefont
  {Hinderer}}, \bibinfo {author} {\bibfnamefont {S.~M.}\ \bibnamefont
  {Nissanke}}, \bibinfo {author} {\bibfnamefont {N.}~\bibnamefont {Ortiz}},\
  and\ \bibinfo {author} {\bibfnamefont {H.}~\bibnamefont {Witek}},\ }\href
  {https://doi.org/10.1088/1361-6382/ac4196} {\bibfield  {journal} {\bibinfo
  {journal} {Class. Quant. Grav.}\ }\textbf {\bibinfo {volume} {39}},\ \bibinfo
  {pages} {035002} (\bibinfo {year} {2022})},\ \Eprint
  {https://arxiv.org/abs/2105.13972} {arXiv:2105.13972 [gr-qc]} \BibitemShut
  {NoStop}%
\bibitem [{\citenamefont {Brax}\ \emph {et~al.}(2021)\citenamefont {Brax},
  \citenamefont {Davis}, \citenamefont {Melville},\ and\ \citenamefont
  {Wong}}]{Brax:2021qqo}%
  \BibitemOpen
  \bibfield  {author} {\bibinfo {author} {\bibfnamefont {P.}~\bibnamefont
  {Brax}}, \bibinfo {author} {\bibfnamefont {A.-C.}\ \bibnamefont {Davis}},
  \bibinfo {author} {\bibfnamefont {S.}~\bibnamefont {Melville}},\ and\
  \bibinfo {author} {\bibfnamefont {L.~K.}\ \bibnamefont {Wong}},\ }\href
  {https://doi.org/10.1088/1475-7516/2021/10/075} {\bibfield  {journal}
  {\bibinfo  {journal} {JCAP}\ }\textbf {\bibinfo {volume} {10}},\ \bibinfo
  {pages} {075}},\ \Eprint {https://arxiv.org/abs/2107.10841} {arXiv:2107.10841
  [gr-qc]} \BibitemShut {NoStop}%
\bibitem [{\citenamefont {Tichy}(2012)}]{Tichy:2012rp}%
  \BibitemOpen
  \bibfield  {author} {\bibinfo {author} {\bibfnamefont {W.}~\bibnamefont
  {Tichy}},\ }\href {https://doi.org/10.1103/PhysRevD.86.064024} {\bibfield
  {journal} {\bibinfo  {journal} {Phys. Rev. D}\ }\textbf {\bibinfo {volume}
  {86}},\ \bibinfo {pages} {064024} (\bibinfo {year} {2012})},\ \Eprint
  {https://arxiv.org/abs/1209.5336} {arXiv:1209.5336 [gr-qc]} \BibitemShut
  {NoStop}%
\bibitem [{\citenamefont {Kyutoku}\ \emph {et~al.}(2014)\citenamefont
  {Kyutoku}, \citenamefont {Shibata},\ and\ \citenamefont
  {Taniguchi}}]{Kyutoku:2014yba}%
  \BibitemOpen
  \bibfield  {author} {\bibinfo {author} {\bibfnamefont {K.}~\bibnamefont
  {Kyutoku}}, \bibinfo {author} {\bibfnamefont {M.}~\bibnamefont {Shibata}},\
  and\ \bibinfo {author} {\bibfnamefont {K.}~\bibnamefont {Taniguchi}},\ }\href
  {https://doi.org/10.1103/PhysRevD.90.064006} {\bibfield  {journal} {\bibinfo
  {journal} {Phys. Rev. D}\ }\textbf {\bibinfo {volume} {90}},\ \bibinfo
  {pages} {064006} (\bibinfo {year} {2014})},\ \Eprint
  {https://arxiv.org/abs/1405.6207} {arXiv:1405.6207 [gr-qc]} \BibitemShut
  {NoStop}%
\bibitem [{\citenamefont {Tsokaros}\ \emph {et~al.}(2015)\citenamefont
  {Tsokaros}, \citenamefont {Ury\={u}},\ and\ \citenamefont
  {Rezzolla}}]{Tsokaros:2015fea}%
  \BibitemOpen
  \bibfield  {author} {\bibinfo {author} {\bibfnamefont {A.}~\bibnamefont
  {Tsokaros}}, \bibinfo {author} {\bibfnamefont {K.}~\bibnamefont {Ury\={u}}},\
  and\ \bibinfo {author} {\bibfnamefont {L.}~\bibnamefont {Rezzolla}},\ }\href
  {https://doi.org/10.1103/PhysRevD.91.104030} {\bibfield  {journal} {\bibinfo
  {journal} {Phys. Rev. D}\ }\textbf {\bibinfo {volume} {91}},\ \bibinfo
  {pages} {104030} (\bibinfo {year} {2015})},\ \Eprint
  {https://arxiv.org/abs/1502.05674} {arXiv:1502.05674 [gr-qc]} \BibitemShut
  {NoStop}%
\bibitem [{\citenamefont {Stanzione}\ \emph {et~al.}(2020)\citenamefont
  {Stanzione}, \citenamefont {West}, \citenamefont {Evans}, \citenamefont
  {Minyard}, \citenamefont {Ghattas},\ and\ \citenamefont
  {Panda}}]{10.1145/3311790.3396656}%
  \BibitemOpen
  \bibfield  {author} {\bibinfo {author} {\bibfnamefont {D.}~\bibnamefont
  {Stanzione}}, \bibinfo {author} {\bibfnamefont {J.}~\bibnamefont {West}},
  \bibinfo {author} {\bibfnamefont {R.~T.}\ \bibnamefont {Evans}}, \bibinfo
  {author} {\bibfnamefont {T.}~\bibnamefont {Minyard}}, \bibinfo {author}
  {\bibfnamefont {O.}~\bibnamefont {Ghattas}},\ and\ \bibinfo {author}
  {\bibfnamefont {D.~K.}\ \bibnamefont {Panda}},\ }in\ \href
  {https://doi.org/10.1145/3311790.3396656} {\emph {\bibinfo {booktitle}
  {Practice and Experience in Advanced Research Computing}}},\ \bibinfo {series
  and number} {PEARC ’20}\ (\bibinfo  {publisher} {Association for Computing
  Machinery},\ \bibinfo {address} {New York, NY, USA},\ \bibinfo {year}
  {2020})\ p.\ \bibinfo {pages} {106–111}\BibitemShut {NoStop}%
\bibitem [{\citenamefont {Arnowitt}\ \emph {et~al.}(2008)\citenamefont
  {Arnowitt}, \citenamefont {Deser},\ and\ \citenamefont
  {Misner}}]{Arnowitt:1962hi}%
  \BibitemOpen
  \bibfield  {author} {\bibinfo {author} {\bibfnamefont {R.~L.}\ \bibnamefont
  {Arnowitt}}, \bibinfo {author} {\bibfnamefont {S.}~\bibnamefont {Deser}},\
  and\ \bibinfo {author} {\bibfnamefont {C.~W.}\ \bibnamefont {Misner}},\
  }\href {https://doi.org/10.1007/s10714-008-0661-1} {\bibfield  {journal}
  {\bibinfo  {journal} {Gen. Rel. Grav.}\ }\textbf {\bibinfo {volume} {40}},\
  \bibinfo {pages} {1997} (\bibinfo {year} {2008})},\ \Eprint
  {https://arxiv.org/abs/gr-qc/0405109} {arXiv:gr-qc/0405109} \BibitemShut
  {NoStop}%
\bibitem [{\citenamefont {Gourgoulhon}(2007)}]{Gourgoulhon:2007ue}%
  \BibitemOpen
  \bibfield  {author} {\bibinfo {author} {\bibfnamefont {E.}~\bibnamefont
  {Gourgoulhon}},\ }\href@noop {} {\  (\bibinfo {year} {2007})},\ \Eprint
  {https://arxiv.org/abs/gr-qc/0703035} {arXiv:gr-qc/0703035} \BibitemShut
  {NoStop}%
\bibitem [{\citenamefont {Baumgarte}\ and\ \citenamefont
  {Shapiro}(2003)}]{Baumgarte:2002jm}%
  \BibitemOpen
  \bibfield  {author} {\bibinfo {author} {\bibfnamefont {T.~W.}\ \bibnamefont
  {Baumgarte}}\ and\ \bibinfo {author} {\bibfnamefont {S.~L.}\ \bibnamefont
  {Shapiro}},\ }\href {https://doi.org/10.1016/S0370-1573(02)00537-9}
  {\bibfield  {journal} {\bibinfo  {journal} {Phys. Rept.}\ }\textbf {\bibinfo
  {volume} {376}},\ \bibinfo {pages} {41} (\bibinfo {year} {2003})},\ \Eprint
  {https://arxiv.org/abs/gr-qc/0211028} {arXiv:gr-qc/0211028} \BibitemShut
  {NoStop}%
\bibitem [{\citenamefont {Alic}\ \emph {et~al.}(2012)\citenamefont {Alic},
  \citenamefont {Bona-Casas}, \citenamefont {Bona}, \citenamefont {Rezzolla},\
  and\ \citenamefont {Palenzuela}}]{Alic:2011gg}%
  \BibitemOpen
  \bibfield  {author} {\bibinfo {author} {\bibfnamefont {D.}~\bibnamefont
  {Alic}}, \bibinfo {author} {\bibfnamefont {C.}~\bibnamefont {Bona-Casas}},
  \bibinfo {author} {\bibfnamefont {C.}~\bibnamefont {Bona}}, \bibinfo {author}
  {\bibfnamefont {L.}~\bibnamefont {Rezzolla}},\ and\ \bibinfo {author}
  {\bibfnamefont {C.}~\bibnamefont {Palenzuela}},\ }\href
  {https://doi.org/10.1103/PhysRevD.85.064040} {\bibfield  {journal} {\bibinfo
  {journal} {Phys. Rev. D}\ }\textbf {\bibinfo {volume} {85}},\ \bibinfo
  {pages} {064040} (\bibinfo {year} {2012})},\ \Eprint
  {https://arxiv.org/abs/1106.2254} {arXiv:1106.2254 [gr-qc]} \BibitemShut
  {NoStop}%
\bibitem [{\citenamefont {Bona}\ \emph {et~al.}(1995)\citenamefont {Bona},
  \citenamefont {Masso}, \citenamefont {Seidel},\ and\ \citenamefont
  {Stela}}]{Bona:1994dr}%
  \BibitemOpen
  \bibfield  {author} {\bibinfo {author} {\bibfnamefont {C.}~\bibnamefont
  {Bona}}, \bibinfo {author} {\bibfnamefont {J.}~\bibnamefont {Masso}},
  \bibinfo {author} {\bibfnamefont {E.}~\bibnamefont {Seidel}},\ and\ \bibinfo
  {author} {\bibfnamefont {J.}~\bibnamefont {Stela}},\ }\href
  {https://doi.org/10.1103/PhysRevLett.75.600} {\bibfield  {journal} {\bibinfo
  {journal} {Phys. Rev. Lett.}\ }\textbf {\bibinfo {volume} {75}},\ \bibinfo
  {pages} {600} (\bibinfo {year} {1995})},\ \Eprint
  {https://arxiv.org/abs/gr-qc/9412071} {arXiv:gr-qc/9412071} \BibitemShut
  {NoStop}%
\bibitem [{\citenamefont {Baker}\ \emph {et~al.}(2006)\citenamefont {Baker},
  \citenamefont {Centrella}, \citenamefont {Choi}, \citenamefont {Koppitz},\
  and\ \citenamefont {van Meter}}]{Baker:2005vv}%
  \BibitemOpen
  \bibfield  {author} {\bibinfo {author} {\bibfnamefont {J.~G.}\ \bibnamefont
  {Baker}}, \bibinfo {author} {\bibfnamefont {J.}~\bibnamefont {Centrella}},
  \bibinfo {author} {\bibfnamefont {D.-I.}\ \bibnamefont {Choi}}, \bibinfo
  {author} {\bibfnamefont {M.}~\bibnamefont {Koppitz}},\ and\ \bibinfo {author}
  {\bibfnamefont {J.}~\bibnamefont {van Meter}},\ }\href
  {https://doi.org/10.1103/PhysRevLett.96.111102} {\bibfield  {journal}
  {\bibinfo  {journal} {Phys. Rev. Lett.}\ }\textbf {\bibinfo {volume} {96}},\
  \bibinfo {pages} {111102} (\bibinfo {year} {2006})},\ \Eprint
  {https://arxiv.org/abs/gr-qc/0511103} {arXiv:gr-qc/0511103} \BibitemShut
  {NoStop}%
\bibitem [{\citenamefont {Campanelli}\ \emph {et~al.}(2006)\citenamefont
  {Campanelli}, \citenamefont {Lousto}, \citenamefont {Marronetti},\ and\
  \citenamefont {Zlochower}}]{Campanelli:2005dd}%
  \BibitemOpen
  \bibfield  {author} {\bibinfo {author} {\bibfnamefont {M.}~\bibnamefont
  {Campanelli}}, \bibinfo {author} {\bibfnamefont {C.~O.}\ \bibnamefont
  {Lousto}}, \bibinfo {author} {\bibfnamefont {P.}~\bibnamefont {Marronetti}},\
  and\ \bibinfo {author} {\bibfnamefont {Y.}~\bibnamefont {Zlochower}},\ }\href
  {https://doi.org/10.1103/PhysRevLett.96.111101} {\bibfield  {journal}
  {\bibinfo  {journal} {Phys. Rev. Lett.}\ }\textbf {\bibinfo {volume} {96}},\
  \bibinfo {pages} {111101} (\bibinfo {year} {2006})},\ \Eprint
  {https://arxiv.org/abs/gr-qc/0511048} {arXiv:gr-qc/0511048} \BibitemShut
  {NoStop}%
\bibitem [{\citenamefont {van Meter}\ \emph {et~al.}(2006)\citenamefont {van
  Meter}, \citenamefont {Baker}, \citenamefont {Koppitz},\ and\ \citenamefont
  {Choi}}]{vanMeter:2006vi}%
  \BibitemOpen
  \bibfield  {author} {\bibinfo {author} {\bibfnamefont {J.~R.}\ \bibnamefont
  {van Meter}}, \bibinfo {author} {\bibfnamefont {J.~G.}\ \bibnamefont
  {Baker}}, \bibinfo {author} {\bibfnamefont {M.}~\bibnamefont {Koppitz}},\
  and\ \bibinfo {author} {\bibfnamefont {D.-I.}\ \bibnamefont {Choi}},\ }\href
  {https://doi.org/10.1103/PhysRevD.73.124011} {\bibfield  {journal} {\bibinfo
  {journal} {Phys. Rev. D}\ }\textbf {\bibinfo {volume} {73}},\ \bibinfo
  {pages} {124011} (\bibinfo {year} {2006})},\ \Eprint
  {https://arxiv.org/abs/gr-qc/0605030} {arXiv:gr-qc/0605030} \BibitemShut
  {NoStop}%
\bibitem [{\citenamefont {Andrade}\ \emph {et~al.}(2021)\citenamefont {Andrade}
  \emph {et~al.}}]{Andrade:2021rbd}%
  \BibitemOpen
  \bibfield  {author} {\bibinfo {author} {\bibfnamefont {T.}~\bibnamefont
  {Andrade}} \emph {et~al.},\ }\href {https://doi.org/10.21105/joss.03703}
  {\bibfield  {journal} {\bibinfo  {journal} {J. Open Source Softw.}\ }\textbf
  {\bibinfo {volume} {6}},\ \bibinfo {pages} {3703} (\bibinfo {year} {2021})},\
  \Eprint {https://arxiv.org/abs/2201.03458} {arXiv:2201.03458 [gr-qc]}
  \BibitemShut {NoStop}%
\bibitem [{\citenamefont {Clough}\ \emph {et~al.}(2015)\citenamefont {Clough},
  \citenamefont {Figueras}, \citenamefont {Finkel}, \citenamefont {Kunesch},
  \citenamefont {Lim},\ and\ \citenamefont {Tunyasuvunakool}}]{Clough:2015sqa}%
  \BibitemOpen
  \bibfield  {author} {\bibinfo {author} {\bibfnamefont {K.}~\bibnamefont
  {Clough}}, \bibinfo {author} {\bibfnamefont {P.}~\bibnamefont {Figueras}},
  \bibinfo {author} {\bibfnamefont {H.}~\bibnamefont {Finkel}}, \bibinfo
  {author} {\bibfnamefont {M.}~\bibnamefont {Kunesch}}, \bibinfo {author}
  {\bibfnamefont {E.~A.}\ \bibnamefont {Lim}},\ and\ \bibinfo {author}
  {\bibfnamefont {S.}~\bibnamefont {Tunyasuvunakool}},\ }\href
  {https://doi.org/10.1088/0264-9381/32/24/245011} {\bibfield  {journal}
  {\bibinfo  {journal} {Class. Quant. Grav.}\ }\textbf {\bibinfo {volume}
  {32}},\ \bibinfo {pages} {245011} (\bibinfo {year} {2015})},\ \Eprint
  {https://arxiv.org/abs/1503.03436} {arXiv:1503.03436 [gr-qc]} \BibitemShut
  {NoStop}%
\end{thebibliography}%

\newpage
\clearpage
\appendix

\section*{Numerical implementation, diagnostic quantities and convergence tests}

We use the $3+1$ formalism \cite{Arnowitt:1962hi,Gourgoulhon:2007ue,York:1978gql,Baumgarte:2002jm,AlcubierreMiguel2008It3N} to foliate the four dimensional spacetime metric into three-dimensional spacelike hypersurfaces indexed by a time coordinate $t$ with line element
\begin{equation}
ds^2=-\alpha^2\,dt^2+\gamma_{ij}(dx^i + \beta^i\,dt)(dx^j + \beta^j\,dt),
\label{ADM}
\end{equation}
where $\alpha$ is called the lapse, $\beta^i$ the shift and $\gamma_{ij}$ the spacial metric. The normal to the hypersurface is given by $n^{\mu} = (-1/\alpha, \beta^i/\alpha)$. 
\begin{table}[b]
\begin{center}
\setlength{\tabcolsep}{1em}
\begin{tabular}{|c|p{3cm}|} 
     \hline \hline
     $d/M$  &   $12.21358$ \\
     $\mbh/M$   &   $0.48847892320123$ \\
     $\vert p_x\vert/M$   &     $5.10846\times 10^{-4}$  \\
     $\vert p_y\vert/M$   &     $8.41746 \times 10^{-2}$   \\
     $\vert p_z\vert/M$   &      $0$ \\
     $T/M$     &      $271.34$ \\
     \hline \hline
\end{tabular}
\end{center}
\caption{Black hole binary initial parameters\footnote{We are grateful to Sebastian Khan for sharing these parameters with us.}. The black holes are initially aligned along the $x$ axis in the $z=0$ plane, with initial momenta $\vec{p}_1 = (-\vert p_x\vert,+\vert p_y\vert,0)$ for the BH with initial position $\vec{r}_1=(d/2,0,0)$ and $\vec{p}_2 = (+\vert p_x\vert ,-\vert p_y\vert,0)$ for the one at $\vec{r}_2=(-d/2,0,0)$.}
\label{table:BH_params}
\end{table}
In addition, we decompose the complex scalar field into two real scalar fields $\varphi_a=(\varphi_1,\varphi_2)$ as $\varphi = \varphi_1 + i\varphi_2$, and reduce the second order Klein-Gordon equation to two first order differential equations
\begin{align}
\partial_t \varphi_a &= \alpha \Pi_a +\beta^i\partial_i \varphi_a \label{eqn:dtphi} ~ , \\ 
\partial_t \Pi_a &= \alpha \gamma^{ij}\partial_i\partial_j \varphi_a +\alpha\left(K\Pi_a -\gamma^{ij}\Gamma^k_{ij}\partial_k \varphi_a -\frac{dV(\varphi)}{d\varphi_a}\right)\nonumber  \\
& + \partial_i \varphi_a \partial^i \alpha + \beta^i\partial_i \Pi_a \label{eqn:dtPi} ~ ,
\end{align}
where $\Pi_a = (\Pi_1, \Pi_2)$ is the conjugate momentum of $\varphi_a$ defined via \eqref{eqn:dtphi}, $K$ is the trace of the extrinsic curvature $K_{ij} = \frac{1}{2\alpha} \left( -\partial_t \gamma_{ij} + D_i\beta_j + D_j \beta_i \right)$, and $D_i$ the covariant derivative associated to the spatial metric $\gamma_{ij}$.\\

\begin{figure}[t]
    \centering
    \includegraphics[width=\linewidth]{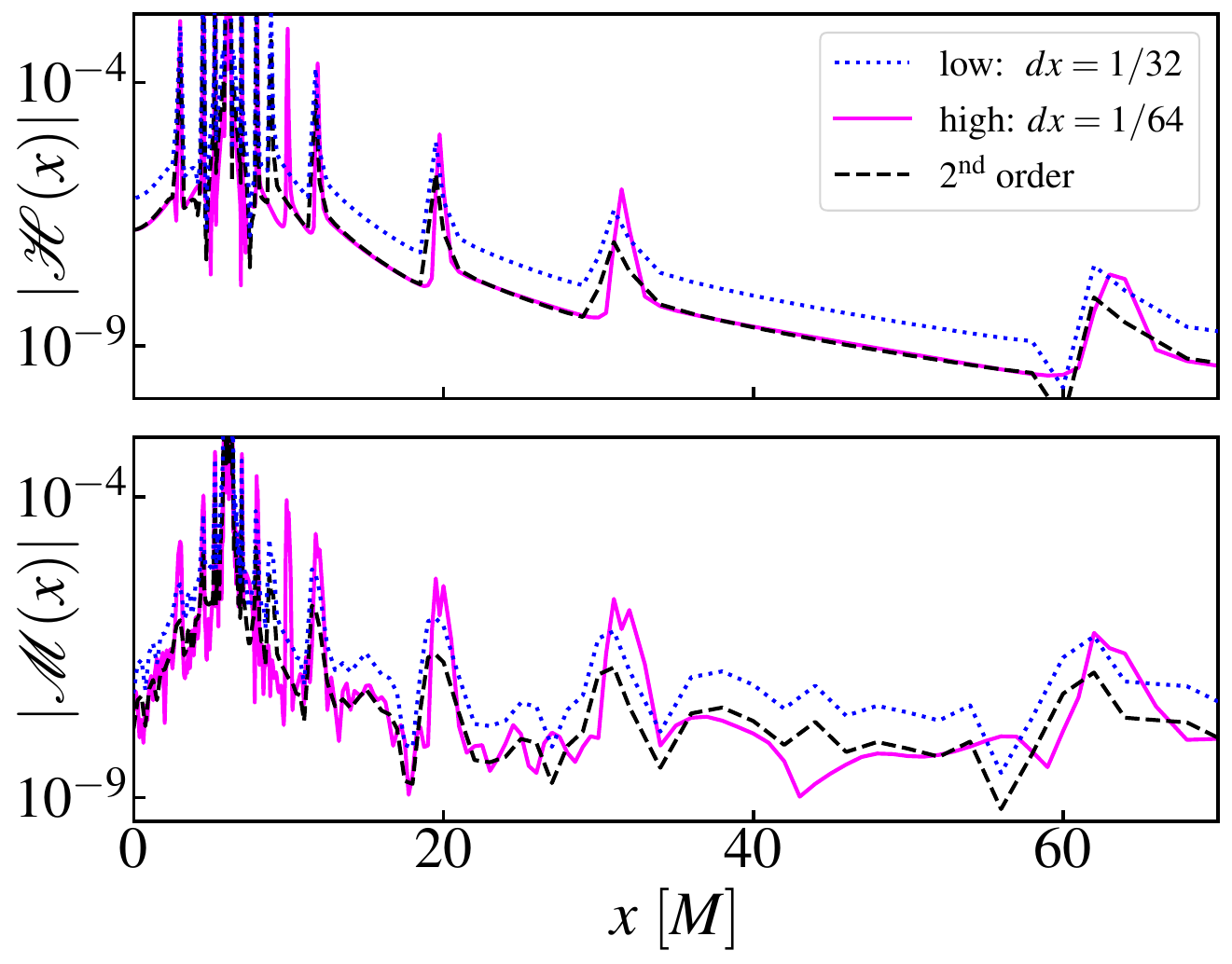}
    \vspace{-20pt}
    \caption{Convergence in the absolute value of the error in the Hamiltonian and momentum constraints for the the fixed orbit initial data (the most sharply peaked, and therefore most difficult, choice of initial scalar field data) at time $t=0$. The error is consistent with the $2^\mathrm{nd}$ order finite difference stencils used in the CTTK solver \cite{Aurrekoetxea:2022mpw}. }
    \label{fig:ICS_Ham_err}
\end{figure}

\begin{figure}[b]
    \centering
    \includegraphics[width=\linewidth]{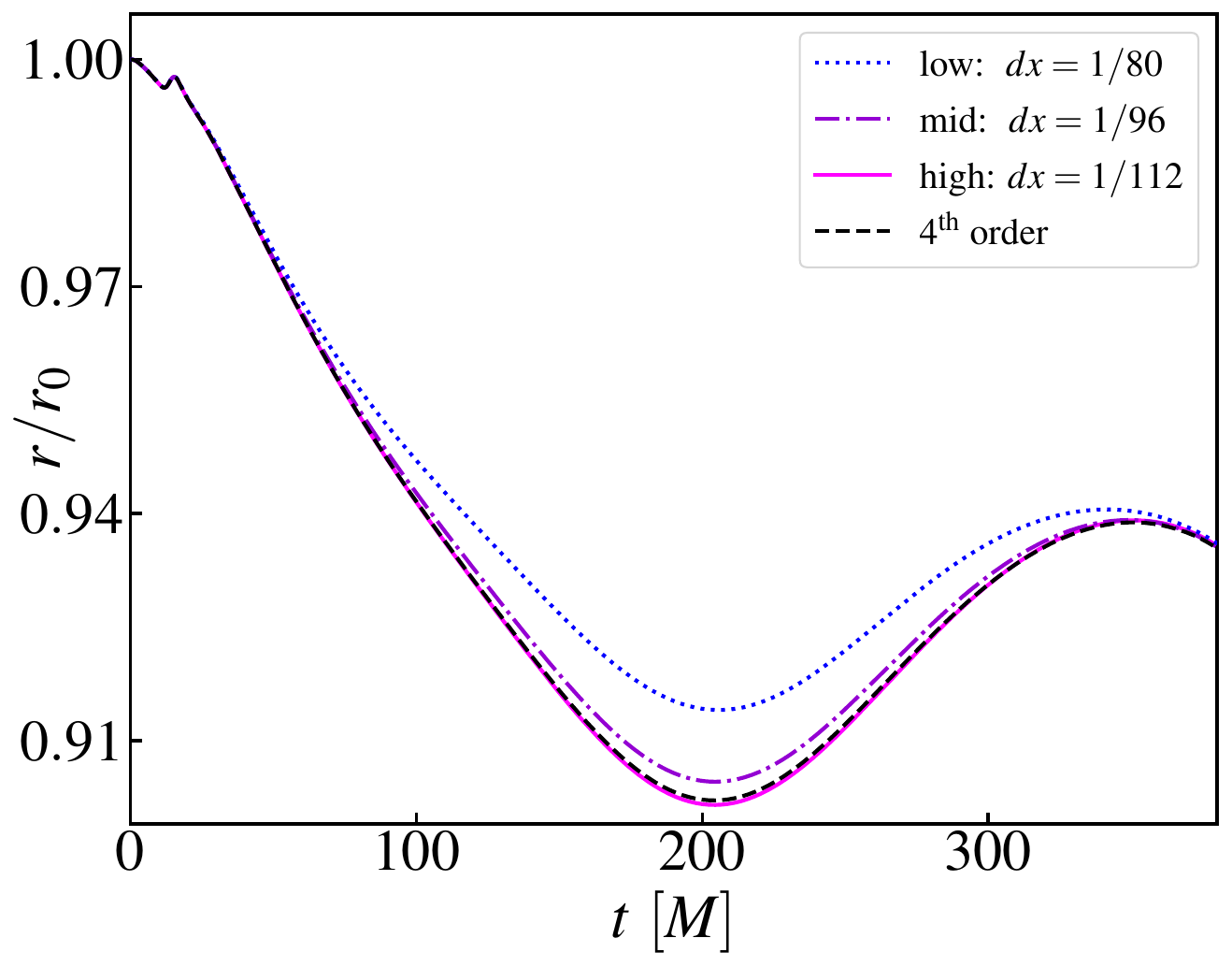}
    \vspace{-20pt}
    \caption{Convergence in radial position of one of the black hole punctures for a BBH with $G=1$ and a scalar cloud of $M_\mathrm{cloud}/M = 10^{-5}$. The black dashed line represents the expected values for $dx = 1/112$ if the simulation had 4th order convergence, which match the measured pink solid line and hence infer that the decrease in the error is consistent with $4^\mathrm{th}$ order stencils used in the evolution code.}
    \label{fig:Relativistic_conv_test}
    \vspace{-5pt}
\end{figure}

For the simple fixed orbit model described in section \ref{sec-fixed_orbits} we only solve equations \eqref{eqn:dtphi} and \eqref{eqn:dtPi} in such a background. We excise (set the evolution variables to zero) within a small region around each black hole centre, inside the horizon, to avoid numerical errors, and evaluate the values of the metric components and their derivatives analytically at each point on the grid. For the relativistic merger, in addition to solving Eqns. \eqref{eqn:dtphi} and \eqref{eqn:dtPi}, we solve the full Einstein equations numerically. For this we use the CCZ4 formalism \cite{Alic:2011gg} with the moving puncture gauge \cite{Bona:1994dr,Baker:2005vv,Campanelli:2005dd,Hannam:2006xw,vanMeter:2006vi}. In both cases we use the open-source numerical relativity code \textsc{grchombo} \cite{Andrade:2021rbd,Clough:2015sqa} with adaptive mesh refinement \cite{Radia:2021smk}. 

We use a simulation box length $L=512M$ and $8$ levels of mesh refinement (See Figs. \ref{fig:ICS_Ham_err} and \ref{fig:Relativistic_conv_test} for convergence tests). Taking advantage of the symmetry in the $xy$ plane we impose reflecting boundary conditions at $z=0$, while for the other boundaries we impose either first order extrapolating boundary conditions (matching the first derivative on the exterior ghost cells to that inside the simulation grid) or Sommerfeld boundary conditions.  

As a diagnostic quantity, we define the effective energy density $\rho_E$ measured by timelike observers as
\begin{equation}
    \rho_E := -\alpha T^0_0 = \alpha \rho - \beta_i S^i,
\end{equation}
with ADM quantities $\rho = -n^{\mu} n^{\mu} T_{\mu\nu}$ and $S^i = -n_a T^{ia}$, such that the total energy in a 3D volume is $E = \int \rho_E \dd V$.

\end{document}